\documentclass[a4paper,11pt]{article}
\pdfoutput=1
\usepackage{jheppubMod}
\usepackage{amsmath,amsfonts,amssymb,array}
\usepackage{bm,bbm,bbm,bbold}
\usepackage{datetime,dsfont}
\usepackage{enumerate}
\usepackage{float,fancyvrb}
\usepackage[T1]{fontenc}
\usepackage{graphicx}
\usepackage{lineno,lipsum,longtable}
\usepackage{multirow,multicol}
\usepackage{subfigure,slashed}
\usepackage{tabu}
\usepackage{ulem}
\usepackage{pdflscape}
\usepackage{soul}
\usepackage[toc,page]{appendix}
\usepackage{url}
\hypersetup{bookmarks=true,unicode=true,pdftoolbar=true,pdfmenubar=true,
  pdffitwindow=false,pdfstartview={FitH},
  pdftitle={Title},pdfauthor={Names},
  pdfsubject={Subject},pdfcreator={Creator},pdfproducer={Producer},
  pdfkeywords={Helicity Polarization}{Monte Carlo Simulations}{Vector Boson Scattering}{Composite Higgs},
  pdfnewwindow=true,colorlinks=true
}

\providecommand{\tabularnewline}{\\}

\newcolumntype{x}[1]{
{\centering\hspace{0pt}}p{#1}}
\def\bea{\begin{eqnarray}}
\def\eea{\end{eqnarray}}
\def\ths{{\theta}}
\def\spa#1.#2{\left\langle#1\,#2\right\rangle}

\newcommand{\GeV}{{\rm ~GeV}}
\newcommand{\TeV}{{\rm ~TeV}}

\newcommand{\invfb}{{\rm ~fb^{-1}}}
\newcommand{\invab}{{\rm ~ab^{-1}}}
\newcommand{\mgFull}{\texttt{MadGraph5\_aMC@NLO}}
\newcommand{\mg}{\texttt{MadGraph}}
\newcommand{\mgFive}{\texttt{MadGraph5}}
\newcommand{\mgamc}{\texttt{mg5amc}}
\newcommand{\mgms}{\texttt{mg5amc+MadSpin}}

\newcommand{\ms}{\texttt{MadSpin}}
\def\eq#1{{eq.~(\ref{#1})}}
\def\eqs#1#2{{eqs.~(\ref{#1})--(\ref{#2})}}
\def\fig#1{{figure~(\ref{#1})}}
\def\sec#1{{section~(\ref{#1})}}

\def\tab#1{{table~(\ref{#1})}}

\def\app#1{{appendix~(\ref{#1})}}

\definecolor{darkgreen}{rgb}{0.0, 0.2, 0.13}
\definecolor{darkmagenta}{rgb}{0.55, 0.0, 0.55}

\title{\Large Automated Predictions from Polarized Matrix Elements}

\author[a]{Diogo Buarque Franzosi,}
\author[b]{Olivier Mattelaer,}
\author[b]{Richard Ruiz,}
\author[c,d]{Sujay Shil}

\affiliation[a]{Department of Physics,
Chalmers University of Technology,\\
Fysikg{\aa}rden 1, 41296 G\"oteborg, Sweden}
\affiliation[b]{Centre for Cosmology, Particle Physics and Phenomenology {\rm (CP3)},\\
Institut de Recherche en Math\'ematique et Physique {\rm (IRMP)},\\
Universit\'e Catholique de Louvain, Chemin du Cyclotron, B-1348 Louvain-la-neuve, Belgium}
\affiliation[c]{Institute of Physics, Sachivalaya Marg, Bhubaneswar, Odisha 751005, India}
\affiliation[d]{Homi Bhabha National Institute, Training School Complex,\\ Anushakti Nagar, Mumbai 400085, India}

\emailAdd{buarque@chalmers.se}
\emailAdd{olivier.mattelaer@uclouvain.be}
\emailAdd{richard.ruiz@uclouvain.be}
\emailAdd{sujayshil1@gmail.com}

\abstract{
The anticipated experimental resolution and data cache of the High Luminosity Large Hadron Collider
will enable precision investigations of polarization in multiboson processes.
This includes, for the first time, vector boson scattering.
To facilitate such studies, we report the  
automation of polarized matrix element computations
in the publicly available Monte Carlo tool suite, MadGraph5\_aMC@NLO.
This enables scattering and decay simulations involving helicity-polarized asymptotic or intermediate states, preserving both spin-correlation and off-shell effects.
As demonstrations of the method, we investigate the leading order production and decay of polarized weak gauge bosons in the process $pp \to j j W^+_\lambda W^-_{\lambda'}$,
with helicity eigenstates $(\lambda,\lambda')$ defined in various reference frames.
We consider the Standard Model at both $\mathcal{O}(\alpha^4)$ and $\mathcal{O}(\alpha^2 \alpha_s^2)$ as well as a benchmark composite Higgs scenario.
We report good agreement with polarization studies based on the On-Shell Projection (OSP) technique.
Future capabilities are discussed.
}

\preprint{CP3-19-52, VBSCan-PUB-09-19, MCnet-19-24, IP/BBSR/2019-10}
\arxivnumber{1912.01725}
\begin{document}
\maketitle
\setcounter{page}{2}
\flushbottom

%%=====================================================================
% !TEX root = mgPolarization_main.tex
\section{Introduction}\label{sec:intro}

The production of asymmetrically polarized fermions and Weak gauge bosons in high energy scattering processes 
is a defining feature of the Standard Model of particle physics (SM)~\cite{Lee:1977yc,Lee:1977eg,Chanowitz:1985hj,Chanowitz:1984ne,Aguilar-Saavedra:2017nik}.
It is also a key indicator of many new physics models that address experimental and theoretical challenges to the SM,
a collection that includes extended gauge theories~\cite{Pati:1974yy,Mohapatra:1974hk,Mohapatra:1974gc,Senjanovic:1975rk,Senjanovic:1978ev}, 
models with extra spatial dimensions~\cite{ArkaniHamed:1998rs,ArkaniHamed:1998nn,Randall:1999ee,Appelquist:2000nn}, 
supersymmetry~\cite{Nilles:1983ge,Haber:1984rc},
as well as composite Higgs (CH)
models~\cite{Kaplan:1983fs,Kaplan:1983sm,Georgi:1984af,Dugan:1984hq,Contino:2003ve,Agashe:2004rs,Contino:2006qr,Agashe:2006at,Bellazzini:2014yua,Panico:2015jxa}.
Even in the decoupling limit~\cite{Appelquist:1974tg} of these scenarios,
their existence generically manifest as polarization-inducing, higher-dimension operators of an effective field theory (EFT).
Consequently, searches for the anomalous polarization of SM particles 
at the Large Hadron Collider (LHC) and {future experiments}~\cite{Baer:2013cma,Boehm:2017nrl,CEPCStudyGroup:2018ghi,Abada:2019lih,Boehm:2019yit}
are motivated as their discovery would have profound implications on our understanding of nature.

With nearly $\mathcal{L}=140\invfb$ of $\sqrt{s}=13\TeV$ collision data after Run II, the LHC experiments have made significant headway in investigating 
rare processes that are sensitive to anomalous chiral couplings, and hence anomalous helicity polarizations.
Among these special channels are associated single top quark production modes~\cite{Aaboud:2017yqf,Aaboud:2017ylb,Sirunyan:2018zgs,Aaboud:2019pkc}, 
EW diboson~\cite{Khachatryan:2015sga,Sirunyan:2019bez,Aaboud:2019gxl,Aaboud:2019nkz} 
and triboson production~\cite{Aad:2019udh,CMS:2019mpq}, 
and, for the first time, EW vector boson scattering (VBS)~\cite{Sirunyan:2017ret,Sirunyan:2017fvv,Aaboud:2018ddq,Sirunyan:2019ksz,Aaboud:2019nmv}.
At the High Luminosity-Large Hadron Collider (HL-LHC) in particular, 
the anticipated experimental resolution and $\mathcal{L}=3-5\invab$ data cache will allow these processes to be measured with unprecedented precision.
For quantitative assessments of the HL-LHC's potential, see Refs.~\cite{CMSCollaboration:2015zni,Atlas:2019qfx} and references therein.

An impeding factor to fully utilizing these data, however, are the limited number fully differential, SM and beyond the SM (BSM) predictions for polarization observables.
While incredible efforts are underway to develop precise predictions up to next-to-next-to-leading order (NNLO) in QCD and/or next-to-leading order (NLO) in EW,
these are largely restricted to only a handful of SM processes~\cite{Melia:2011tj,Baglio:2018rcu}.
Likewise, the direct simulation of polarized parton scattering in hadron collisions using public Monte Carlo (MC) tools is found almost exclusively
at leading order (LO) and again restricted to certain scattering topologies~\cite{Ballestrero:2007xq}.
Though the availability of such public tools has led to a number of complimentary investigations 
on the production of polarized EW bosons at the LHC~\cite{Ballestrero:2007xq,Ballestrero:2011pe,Ballestrero:2017bxn,Ballestrero:2018anz,Ballestrero:2019qoy,Baglio:2019nmc,Rahaman:2019lab}.

In the present work, we report the development of a scheme to model polarized parton scattering in hadron collisions and its implementation into the publicly 
available\footnote{Available from version 2.7.0 at the URL \href{https://launchpad.net/mg5amcnlo}{launchpad.net/mg5amcnlo}.}
event generator~\mgFull~{(dubbed \mgamc)}. 
By ``polarized parton scattering'' we specifically mean $2\to n$-scattering and $1\to n$-decay processes that are determined from polarized matrix elements (MEs).
That is to say, matrix elements where some or all external states are in a definite helicity eigenstates and where spin-averaging or spin-summing is truncated or not present.
(For simplicity, we refer to all short-distance particles, including massive, colorless EW states, as partons throughout this work.)
The method enables the LO simulation of tree-level scattering and decay processes involving external states in fixed helicity eigenstates in an arbitrary reference frame. 
This includes massless QCD partons, heavy quarks, all leptons, the EW gauge sector, and states up to spin 3/2 and 2.
When using the narrow width approximation (NWA), 
spin correlations of decaying polarized resonances are maintained through the decomposition of fermionic and bosonic propagators 
into their respective transverse, longitudinal, and auxiliary~(``scalar'') components; the last of which necessarily vanishes in the on-shell limit.
Extension to new physics scenarios is achieved when used with {\texttt{Universal FeynRules Object} (UFO)~\cite{Christensen:2008py,Degrande:2011ua,Alloul:2013bka} libraries.}

Our work continues as follows:
In \sec{sec:setup} we summarize our computational setup.
In \sec{sec:polarDef} we describe our 
formalism for constructing polarized MEs
and its implementation into the program~\mgamc.
(More technical, implementation details and checks are reported in \app{sec:codingBits}.)
We then investigate in \sec{sec:vbs}  
the production and decay of polarized $W^+W^-$ pairs from EW (\sec{sec:vbs_ew}) and mixed EW-QCD (\sec{sec:vbs_qcd}) processes $pp\to j j  W^+W^-$ 
in the SM as well as a benchmark CH scenario (\sec{sec:vbs_bsm}).
There we compare our methods to the so-called on-shell projection (OSP) technique~\cite{Aeppli:1993cb,Aeppli:1993rs,Denner:2000bj} 
and report good agreement with past studies~\cite{Ballestrero:2007xq,Ballestrero:2011pe,Ballestrero:2017bxn,Ballestrero:2018anz,Ballestrero:2019qoy}.
In Sec.~\ref{sec:conclusions} we summarize our results.

Throughout this study, we focus on EW and QCD processes at LO.
{We also report that fully differential event simulation up to NLO in QCD with parton shower matching is also possible for processes involving polarized, color-singlet final states.}
However, we report such investigations in a companion paper~\cite{COCITE}.

%%=====================================================================
% !TEX root = mgPolarization_main.tex

\section{Computational Setup}\label{sec:setup}
In this section, we briefly summarize the computational framework used in our study.
In particular, we describe the components of our MC tool chain and their relevant tunings needed for reproducibility.
Standard Model and CH model input parameters used in our case studies are also listed.
While we heavily utilize and partially expand on the MC suite~\mgamc, 
a full characterization of it is outside the scope of this work and is available elsewhere~\cite{Alwall:2014hca,Frederix:2018nkq}.
The description of our method for constructing polarized MEs in hadron collisions and its implementation are given in \sec{sec:polarDef} and \app{sec:codingBits}.

\subsection*{Monte Carlo Framework and Tuning}\label{sec:setup_mc}
We simulate parton scattering with
polarized and unpolarized MEs in $\sqrt{s}=13\TeV$ $pp$ collisions at LO {in perturbation theory}
using the software suite~\mgamc~\cite{Alwall:2014hca}.
Working in the so-called \texttt{HELAS} basis~\cite{Murayama:1992gi}, 
tree-level matrix elements are evaluated numerically using helicity amplitudes that are capable of handling massive states~\cite{Maltoni:2002qb,deAquino:2011ub},
and with QCD color decomposition based on color flow~\cite{Maltoni:2002mq}.
Decays of unstable, resonant states are handled using the spin-correlated NWA, as implemented in \texttt{MadSpin}~\cite{Artoisenet:2012st}.

\subsection*{Standard Model Inputs}\label{sec:setup_inputs_SM}
For SM inputs, we work in the $n_f=4$ massless quark scheme, approximate the Cabbibo-Kobayashi-Masakawa (CKM) matrix to be diagonal with unit {entries}, and take
\begin{eqnarray}
M_W = 80.419\GeV, \quad 
M_Z = 91.188\GeV, \quad
G_F = 1.16639\times10^{-5}{\GeV}^2.
\end{eqnarray}
In \mgamc, this corresponds to importing the internal \texttt{sm} model library.
For SM case studies, we use the NNPDF 2.3 LO parton distribution function (PDF) set with $\alpha_s(\mu)=0.119$ (\texttt{lhaid=246800})~\cite{Ball:2013hta}.
 We set our collinear factorization $(\mu_f)$ scale to the \mgFive~default.

\subsection*{Composite Higgs Inputs}\label{sec:setup_inputs_CH}
Besides the SM, we also investigate VBS in the context of a CH scenario.
For that we use the Higgs Characterization Model libraries of Ref.~\cite{Artoisenet:2013puc}, which provides a general parametrization of the Higgs boson's spin and couplings.
We limit ourselves to a somewhat generic CH situation, where the couplings of a SM-like Higgs are rescaled by an overall factor.
In Ref.~\cite{Artoisenet:2013puc}, this factor is identified as \texttt{kSM}  with \texttt{kSM=1} denoting the SM limit. 
Throughout this work, we fix the SM (or SM-like) Higgs mass to $m_H = 125\GeV$.
All the other parameters are set as described above with the exception of using the NNPDF 3.1 NLO+LUXqed (\texttt{lhaid=324900}) PDF set~\cite{Bertone:2017bme},
with PDF running handled using LHAPDF v6.1.6~\cite{Buckley:2014ana}.

%%=====================================================================
% !TEX root = mgPolarization_main.tex
\section{Parton and Hadron Scattering with Polarized Matrix Elements}\label{sec:polarDef}

We now describe the scattering formalism  underlying our implementation of polarized parton scattering into the event generator \mgamc.
We start in \sec{sec:polarDef_scattering} with building meaningful definitions 
of parton\footnote{To reiterate: throughout this text, we use the term ``parton'' for any external, short-distance particle, including massive, colorless EW states.} 
scattering with polarized MEs in unpolarized hadron collisions,
noting instances of reference frame-dependence that are not usually present in standard MC computations.
In \sec{sec:polarDef_decay}, we describe our treatment of decaying polarized resonances.
Additional details related to technical implementation and usage are reported in \app{sec:codingBits}.
Physics demonstrations are deferred to \sec{sec:vbs}.

\subsection{Scattering Formalism with Polarized Matrix Elements}\label{sec:polarDef_scattering}
\subsubsection*{Preliminaries: Scattering with Unpolarized Matrix Elements}
In unpolarized proton collisions,
the scattering observable $\tilde{\mathcal{O}}$ built from the $n$-body final state $\mathcal{B}$ 
at  momentum transfers $(\sqrt{Q^2})$ much larger than the nonperturbative QCD scale $(\Lambda_{\rm NP})$
is governed by the Collinear Factorization Theorem~\cite{Collins:1984kg,Collins:1985ue,Stewart:2009yx,Collins:2011zzd,Becher:2012qa},
\begin{eqnarray}
\cfrac{d\sigma(pp\to\mathcal{B}+X)}{d\tilde{\mathcal{O}}}\Bigg\vert_{\tilde{\mathcal{O}}=\tilde{\mathcal{O}}_0} &=& f \otimes f \otimes \Delta \otimes 
\cfrac{d\hat{\sigma}}{d\tilde{\mathcal{O}}}\Bigg\vert_{\tilde{\mathcal{O}}=\tilde{\mathcal{O}}_0} 
 + \mathcal{O}\left(\frac{\Lambda^t_{\rm NP}}{Q^{t+2}}\right) 
\label{eq:factThemShort}\\
&=& 
 \sum_{i,j=q,g,\gamma} 
\int_{\tau_0}^1 d\tau ~\int_{\tau}^1 \frac{d\xi_1}{\xi_1} ~\int_{\tau/\xi_1}^1 \frac{dz}{z} 
\frac{1}{(1+\delta_{ij})}
\nonumber\\
& &
\times ~ \left[ f_{i/p}(\xi_1,\mu_f)f_{j/p}(\xi_2,\mu_f)+ (1\leftrightarrow2)\right] ~\times ~ \Delta_{ij}(z,\mu_f,\mu_r,\mu_s)
\nonumber\\
& &
\times ~ 
\cfrac{d\hat{\sigma}(ij\to \mathcal{B}; \{Q^2,s,\mu_f,\mu_r,\mu_s\})}{d\tilde{\mathcal{O}}}
\Bigg\vert_{\tilde{\mathcal{O}}=\tilde{\mathcal{O}}_0} 
~+~ \mathcal{O}\left(\frac{\Lambda^t_{\rm NP}}{Q^{t+2}}\right). ~\quad
\label{eq:factThemLong}
\end{eqnarray}

For protons $m=1,2$, with 4-momenta $P_m=(\sqrt{s}/2)(1,0,0,\pm1)$,
the above stipulates that inclusive, hadron-level observables $(d\sigma/d\tilde{\mathcal{O}})$ that are functions of external momenta, i.e., $\tilde{\mathcal{O}} = g(p_1,\dots,p_n)$, 
can be expressed as the product of probabilities (convolution) for 
(a) finding partons $i$ and $j\in\{q,\overline{q},g,\gamma\}$,  with $q\in\{u,d,c,s\}$, in proton $m$, which is described by $f$;
(b) the renormalization group (RG) scale evolution of $i$ and $j$ from a proton to the hard scattering process, described by $\Delta$;
and (c) the exclusive, parton-level hard scattering process $ij \to \mathcal{B}$, governed by $d\hat{\sigma}/d\tilde{\mathcal{O}}$.
Here, $\tau=Q^2/s$ is the hard threshold at which $ij\to\mathcal{B}$ proceeds,
and for $\tau < \tau_0=\min\{Q^2\}/s$, the production of $\mathcal{B}$ is kinematically forbidden.

More specifically, $f_{k/p}(\xi_m,\mu_f)$ is the collinear PDF,
which for momentum fraction $0 < \xi_m < 1$, represents the likelihood of parton $k$ in proton $m$ possessing a momentum  $p_k = (\xi_m E_m,0,0,\pm\xi_m E_m)$.
Using the DGLAP evolution equations~\cite{Gribov:1972ri, Dokshitzer:1977sg,Altarelli:1977zs}, $f$ can be RG-evolved to the collinear cutoff / factorization scale $\mu_f$.
This accounts for (resums) an arbitrary number of initial-state emissions that are produced in association with $k$ and carry a relative transverse momentum $p_T<\mu_f$.
Factors of $(1\leftrightarrow2)$ and $(1+\delta_{ij})$ account for identical beam and identical initial parton symmetrization.

The Sudakov factor $\Delta_{ij}(z,\mu_f,\mu_r,\mu_s)$ accounts for (resums) soft and/or collinear emissions of massless partons 
carrying a momentum fraction $z = Q^2 / \xi_1\xi_2 s$, away from the $(ij)$ system prior to the hard $ij\to \mathcal{B}$ scattering process.
Through various RG evolutions between $\mu_f$, the UV renormalization scale $\mu_r$, and the Sudakov cutoff/ factorization scale $\mu_s$, 
 $\Delta$ ensures that \eq{eq:factThemShort} remains RG scale-independent~\cite{Contopanagos:1996nh}.
 In our notation, $\Delta$ additionally records $i\to k$ parton depletion and $k\to i$ parton buildup for hard scattering partons $i$ and $j$.
 In general-purpose, MC event generators, $\Delta_{ij}(z)\approx \delta_{ij}\delta(1-z) + \mathcal{O}(\alpha_s)$ can be identified as the parton shower
 and principally models collinear parton emissions, though developments to expand this domain are ongoing~\cite{Platzer:2018pmd,Gieseke:2018gff,Forshaw:2019ver}.

When built from the $n$-body final state $\mathcal{B}$, parton-level scattering observables $d\hat{\sigma}/d\tilde{\mathcal{O}}$
are derivable from the fully differentiated scattering rate $d\hat{\sigma}/dPS_n$,
\begin{eqnarray}
\cfrac{d\hat{\sigma}(ij\to \mathcal{B})}{d\tilde{\mathcal{O}}}
\Bigg\vert_{\tilde{\mathcal{O}}=\tilde{\mathcal{O}}_0} 
&=&
~ \int dPS_{n} ~ \delta(\tilde{\mathcal{O}}-\tilde{\mathcal{O}}_0) ~  \frac{d\hat{\sigma}(ij\to \mathcal{B})}{dPS_n},
\label{eq:diffXSec}
\end{eqnarray}
 where $dPS_n$ is the separately Lorentz-invariant, $n$-body phase space measure given by
  \begin{eqnarray}
dPS_n(p_i+p_j; p_{f=1},\dots,p_{f=n}) = (2\pi)^4 \delta^4\left(p_i + p_j - \sum_{f=1}^n p_f\right)\prod^{n}_{f} \cfrac{d^3 p_f}{(2\pi)^3 2E_f}.
\label{eq:dPS}
\end{eqnarray}
Eq.~(\ref{eq:diffXSec})  can be expressed in terms of perturbative matrix elements by the usual expression:
\begin{eqnarray}
 \frac{d\hat{\sigma}(ij\to \mathcal{B})}{dPS_n} = \frac{1}{2Q^2}\frac{1}{(2s_i+1)(2s_j+1)N_c^i N_c^j} \sum_{\rm dof} \vert \mathcal{M}(ij\to \mathcal{B})\vert^2.
 \label{eq:partonScatt_unpol}
 \end{eqnarray}
Here  $s_k=1/2$ and $N_c^k$ are the helicity and SU$(3)_c$ color symmetrization factors for massless parton $k=i,j$.
For massive spin-1 states, $s_k=1$ and the $2Q^2$ flux factor is scaled by the kinematic K\"allen function.
After summing over all external helicity and color polarizations (dof), $\sum\vert\mathcal{M}\vert^2$ is the (squared) Lorentz-invariant matrix describing $ij \to \mathcal{B}$ scattering.
The total parton-level $ij\to \mathcal{B}$ cross section $(\hat{\sigma})$ is recoverable upon integration over $dPS_n$
\begin{eqnarray}
\hat{\sigma} = \int d\hat{\sigma} &=& \int dPS_n~\frac{d\hat{\sigma}}{dPS_n}.
\end{eqnarray}

While \eq{eq:factThemShort} is formally proved for only a handful of processes~\cite{Collins:2011zzd},
we make the strong but standard assumption that the relation, with appropriate modifications, 
broadly holds for other processes, including heavy quark and multijet production.
For conciseness, we omit insertion of fragmentation functions $(J)$ into \eq{eq:factThemShort} for exclusive such final states.

\subsubsection*{Scattering Helicity-Polarized Partons at the Parton Level}
In building the expression for unpolarized parton scattering in \eq{eq:partonScatt_unpol},
one takes the crucial step of 
averaging over discrete spacetime and internal quantum numbers
 for initial-state (IS) partons but only sum  discrete degrees of freedom (dof) for final-state (FS) partons.
This leads to the familiar IS dof-averaged and FS dof-summed, squared matrix element
\begin{eqnarray}
\overline{ \vert \mathcal{M}(ij\to \mathcal{B})\vert^2} ~\equiv~  \frac{1}{(2s_i+1)(2s_j+1)N_c^i N_c^j} \sum_{\rm dof} \vert \mathcal{M}(ij\to \mathcal{B})\vert^2.
\label{eq:ME_unpol}
\end{eqnarray}
In dropping all summations over all external helicity eigenstates and fixing the helicities of all external partons in the $ij\to\mathcal{B}$ process, which we denote generically as
\begin{equation}
i_\lambda + j_{\lambda'} \to \mathcal{B}_{\tilde{\lambda}},
\label{eq:polPartonScat}
\end{equation}
with $\tilde{\lambda}$ representing the set of $n$ helicity eigenstates,
one can define 
the totally helicity-polarized, IS color-averaged and FS 
color-summed
squared matrix element as
\begin{eqnarray}
\overline{ \vert \mathcal{M}(i_\lambda j_{\lambda'}\to \mathcal{B}_{\tilde{\lambda}})\vert^2} 
~\equiv~  \frac{1}{N_c^{i_\lambda} N_c^{j_{\lambda'}}} \sum_{\rm color} \vert \mathcal{M}(i_\lambda j_{\lambda'}\to \mathcal{B}_{\tilde{\lambda}})\vert^2.
\label{eq:ME_pol}
\end{eqnarray}
Generically, we use the terms ``polarized matrix elements'' and ``helicity-polarized matrix elements'' interchangeably 
to mean $\mathcal{M}(i_\lambda j_{\lambda'}\to \mathcal{B}_{\tilde{\lambda}})$.
The label ``totally helicity-polarized'' qualifies that all external partons are in a fixed helicity state,
as oppose to instances where only a subset of external partons are in a fixed helicity state.
Such configurations corresponds to ``partially helicity-polarized'' matrix elements and can be constructed analogously.
For example: for an unpolarized $i,j$, and a totally polarized $\mathcal{B}$, 
the dof-averaged and color-summed, squared matrix element is
\begin{eqnarray}
\overline{ \vert \mathcal{M}(i j\to \mathcal{B}_{\tilde{\lambda}})\vert^2} ~\equiv~  \frac{1}{(2s_i+1)(2s_j+1)N_c^{i_\lambda} N_c^{j_{\lambda'}}} 
\sum_{\rm color, \lambda,\lambda'} \vert \mathcal{M}(i_\lambda j_{\lambda'} \to \mathcal{B}_{\tilde{\lambda}})\vert^2.
\label{eq:ME_mixedpol}
\end{eqnarray}
Unambiguously, \eqs{eq:ME_unpol}{eq:ME_mixedpol} are related by reintroducing helicity averaging / summing:
\begin{eqnarray}
\overline{ \vert \mathcal{M}(ij\to \mathcal{B})\vert^2} &=&  \frac{1}{(2s_i+1)(2s_j+1)} \sum_{\rm \lambda,\lambda',\tilde{\lambda}} \overline{ \vert \mathcal{M}(i_\lambda j_{\lambda'}\to \mathcal{B}_{\tilde{\lambda}})\vert^2}
\\
&=& \sum_{\tilde{\lambda}} \overline{ \vert \mathcal{M}(ij\to \mathcal{B}_{\tilde{\lambda}})\vert^2}
\label{eq:ME_relationships}
\end{eqnarray}
Other configurations, such as with totally or partially polarized IS partons with unpolarized FS partons, can be also constructed so long as helicity averaging factors are consistently accounted.
Subsequently, these permutations need not be discussed further.

Given a definition for squared matrix elements describing 
parton scattering with fixed, external helicity polarizations as in ~\eq{eq:ME_pol},
 one can construct scattering observables as done for unpolarized parton scattering.
 To do this, we promote the fully differentiated scattering rate $d\hat{\sigma}/dPS_n$ for unpolarized parton scattering in \eq{eq:partonScatt_unpol}
 by using instead the totally polarized squared matrix elements in \eq{eq:ME_pol}.
 Explicitly, the fully differentiated scattering rate for the totally polarized partonic process $i_\lambda + j_{\lambda'} \to \mathcal{B}_{\tilde{\lambda}}$ is
 \begin{eqnarray}
 \frac{d\hat{\sigma}(i_\lambda + j_{\lambda'} \to \mathcal{B}_{\tilde{\lambda}})}{dPS_n} &=&
\frac{1}{2Q^2}\overline{ \vert \mathcal{M}(i_\lambda j_{\lambda'}\to \mathcal{B}_{\tilde{\lambda}})\vert^2} 
 \\
 &=& \frac{1}{2Q^2} \frac{1}{N_c^{i_\lambda} N_c^{j_{\lambda'}}} \sum_{\rm color} \vert \mathcal{M}(i_\lambda j_{\lambda'}\to \mathcal{B}_{\tilde{\lambda}})\vert^2.
 \label{eq:partonScatt_pol}
 \end{eqnarray}
 Likewise, for unpolarized IS partons but a totally polarized FS $\mathcal{B}$,  
 the fully differentiated scattering rate is given by
 \begin{eqnarray}
 \frac{d\hat{\sigma}(i j\to \mathcal{B}_{\tilde{\lambda}})}{dPS_n} &=&
\frac{1}{2Q^2}\overline{ \vert \mathcal{M}(i j\to \mathcal{B}_{\tilde{\lambda}})\vert^2}
 \\
 &=& \frac{1}{2Q^2}  \frac{1}{(2s_i+1)(2s_j+1)N_c^{i_\lambda} N_c^{j_{\lambda'}}} 
\sum_{\rm color, \lambda,\lambda'} \vert \mathcal{M}(i_\lambda j_{\lambda'} \to \mathcal{B}_{\tilde{\lambda}})\vert^2.
 \label{eq:partonScatt_mixedpol}
 \end{eqnarray}
It is clear that the relation among \eq{eq:partonScatt_pol}, \eq{eq:partonScatt_mixedpol}, and the unpolarized case proceeds identically to 
that established in \eq{eq:ME_relationships}.
Finally, upon phase space integration one obtains parton-level, total cross sections and differential observables as defined in \eq{eq:diffXSec}.

Unlike \eq{eq:ME_unpol}, the polarized expressions of \eq{eq:ME_pol} and \eq{eq:ME_mixedpol} are \textit{not} guaranteed to be Lorentz-invariant quantities as they technically possesses uncontracted Lorentz indices.
To be precise: in standard construction of helicity amplitudes within the SM, e.g.~Ref.~\cite{Weinberg:1995mt},
spin-$1/2$ spinors $u^m(p,\lambda_f),~v^m(p,\lambda_{\bar{f}})$, spin-$1$ polarization vectors $\varepsilon^\rho(p,\lambda_V)$, and their conjugations, each carry two 
indices\footnote{The two do not have a one-to-one correspondence. For example: for a scalar field $\phi$, the field operator $\partial_\mu\phi$ is in a vector representation but possesses only a single (trivial) helicity state.}: 
one to denote a component within their Lorentz group representation, e.g., $m=1,\dots,4$, and $\rho=0,\dots,3$,
and
a second to denote their helicity polarization, e.g., $\lambda_f, \lambda_{\bar{f}}=\pm1$ and $\lambda_V=0,\pm1$.
(Such statements hold also for tensor fields $h_{\rho\sigma}$, etc.,  but need not to be discussed further as the conclusions are the same.)
The Lagrangian-based formulation of quantum field theory, and hence Feynman rules, 
leads to scattering amplitudes that are manifestly reference frame-independent for only the first type of index when all such indices are contracted.
That is to say, when all $u,v$ spinors are acted upon by $\bar{u},\bar{v}$, and all $\varepsilon^\mu,\partial^\mu,\sigma^{\mu\nu},\dots$ are acted upon by $\varepsilon_\mu,\partial_\mu,\sigma_{\mu\nu},\dots$,
in some appropriate permutation.
Lorentz invariance is only achieved when all indices of the first type are contracted and all indices of the second type are summed.
Since helicity polarizations are reference frame-dependent, one must stipulate a reference frame when using \eq{eq:ME_pol} or its variations.
While conceptually simple, for MC event generators this introduces a technical restriction on exploiting Lorentz invariance 
that is often used in computing matrix elements for unpolarized parton scattering.

In~\mgamc, this technicality is managed by exploiting the separately Lorentz-invariant nature of the phase space volume measure given in \eq{eq:dPS}.
To summarize: 
A point in phase space is first generated for computing a polarized ME in the same manner as for an unpolarized ME.
External momenta are then Lorentz boosted to a definite reference stipulated by the user or to a default option; see \app{sec:codingBits_evtgen}.
Helicity amplitudes are then evaluated numerically in this frame.
Upon completion of phase space integration, weighted or unweighted events are written to file in standard Les Houches format~\cite{Alwall:2006yp}.
In the present implementation, phase space cuts on momenta are applied in the partonic c.m.~frame, 
{with the exception of rapidity cuts, which are applied in the lab~frame}{\footnote{Note that most of the observables defined in \texttt{run\_card.dat} are invariant under boosts along the $z$-direction and are thus the same in the lab frame or the partonic c.m. frame. }}.
In principle, it is also possible to apply phase space cuts in an reconstructable reference frame in \mgamc~using the \texttt{dummy\_fct.f}
capabilities.

\subsubsection*{Scattering Helicity-Polarized Partons at the Hadron Level}
To finally define a version of polarized parton scattering in unpolarized hadron collisions that can be implemented in MC event generators, 
we argue that the Factorization Theorem of \eq{eq:factThemShort} can be extended as desired.
While a full, field-theoretic derivation is beyond this work, 
principle tenets are already established\footnote{More specifically, established for only a few inclusive processes at leading power approximations.
We assume consistently that the theorem holds for other processes in which perturbative QCD is valid.}
 in Ref.~\cite{Collins:2011zzd} and references therein.

We start by noting that PDFs describing unpolarized partons $k$ out of unpolarized hadrons $\mathcal{P}$
can be defined to all orders in $\alpha_s$ as a transition amplitude given by~\cite{Collins:1985ue,Collins:2011zzd}:
\begin{equation}
f_{k/p}(\xi) = \frac{1}{2\pi}\int_{-\infty}^{\infty} dt ~ e^{-ix\cdot p}  \langle \mathcal{P}(P)  \vert \hat{\mathcal{O}}_{k\mathcal{P}}(x) \vert \mathcal{P}(P) \rangle.
\label{eq:unpolPDFdef}
\end{equation}
Here, $\hat{\mathcal{O}}_{k\mathcal{P}}(x)$ denotes the composite field operator that extracts parton $k$ with momentum $p_k = \xi P$ from $\mathcal{P}$. 
The integral is a Fourier integral that 	takes the amplitude for $\hat{\mathcal{O}}_{k\mathcal{P}}(x)$ into momentum space.
As \eq{eq:unpolPDFdef} is defined at the momentum transfer scale $\Lambda_{\rm NP}$, 
a first principle determination of $\langle \hat{\mathcal{O}}_{k\mathcal{P}}\rangle$, and hence $f$, is not possible with perturbative methods.
That said, in real scattering experiments, massless, initial-state partons are very nearly on their mass shell, indicating that the underlying dynamics of \eq{eq:unpolPDFdef}
occur on a different time scale, $\tau\sim 1/\Lambda_{\rm NP}$, and not on the time scale of the hard process, $\tau\sim 1/Q$.
Hence, the dynamics of IS partons $i$ and $j$ are effectively decoupled from the hard scattering process $ij\to\mathcal{B}$.
Thus, $f$ are factorizable, i.e., can be written as \eq{eq:factThemShort}, up to corrections of the order $\mathcal{O}(\Lambda^t_{\rm NP}/Q^{t+2})$, for $t>0$.
Since $f$ are factorizable, they can be RG-evolved~\cite{Contopanagos:1996nh} to a cutoff / factorization scale $\mu_f \gg  \Lambda_{\rm NP}$, using perturbative methods (DGLAP evolution),
and subsequently entered into real scattering computations.

It follows then that IS partons can be approximated as asymptotic states in definite helicity eigenstates.
For massless partons, this becomes a matter of splitting the operator $\hat{\mathcal{O}}_{k\mathcal{P}}(x)$ for unpolarized $k$
into two orthogonal pieces using chiral projection operators:
\begin{equation}
\hat{\mathcal{O}}_{kP}(x) = \hat{\mathcal{O}}_{k_L \mathcal{P}}(x) + \hat{\mathcal{O}}_{k_R\mathcal{P}}(x).
\end{equation}
Here, $\hat{\mathcal{O}}_{k_\lambda \mathcal{P}}$ is the operator that extracts $k$ with helicity $\lambda$ from (unpolarized) $\mathcal{P}$.
Such partitioning is possible since chiral and helicity eigenstates are identical for massless particles.
Consistently, one can decompose the PDF  in \eq{eq:unpolPDFdef} into {left-handed} (LH) and {right-handed} (RH) helicity components:
\begin{eqnarray}
f_{k/p}(\xi) 		&=& f_{k_L/p}(\xi) + f_{k_R/p}(\xi), \quad\text{with} \\
f_{k_\lambda/p}(\xi) 	&\equiv& \frac{1}{2\pi}\int_{-\infty}^{\infty} dt ~ e^{-ix\cdot p}  \langle \mathcal{P}(P)  \vert \hat{\mathcal{O}}_{k_\lambda\mathcal{P}}(x) \vert \mathcal{P}(P) \rangle.
\label{eq:polPDFdef}
\end{eqnarray}
The PDF of \eq{eq:polPDFdef} describes the density of a hadron that effectively contains twice as many helicity-polarized parton species as an ``unpolarized'' PDF.
As DGLAP evolution is derivable from perturbative methods,
it is possible to decompose them into helicity components as done, for example, in Refs.~\cite{Ciafaloni:2005fm,Ciafaloni:2009tf,Chen:2016wkt,Bauer:2017isx,Manohar:2018kfx}.
Alternatively, one can pragmatically bypass a numerical extraction of $f_{k_\lambda/p}(\xi,\mu_f)$ by noting that massless SU$(3)_c\otimes$U$(1)_{\rm QED}$ is a parity-invariant theory.
In such theories and for unpolarized hadrons, PDFs  of partons with opposite helicities are equal, i.e., $f_{k_L/p}(\xi,\mu_f)=f_{k_R/p}(\xi,\mu_f)$~\cite{Collins:2011zzd}.
In other words: while it is possible for, say, an $u_L$ quark to split into a gluon that splits into an $u_R$ quark, such helicity depletion and buildup wash out.
One can then introduce a normalization factor $\mathcal{N}=1/2$, and extract PDFs for polarized partons from PDFs for unpolarized partons using:
\begin{eqnarray}
f_{k_\lambda/p}(\xi,\mu_f) = f_{k_{-\lambda}/p}(\xi,\mu_f) = \mathcal{N} \times  f_{k/p}(\xi,\mu_f).
\label{eq:polPDFredef}
\end{eqnarray}
Imposing parity invariance means that the generation of new polarized PDF sets are not needed in real MC simulations.
One only needs a wrapper routine to implement \eq{eq:polPDFredef}.

Using identical arguments for splitting the DGLAP evolution equations into orthogonal helicity components, 
the perturbative (in the coupling sense) component of the Sudakov factor $\Delta$ can also be split into permutations of $i$'s and $j$'s helicities $\lambda$ and $\lambda'$:
\begin{eqnarray}
\Delta_{ij}(z,\mu_f,\mu_r,\mu_s) = \sum_{\lambda,\lambda'} \Delta_{i_\lambda j_{\lambda'}}(z,\mu_f,\mu_r,\mu_s). 
\label{eq:polSudakov}
\end{eqnarray}
A generic implementation of \eq{eq:polSudakov} is less clearcut than for PDFs.
The difference stems from the fact that IS partons, before Sudakov evolution, propagate along the beam axis and therefore possess an azimuthal rotation symmetry.
This means that IS partons from polarized and unpolarized parton PDFs transmit the entirety of their polarization information along the beam axis and is captured entirely by matrix elements.
However, Sudakov evolution, particularly as implemented via parton showers, injects relative transverse momentum into external partons through IS and FS radiation.
In general, this ``kick'' breaks preexisting rotational symmetry and induces azimuthal spin correlation.
Proposals for how to enforce azimuthal spin correlation in MC simulations appear throughout the literature~\cite{Collins:1987cp,Knowles:1988vs,Richardson:2001df},
and their implementation are under active investigation~\cite{Frederix:2009yq,Richardson:2018pvo,Cormier:2018tog}.

Taken all together, a consistent description of 
 parton scattering via polarized ME in unpolarized hadron collisions emerges.
Combining the totally polarized and fully differentiated, parton-level scattering rate for $i_\lambda + j_{\lambda'} \to \mathcal{B}_{\tilde{\lambda}}$ in \eq{eq:partonScatt_pol},
with the polarized PDFs of \eq{eq:polPDFdef} and the polarized Sudakov factor in \eq{eq:polSudakov},
the fully differentiated, hadron-level scattering rate for the production of $\mathcal{B}_{\tilde{\lambda}}$ 
from partons $i_\lambda$ and $j_{\lambda'}$ is
\begin{eqnarray}
\cfrac{d\sigma(pp\to\mathcal{B}_{\tilde{\lambda}}+X)}{dPS_n}\Bigg\vert_{i_\lambda,j_{\lambda'}}
&=& f_{i_\lambda} \otimes f_{j_{\lambda'}} \otimes \Delta_{i_\lambda,j_{\lambda'}} \otimes 
\cfrac{d\hat{\sigma}_{i_\lambda,j_{\lambda'}}}{dPS_n} 
 + \mathcal{O}\left(\frac{\Lambda^t_{\rm NP}}{Q^{t+2}}\right) 
\label{eq:polfactThemShort}\\
&=& 
\int_{\tau_0}^1 d\tau ~\int_{\tau}^1 \frac{d\xi_1}{\xi_1} ~\int_{\tau/\xi_1}^1 \frac{dz}{z} 
\frac{1}{(1+\delta_{i_\lambda,j_{\lambda'}})}
\nonumber\\
& &
\times ~ \left[ f_{i_\lambda/p}(\xi_1,\mu_f)f_{j_{\lambda'}/p}(\xi_2,\mu_f)+ (1\leftrightarrow2)\right] ~\times ~ \Delta_{i_\lambda,j_{\lambda'}}(z,\mu_f,\mu_r,\mu_s)
\nonumber\\
& &
\times ~ 
\cfrac{d\hat{\sigma}(i_\lambda + j_{\lambda'}\to\mathcal{B}_{\tilde{\lambda}}; \{Q^2,s,\mu_f,\mu_r,\mu_s\})}{dPS_n}
~+~ \mathcal{O}\left(\frac{\Lambda^t_{\rm NP}}{Q^{t+2}}\right). ~\quad
\end{eqnarray}
Accounting for all parton species, including those in different helicity states,
the production of $\mathcal{B}_{\tilde{\lambda}}$ from spin-averaged IS partons, in terms of IS states in definite helicity states is
{\small
\begin{eqnarray}
\cfrac{d\sigma(pp\to\mathcal{B}_{\tilde{\lambda}}+X)}{dPS_n} &=& 
\sum_{i_\lambda,j_{\lambda'}=q_L,g_R,\dots}
\cfrac{d\sigma(pp\to\mathcal{B}_{\tilde{\lambda}}+X)}{dPS_n}\Bigg\vert_{i_\lambda,j_{\lambda'}}
\\
&=& 
\sum_{i_\lambda,j_{\lambda'}=q_L,g_R,\dots}
f_{i_\lambda} \otimes f_{j_{\lambda'}} \otimes \Delta_{i_\lambda,j_{\lambda'}} \otimes 
\cfrac{d\hat{\sigma}_{i_\lambda,j_{\lambda'}}}{dPS_n} 
 + \mathcal{O}\left(\frac{\Lambda^t_{\rm NP}}{Q^{t+2}}\right) 
\label{eq:polFactThm}
\\
&=& 
\sum_{i,j=q,g,\dots}
f_i  \otimes f_j \otimes \Delta_{ij}  \otimes 
 \frac{1}{(2s_i+1)(2s_j+1)}
\sum_{\lambda,\lambda'}
\cfrac{d\hat{\sigma}_{i_\lambda,j_{\lambda'}}}{dPS_n} 
 + \mathcal{O}\left(\frac{\Lambda^t_{\rm NP}}{Q^{t+2}}\right). \qquad 
 \label{eq:mixedpolFactThm}
\end{eqnarray}
}
Between the second and third lines, we split the single summation over helicity-polarized parton species into a double sum over unpolarized parton species and parton helicities.
We then exploit that massless, IS parton species cannot contribute to a scattering requiring the opposite helicity, e.g., 
$f_{u_R}f_{\bar{u}_R}\otimes\Delta_{u_R\bar{u}_R}\otimes\hat{\sigma}_{u_L\bar{u}_R}=0$.
Such helicity inversion is proportional to parton masses, and hence vanishing.
(We reiterate that in this notation, factorizable $k_{\lambda''}\to i_\lambda$ parton buildup and $i_\lambda\to k_{\lambda''}$ parton depletion are handled internally 
by polarized Sudakov evolution.)
This allows us to rewrite the helicity-dependent PDFs and Sudakov factor in terms of their helicity-independent counterparts,
and demonstrates that the normalization factor $\mathcal{N}$ for polarized parton densities in \eq{eq:polPDFredef} 
can be identified as the spin-averaging symmetry factor in unpolarized parton scattering.
Moreover, one sees that the  Factorization Theorem of \eq{eq:factThemShort} is recovered after a summation over FS helicity polarizations $\tilde{\lambda}$,
and therefore shows consistency with the above construction.

{For leading order processes, we report the implementation of \eq{eq:polFactThm}, for helicity polarizations defined in a reconstructable reference frame, 
at FO into the event generator~\mgFull.}
Importantly, the FO stipulation implies that the Sudakov factor is expanded to zeroth order, i.e., $\Delta_{ij}(z)\approx\delta_{ij}\delta(1-z)$, leading to the simpler relationship:
\begin{eqnarray}
\cfrac{d\sigma(pp\to\mathcal{B}_{\tilde{\lambda}}+X)}{dPS_n} &\approx& 
\sum_{i_\lambda,j_{\lambda'}=q_L,g_R,\dots}
f_{i_\lambda} \otimes f_{j_{\lambda'}} \otimes  
\cfrac{d\hat{\sigma}_{i_\lambda,j_{\lambda'}}}{dPS_n} 
 + \mathcal{O}\left(\frac{\Lambda^t_{\rm NP}}{Q^{t+2}}\right) 
\label{eq:polFactThmMG5LO}
\\
&=& 
\sum_{i,j=q,g,\dots}
f_i  \otimes f_j \otimes 
 \frac{1}{(2s_i+1)(2s_j+1)}
\sum_{\lambda,\lambda'}
\cfrac{d\hat{\sigma}_{i_\lambda,j_{\lambda'}}}{dPS_n} 
 + \mathcal{O}\left(\frac{\Lambda^t_{\rm NP}}{Q^{t+2}}\right). \qquad ~
 \label{eq:mixedpolFactThmMG5LO}
\end{eqnarray}
{As helicity information is recorded in MC event files at LO, 
 parton-level polarizations
 can then be passed to a parton shower as desired.}
We report also the implementation of \eq{eq:mixedpolFactThm} for polarized, colorless, external states, 
with or without additional, unpolarized QCD partons and heavy quarks, e.g., $pp\to Z_\lambda + nj$, $pp\to W_\lambda Z_{\lambda'}$, or $e^+_Re^-_L \to Z_\lambda + t\overline{t}$, at NLO in QCD.
Details are reported in the companion paper Ref.~\cite{COCITE}.

%%%%%%%%%%%%%%%%%%%%%%%%%%%%%%%%%%%%%%%%%%%%%%%%
\subsection{Decays of Helicity-Polarized Resonances}\label{sec:polarDef_decay}

The polarization features introduced into \mgamc~extend also to unstable resonances.
In the default usage of \mgamc,
the production and decay syntax trigger the so-called spin-correlated NWA~\cite{Alwall:2008pm}.  
Whereas the usual (spin-uncorrelated) NWA factorizes matrix elements, for example, for $q\overline{q},gg \to t \overline{t} \to t \overline{b} W^-$ into the product of two decoupled amplitudes, 
\mgamc~instead first generates the helicity amplitude for the $2\to2$ scattering process $q\overline{q},gg \to t \overline{t}$,
but replaces the outgoing $\overline{v}(p_{\overline{t}},\lambda_{\overline{t}})$ spinor
with a fermionic Breit-Wigner (BW) propagator for the internal $\overline{t}$ and the appropriately contracted $1\to2$ decay current.
Likewise, for $e^+ e^- \to W^+ W^-$ with $W^+ \to e^+ \nu_e$ and $W^- \to e^- \nu_e$,
\mgamc~replaces the outgoing polarization vectors 
$\varepsilon^*_\mu(p_{W^+},\lambda_{W^+})$ and $\varepsilon^*_\nu(p_{W^-},\lambda_{W^-})$,
which describe $W^+$ and $W^-$ respectively in the $2\to2$ scattering amplitude for $e^+e^-\to W^+W^-$,
are each replaced by a bosonic BW propagator and a contracted $1\to2$ decay current.

To propagate the polarization of an unstable resonance to its decay products, 
we consider modifying this procedure by inserting a ``spin-truncated'' propagator \textit{in lieu} of a normal BW propagator.
For fermion $F$ and antifermion $\overline{F}$ with fixed helicity $\lambda$,
new propagators are defined by denominators with a BW pole structure but a numerator given by the outer product of spinors at helicity $\lambda$.
Explicitly, the replacement is
\begin{eqnarray}
S_F(q,m_q,\Gamma_q) &\to& S_F^{\lambda}(q,m_q,\Gamma_q)= \cfrac{i u(q,\lambda)\overline{u}(q,\lambda)}{q^2 - m_q^2 + i m_q \Gamma_q}, 
\label{eq:polarDef_decay_FXProp}
\\
S_{\overline{F}}(q,m_q,\Gamma_q) &\to& S_{\overline{F}}^{\lambda}(q,m_q,\Gamma_q)= \cfrac{-i v(q,\lambda)\overline{v}(q,\lambda)}{q^2 - m_q^2 + i m_q \Gamma_q}.
\label{eq:polarDef_decay_FBProp}
\end{eqnarray}
The origin of this structure stems from the condition that the full propagator is the coherent sum of the spin-truncated propagator over all helicity states $\lambda$. That is, 
 \begin{eqnarray}
S_F(q,m_q,\Gamma_q) &=&  \cfrac{i(\not\!q + m)}{q^2 - m_q^2 + i m_q \Gamma_q} = \cfrac{i\sum_{\lambda\in\{\pm1\}}u(q,\lambda)\overline{u}(q,\lambda)}{q^2 - m_q^2 + i m_q \Gamma_q} 
= \sum_{\lambda\in\{\pm1\}}S_F^{\lambda}(q,m_q,\Gamma_q),
\\
S_{\overline{F}}(q,m_q,\Gamma_q) &=&  \cfrac{-i(\not\!q - m)}{q^2 - m_q^2 + i m_q \Gamma_q} = \cfrac{-i\sum_{\lambda\in\{\pm1\}}v(q,\lambda)\overline{v}(q,\lambda)}{q^2 - m_q^2 + i m_q \Gamma_q} 
= \sum_{\lambda\in\{\pm1\}}S_{\overline{F}}^{\lambda}(q,m_q,\Gamma_q). ~\qquad
\end{eqnarray}

For massive gauge bosons, we introduce a similar spin-truncated propagator given by
\begin{eqnarray}
\Pi_{\mu\nu}(q,M_V,\Gamma_V) &\to& \Pi_{\mu\nu}^\lambda(q,M_V,\Gamma_V)= \cfrac{-i\varepsilon(q,\lambda)\varepsilon^*(q,\lambda)}{q^2 - M_V^2 + i M_V \Gamma_V}. 	
\label{eq:polarDef_decay_VXProp}	
\end{eqnarray}
For gauge bosons, the relation of the spin-truncated propagator to the full propagator is different due to gauge theory redundancies,
i.e., using 4-component vectors to describe quantities possessing only two or three degrees of freedom.

For massive gauge bosons, the full propagator is recovered from $\Pi_{\mu\nu}^\lambda$ by summing over both transverse polarizations,
the longitudinal polarization at a given virtuality, and an auxiliary (or scalar) polarization that rapidly vanishes in the on-shell limit~\cite{Weinberg:1995mt,Halzen:1984mc}.
(For massless gauge bosons, there is a cancellation between the longitudinal and auxiliary components.)
In the unitary gauge for massive spin-1 states, the full and spin-truncated propagators are related explicitly by
\begin{eqnarray}
\Pi_{\mu\nu}(q,M_V,\Gamma_V) &=&  \cfrac{-i\left[g_{\mu\nu} - \frac{q_\mu q_\nu}{M^2_V}\right]}{q^2 - M_V^2 + i M_V \Gamma_V} 
						=	\sum_{\lambda\in\{0,\pm1,A\}}  \Pi_{\mu\nu}^\lambda(q,M_V,\Gamma_V).
\label{eq:polarPropV}						
\end{eqnarray}
We report that these propagator decompositions has been implemented in \mgFive~and \ms.
While the transverse and longitudinal polarization vectors are defined according to the \texttt{HELAS} convention~\cite{Murayama:1992gi},
we set as our auxiliary (or scalar) polarization vector
\begin{equation}
\label{eq:auxprop}
\varepsilon^\mu(q,\lambda=A) = \frac{q^\mu}{M_V}\sqrt{\frac{q^2-M_V^2}{q^2}}\,.
\end{equation}

%%=====================================================================
% !TEX root = mgPolarization_main.tex
\section{Polarized Vector Boson Scattering in the SM and beyond}\label{sec:vbs}

%%%%%%%%%%%%%%%%%%%%%%%%%%%%%%%%%%%%%%%%%%%%%%%%%%%%%%%%%%%%%%
%%%%%%%%%%%%%%%%%%%%%%%%%%%%%%%%%%%%%%%%%%%%%%%%%%%%%%%%%%%%%%

Exploring EW VBS is a key step to understanding the SM, and in particular the underlying mechanism of EW symmetry breaking (EWSB).
More specifically, VBS is sensitive to whether EWSB is described by more than just the SM Higgs sector 
due to inevitable disturbances of strong cancellations in amplitudes involving longitudinally polarized weak bosons~\cite{Lee:1977yc,Lee:1977eg,Chanowitz:1984ne,Chanowitz:1985hj}. 
As the first observations of VBS were at last achieved by the ATLAS and CMS collaborations during Run II of the LHC program~\cite{Sirunyan:2017ret,Sirunyan:2017fvv,Aaboud:2018ddq,Sirunyan:2019ksz,Aaboud:2019nmv},
their use as a direct probe of new physics is now possible.
In a more general setup, 
VBS is sensitive to peculiar new physics that can be described by dimension-6 and dimension-8 effective operators~\cite{Brivio:2013pma,Gomez-Ambrosio:2018pnl,Zhang:2018shp},
assuming the usual decoupling limit~\cite{Appelquist:1974tg}.
Such new physics would manifest as the anomalous production of EW states in specific helicity configurations,
and hence motivates one to investigate means to experimentally disentangle EW boson polarizations.

In this section we investigate VBS production of polarized weak bosons at the $\sqrt{s}=13\TeV$ LHC, within the polarized \mgamc~framework.
We consider EW and mixed EW-QCD production of $W^+W^-$ boson pairs with helicities $(\lambda,\lambda')$ and two partons at LO,
\begin{equation}
q_1 q_2 ~\to~ q_1' q_2' W^+_\lambda W^-_{\lambda'} .
\label{eq:vbs_vbsDef}
\end{equation}
We start in \sec{sec:vbs_bsm} with discussing high energy VBS in the context of Composite Higgs (CH) models. 
This class of models manifest as an enhancement of scattering rates involving longitudinal weak bosons.
Subsequently, we illustrate the \mgamc~polarization framework 
by focusing on the reference frame-dependence of observables, e.g., polarization fractions,  
built from polarized states and with phase space cuts on particle kinematics.

In \sec{sec:vbs_ew} we extend the study to observables built from the decay of $W^+_\lambda\to \mu^+\nu_\mu$ and $W^-_{\lambda'}\to e^-\bar{\nu}_e$,
when \eq{eq:vbs_vbsDef} proceeds at $\mathcal{O}(\alpha^4)$.
To do this, we use the \ms~framework in conjunction with helicity-polarized samples generated from~\mgamc.
We give special attention to angular observables that are sensitive to the polarization of the parent particle $W^\pm$.
The same process was studied by the \texttt{Phantom} MC collaboration~\cite{Ballestrero:2007xq,Ballestrero:2011pe,Ballestrero:2017bxn,Ballestrero:2019qoy}, using the on-shell projection (OSP) technique.
The agreement we find not only serves as a check of the two methodologies but also as a basis for future studies.
In \sec{sec:vbs_qcd} we repeat this exercise but for when \eq{eq:vbs_vbsDef} proceeds at $\mathcal{O}(\alpha^2\alpha_s^2)$.
Throughout this section, we summarize the new or relevant syntax for \mgamc~and~\ms~needed for our study.

%%%%%%%%%%%%%%%%%%%%%%%%%%%%%%%%%%%%%%%%%%%%%%%%%%%%%%%%%%%%%%%%%%%%%%%%%%%%
\subsection{Vector Boson Scattering in Composite Higgs Models}\label{sec:vbs_bsm}

In this section we investigate CH models~\cite{Kaplan:1983fs,Kaplan:1983sm,Georgi:1984af,Dugan:1984hq,Contino:2003ve,Agashe:2004rs,Contino:2006qr,Agashe:2006at}, 
in high energy VBS using polarized parton event generation.
Promising, modern incarnations of these scenarios predict that the Higgs coupling to weak gauge bosons 
are rescaled by a common (dimensionless) factor $a$, and can be described by the effective interaction Lagrangian~\cite{Bellazzini:2014yua,Panico:2015jxa}
\begin{equation}
\mathcal{L} \supset \left(\frac{m_Z^2}{2} Z_\mu Z^\mu + m_W^2 W^+_\mu W^{-\mu}\right)\left(1+2 a \frac{h}{v}+\cdots\right)\,.
\end{equation}
The presence of $a$ away from unity disrupts fine cancellations in SM amplitudes describing longitudinal weak boson scattering,
and leads to amplitudes growing with the invariant mass of the $(VV)$-system (squared), for $V\in (W,Z).$
That is, $\mathcal{M}(V_0 V_0\to V_0 V_0)\sim a M^2(VV)/v^2$,
which can potentially be observed at the LHC.
Direct measurements of Higgs couplings constrain $a$ at the 95\% CL to be $a\gtrsim 0.9$~\cite{Khachatryan:2016vau}.
Indirect EW precision data also require $a\gtrsim 0.98$~\cite{Arbey:2015exa}, 
but can be relaxed if additional assumptions are satisfied~\cite{BuarqueFranzosi:2018eaj}.

\begin{table}[!t]
\begin{center}
\begin{tabular}{ c || c | c | c | c | c | c | c | c |}
\hline
\hline
			&  \multicolumn{2}{c|}{p-CM SM $(a=1)$}	&  \multicolumn{3}{c|}{p-CM CH $(a=0.8)$}	&  \multicolumn{3}{c|}{p-CM CH $(a=0.9)$}			\tabularnewline
Process		&	$\sigma$ [fb]		& $f_{\lambda\lambda'}$	&	$\sigma$ [fb]		& $f_{\lambda\lambda'}$	&  $\sigma^{\rm CH}/\sigma^{\rm SM}$	
															&	$\sigma$ [fb]		& $f_{\lambda\lambda'}$	&  $\sigma^{\rm CH}/\sigma^{\rm SM}$\tabularnewline
\hline
$jjW^+W^-$ 					&  171   		& $\dots$ 	    	&   173   	& $\dots$   	&  1.00  &   172		& $\dots$	&  1.00	\tabularnewline
$jjW_T^+W_T^-$				&  119		& 70\% 		&   116	& 69\% 		&  0.98  &   115		& 69\% 	&  0.96  	\tabularnewline
$jjW_0^+W_T^-$				&  20.6		& 12\%    		&   21.5	& 13\%  		&  1.05  &   22.0	& 13\%  	&  1.07	\tabularnewline
$jjW_T^+W_0^-$				&  23.8		& 14\%  		&   24.1	&14\%   		&  1.01  &   23.9	&14\%   	&  1.01	\tabularnewline
$jjW_0^+W_0^-$				&  5.45		& 3\%	 	&   7.17	& 4\%  		&  1.31  &   6.01		& 4\%  		&  1.10	\tabularnewline
	    \hline
\end{tabular}
\caption{
Generator-level cross section [fb] for the unpolarized, EW process $pp\to jj W^+_\lambda W^-_{\lambda'}$ at LO, 
for the SM limit $(a=1.0)$ and two benchmark Composite Higgs scenarios $a=0.8$ and $a=0.9$,
as well as the same information for various $(\lambda,\lambda')$ helicity configurations
defined in the parton c.m.~frame (p-CM) with their polarization fraction $f_{\lambda\lambda'}$ [\%].
}
\label{tab:CHxsPart}
\end{center}
\end{table}

To quantify the impact of a CH scenario on VBS, we make use of the NLO Higgs Characterization UFO model~\cite{Artoisenet:2013puc} described in \sec{sec:setup} and focus on the LO EW process
\begin{equation}
p p ~\to~ j j W^+_\lambda W^-_{\lambda'},
\end{equation}
for helicity states $\lambda,\lambda'=0,\pm1$.
Throughout this analysis we do not make the so-called {Vector Boson Fusion Approximation}, which considers only genuine $WW/ZZ\to WW$ scattering diagrams,
and is known to neglect significant interference effects~\cite{Accomando:2006mc,Campanario:2013fsa,Campanario:2018ppz}.
We instead include all interfering diagrams at $\mathcal{O}(\alpha^4)$, including other VBS topologies, like $\gamma\gamma\to WW$, and non-VBS contributions.
The \mgamc~syntax to model the production of unpolarized and polarized $W^+W^-$ pairs in unpolarized $pp$ collisions is, respectively,
\begin{verbatim}
import model HC_UFO-CH
generate p p > j j w+ w- QCD=0 QED<=4
generate p p > j j w+{X} w-{Y} QCD=0 QED<=4
\end{verbatim}
Here, one should replace \texttt{X} and \texttt{Y} by all permutations of \texttt{0} (longitudinal helicity) and \texttt{T} 
(transverse helicities)\footnote{
We note that the syntax \texttt{T} coherently sums over both LH and RH helicity states; see \app{sec:codingBits_syntax} for details.
For further details regarding the usage of~\mgamc, we refer readers to Refs.~\cite{Alwall:2011uj,Alwall:2014hca}.}.
For event generation, we consider three benchmark scenarios: $a=0.8$, $a=0.9$, and the SM limit of $a=1$.
In the UFO model,  $a$ is identified as the parameter \texttt{kSM} and can be set in \texttt{param\_card.dat} or at runtime with the command
\begin{verbatim}
set kSM A
\end{verbatim}
To define helicity polarizations for the $W^+_\lambda W^-_{\lambda}$ pairs, we consider two reference frames:
(i) The rest frame defined by the two initial-state partons in the $2\to4$ process, which we label as the partonic c.m.~(p-CM) frame.
(ii) The rest frame of the $W^+W^-$ system, which we label as the $WW$ c.m.~($WW$-CM) frame.
Both frames can be specified using the new \texttt{me\_frame} selector tag in \mgamc's  \texttt{run\_card.dat} input file (see \app{sec:codingBits_evtgen} for details).
By momentum conservation, the p-CM~frame can be built from either summing the two IS partons' momenta or the four FS partons' momenta.
This corresponds to the syntax
\begin{verbatim}
[1, 2]       = me_frame
[3, 4, 5, 6] = me_frame 
\end{verbatim}
The $WW$-CM frame is built most directly from the $W^+W^-$ itself, and corresponds to
\begin{verbatim}
[5, 6]	= me_frame
\end{verbatim}
To obtain total and differential cross sections\footnote{
More specifically, we produced 100k generator-level events per simulation, 
for  five polarization configurations (TT, TL, LT, LL, unpolarized), two rest frames, ($WW$ c.m. and partonic c.m.),  and three parameters benchmarks ($a=0.8,0.9,1.0$).
Each generation required approximately 19 days of CPU time, totaling about 570 CPU-days for the 30 samples,
using a small  heterogenuous cluster with cores of various architectures (from opteron to skylake gold, 2.3 GHz 32GB RAM or 2.6 GHz 64GB RAM).
}, we impose the following generator-level phase space cuts,
which as described in \app{sec:codingBits_evtgen}, are applied in the p-CM~frame:
\begin{eqnarray}
p_T(j)>20\GeV,\quad |\eta(j)|<5,\quad M(jj)>250\GeV,\quad \Delta \eta(jj)>2.5, \nonumber \\
M(W^+W^-)>300\GeV,\quad p_T(W^\pm)>30\GeV,\quad |\eta(W^\pm)|<2.5\,.
\label{eq:CHcuts}
\end{eqnarray}
The cuts serve several purposes:
First, they regulate collinear and soft singularities from interfering VBS and non-VBS diagrams.
Second, they correspond to typical, analysis-level selection cuts that enhance the VBS topology over 
interfering EW diagrams.
Third, they enhance the appearance of new physics.
This follows from nonzero $a$ coefficients leading to an enhancement in the scattering amplitude that grows with increasing $M(WW)$.

\begin{table}[!t]
\begin{center}
\begin{tabular}{ c || c | c | c | c | c | c | c | c |}
\hline
\hline
			&  \multicolumn{2}{c|}{$WW$-CM SM $(a=1)$}	&  \multicolumn{3}{c|}{$WW$-CM CH $(a=0.8)$}	&  \multicolumn{3}{c|}{$WW$-CM CH $(a=0.9)$}			\tabularnewline
Process		&	$\sigma$ [fb]		& $f_{\lambda\lambda'}$	&	$\sigma$ [fb]		& $f_{\lambda\lambda'}$	&  $\sigma^{\rm CH}/\sigma^{\rm SM}$	
															&	$\sigma$ [fb]		& $f_{\lambda\lambda'}$	&  $\sigma^{\rm CH}/\sigma^{\rm SM}$\tabularnewline
\hline
$jjW^+W^-$ 					& 171   			& $\dots$ 	    	&   173   			& $\dots$   	&  1.00  &   172		& $\dots$   		&  1.00	\tabularnewline
$jjW_T^+W_T^-$				&  118			& 69\% 		&   114			& 68\% 		&  0.96  &   118		& 69\% 			&  1.00	\tabularnewline
$jjW_0^+W_T^-$				&  22.2			& 13\%    		&   21.6			& 13\%  		&  0.97  &   21.6	& 12\%  			&  0.97	\tabularnewline
$jjW_T^+W_0^-$				&  24.1			& 14\%  		&   23.6			& 14\%   		&  0.98  &   24.0	& 14\%   			&  0.99	\tabularnewline
$jjW_0^+W_0^-$				&  6.93		& 4\%	 	&  8.96	& 5\%  		&  1.29  &   7.81		& 5\%  	&  1.13	\tabularnewline
	    \hline
\end{tabular}
\caption{
Same as \tab{tab:CHxsPart} but for the $WW$ c.m. frame ($WW$-CM).
}
\label{tab:CHxsWW}
\end{center}
\end{table}

In \tab{tab:CHxsPart} we show the generator-level cross sections [fb] for unpolarized $W^+ W^-$ production 
and each $W^+_\lambda W^-_{\lambda'}$ helicity polarization configuration $(\lambda,\lambda')$,
defined in the p-CM, for the CH benchmark scenarios ($a=0.8$ and $a=0.9$) and the SM limit $(a=1.0)$. 
We also report the ratio between the CH and SM rates as well as 
the polarization fraction $f_{\lambda\lambda'}$, defined as the ratio of the $(\lambda,\lambda')$ helicity configuration to the unpolarized rate:
\begin{equation}
f_{\lambda\lambda'} = \sigma(p p \to j j W^+_{\lambda}W^-_{\lambda'}) ~/~ \sigma(p p \to j j W^+W^-).
\end{equation}
As expected, nonzero $a$ largely impacts the longitudinal $(\lambda,\lambda')=(0,0)$ state, which displays roughly a 30\% (13\%) increase in cross section for $a=0.8$ ($a=0.9$) over the SM prediction.
However, in the absence of more stringent selection cuts, the changes in $f_{\lambda\lambda'}$ indicate that a percent-level determination of polarization fractions would be needed to observe such disturbances.
In \tab{tab:CHxsWW} we show the equivalent results for polarizations defined in the $WW$-CM. 
Only a slight difference is noticed and thus need not be discussed further.

Due to rounding errors, the sum of $f_{\lambda\lambda'}$ obfuscates that the sum of the polarization configurations reproduces the unpolarized rate.
Satisfying closure requirements in $2\to4$ parton scattering in $pp$ collisions represents a highly nontrivial check of our method.

\begin{figure*}[!t]
\begin{center}
\subfigure[]{\includegraphics[width=.48\textwidth]{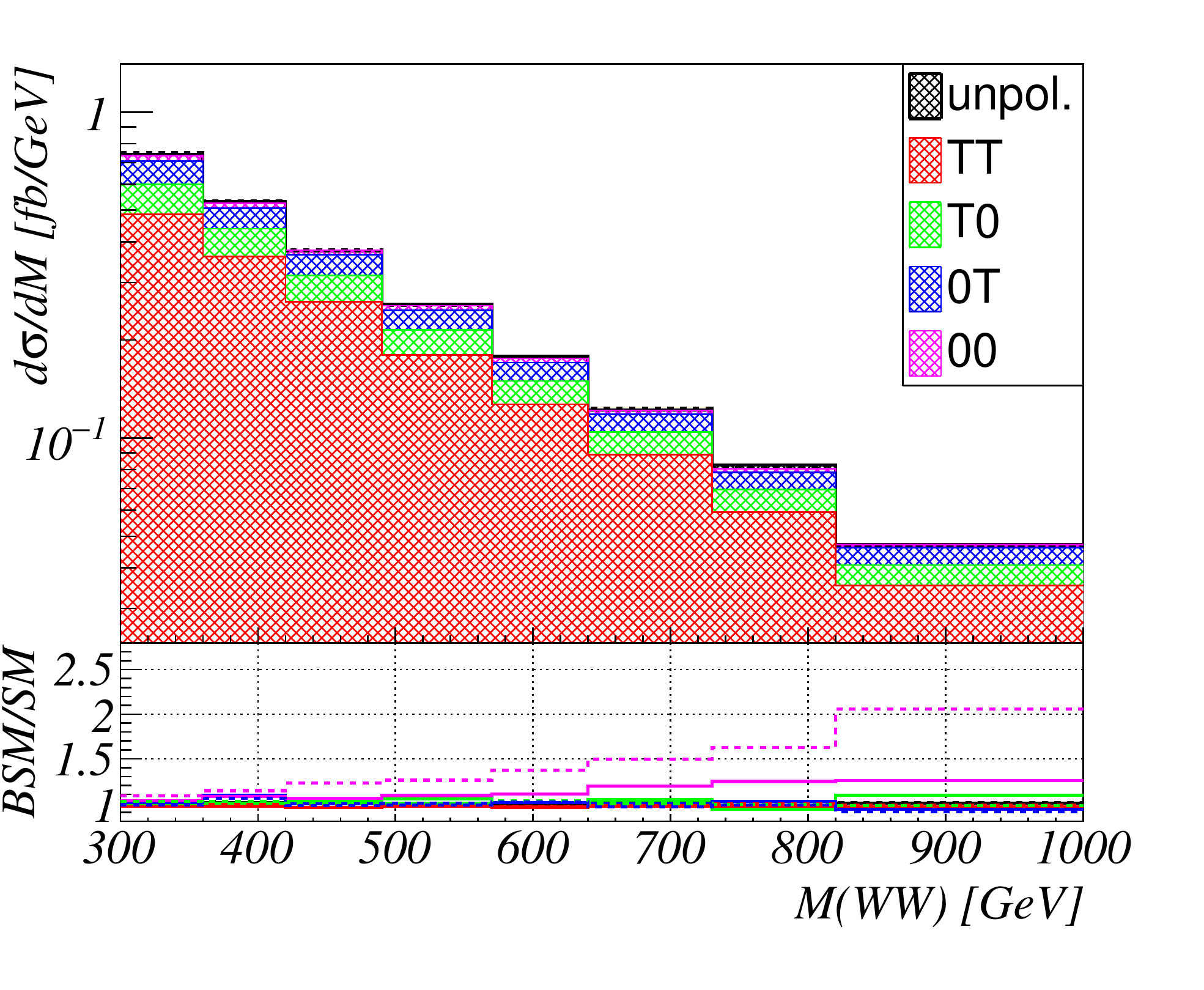}	\label{fig:VBSCH_mww_rf12}		}
\subfigure[]{\includegraphics[width=.48\textwidth]{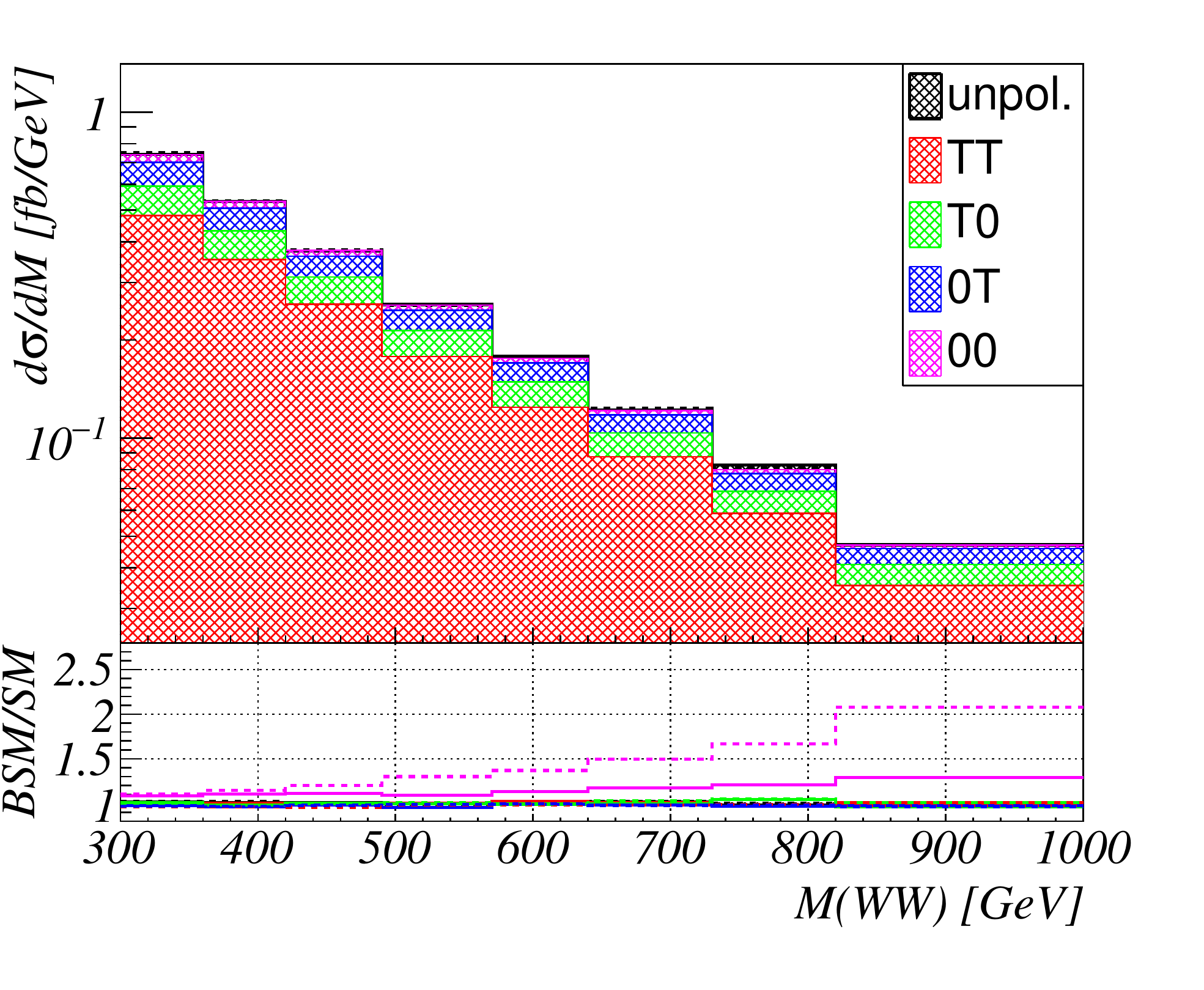}	\label{fig:VBSCH_mww_rf56}		}
\end{center}
\caption{
The $WW$ invariant mass spectrum $(d\sigma/dM)$ for the unpolarized, EW process $pp\to jj W^+_\lambda W^-_{\lambda'}$ at LO, 
in the SM limit $(a=1.0)$. The lower panel shows the ratio $\mathcal{R}[M(WW)]$ \eq{eq:CHratio} of the Composite Higgs scenarios with $a=0.8$ (dashed line) and $a=0.9$ (solid line). The polarization $(\lambda,\lambda')$ is defined 
in the (a) parton c.m.~frame
and (b) $WW$ c.m.~frame. 
}
\label{fig:CHWWdist}
\end{figure*}

Turning to differential information, we show in \fig{fig:CHWWdist} the invariant mass distribution of the $(WW)$-system, $d\sigma/dM(WW)$,
according to the various helicity configurations for the SM.
In the lower panel, we show the differential ratio with respect to the SM, i.e.,
\begin{equation}
\mathcal{R}[M(WW)] = d\sigma^{\rm CH}/dM(WW) ~/~ d\sigma^{\rm SM}/dM(WW),
\label{eq:CHratio}
\end{equation}
for the CH scenarios $a=0.8$ (dashed lines) and $a=0.9$ (solid lines).
In \fig{fig:VBSCH_mww_rf12}, helicity polarizations are defined in the p-CM frame and in the $WW$-CM frame in \fig{fig:VBSCH_mww_rf56}.
We observe explicitly in the lower panel the growing behavior of the CH cross section relative to the SM prediction with increasing $M(WW)$.
We note that tighter selection cuts, such as $M(WW)>625-825\GeV$ can further enhance the ratio $\mathcal{R}[M(WW)]$, though perhaps at a high cross section cost.
 An alternative possibility is to extract the polarizations via observables built from the $W^+W^-$ decay products, which we discuss in the next section.

%%%%%%%%%%%%%%%%%%%%%%%%%%%%%%%%%%%%%%%%%%%%%%%%%%%%%%%%%%%%%%%%%%%%%%%%%%%%
\subsection{Polarized $W$ Bosons in EW Production of $jj W^+ W^-$}\label{sec:vbs_ew}
In weak boson decays to charged leptons, it is well-known that the polarization of a parent boson is imprinted on the kinematics of its decay products.
This follows from stable fermions being effectively massless compared to the EW scale.
This is especially true of angular observables, which also feature particular sensitivity to the $(V-A)$ structure of bosonic couplings to matter.
These observables therefore serve as a test of the SM's chiral structure and, for example, a probe of the coupling structure of new physics.

Here we investigate the production of $W^+ W^-_\lambda$ pairs, via the pure EW process
\begin{equation}
pp ~\to~ jj W^+ W^-_{\lambda}, \quad\text{with}\quad W^+ \to \mu^+\nu_\mu \quad\text{and}\quad W^-_{\lambda}\to e^-\bar{\nu}_e,
\label{eq:vbs_sm_ProcDef}
\end{equation}
at LO.
The process is defined with an unpolarized $W^+$ boson and a polarized $W^-_\lambda$ boson with helicity $\lambda=0,T$.
We propagate the polarization of the $W^-_{\lambda}$ to its decay products using 
\ms~as described in \sec{sec:polarDef_decay}. 
As a high-level check, we also propagate the $W^-_{\lambda}$ polarization using the OSP method~\cite{Aeppli:1993cb,Aeppli:1993rs,Denner:2000bj}.
In the context of VBS, the OSP technique has been used in Refs.~\cite{Ballestrero:2007xq,Ballestrero:2011pe,Ballestrero:2017bxn,Ballestrero:2018anz,Ballestrero:2019qoy} and 
is implemented in \mgamc~ under an unsupported, standalone development branch for the purpose of this work. 
In short, the method amounts to setting the momenta of the $W^\pm$ bosons, $k_\pm$, 
to their mass-shell values $(k^2_\pm =M_W^2)$ in the numerators of matrix elements for the full $2\to6$ EW process $qq' \to q q' e^- \overline{\nu}_e \mu^+ \nu_\mu$.
The virtuality $k_\pm^2$ in the denominator of propagators is allowed to float. 
Non-resonant diagrams are neglected. 
In practice, the $k_\pm^2$ are restricted to the neighborhood of $k^2_\pm =M_W^2$ using phase space cuts,
thereby approximating the spin-correlated, NWA employed by \ms. 

For the process in \eq{eq:vbs_sm_ProcDef}, we define the polar angle $\theta$
as the angle between the $W^-$ flight direction in the p-CM frame and the $e^-$ flight direction in the $W^-$ rest frame, i.e.,
\begin{equation}
\cos\theta = \cfrac{\vec{p}_W\cdot \vec{\widetilde{p_e}}}{\vert\vec{p}_W\vert~\vert \vec{\widetilde{p_e}}\vert}.
\label{eq:vbs_sm_polarDef}
\end{equation}
Here, $\vec{p}_W$ is the 3-momentum of the $W^-$ in the  p-CM frame and $\vec{\widetilde{p_e}}$ is the 3-momentum of the $e^-$ in the $W^-$ rest frame. 
Similarly, an azimuthal angle $\phi$ can be defined as the opening angle between the $W^{-}$ boson's production  plane  (defined by the $W^-$ and beam direction) and its decay plane.
Analytically, this is given by
\begin{equation}
\phi = \tan^{-1}\left[ \frac{\hat{v}_1\cdot \vec{\widetilde{p}}_{e} }{\hat{v}_2\cdot \vec{\widetilde{p}}_{e} }\right],
\label{eq:vbs_sm_aziDef}
\end{equation}
where the two unit vectors $\hat{v}_1$ and $\hat{v}_2$ are defined to be,
\begin{equation}
\hat{v}_1= \frac{\vec{\widetilde{P}_{i}}\times \vec{p}_{W} }{\vert \vec{\widetilde{P}}_{i}\times \vec{p}_{W} \vert}\, \quad\text{and}\quad
\hat{v}_2= \frac{(\vec{\widetilde{P}}_{i}\times \vec{p}_{W})\times \vec{p}_{W}}{\vert (\vec{\widetilde{P}}_{i}\times \vec{p}_{W})\times \vec{p}_{W} \vert}.
\end{equation}
Here $\vec{\widetilde{P}_{i}}$ is the 3-momentum of any of the IS partons in the $W^-$ rest frame.
For definiteness, we choose the $i$ that makes the smallest opening angle with $W^-$'s flight direction in the p-CM frame.
If one defines the $z$-direction along $\vec{p}_{W}$, we can identify $\hat{v}_1=\hat{y}$ as the unit vector in the $y$-direction, $\hat{v}_2=\hat{x}$, 
and  we reproduce the coordinate system of Ref. \cite{Bern:2011ie}.

In this convention, the matrix element $\mathcal{M}_\lambda$, {with $\lambda$ defined in the p-CM frame,} describing \eq{eq:vbs_sm_ProcDef} 
depends on $\theta$ and $\phi$ through the $W^-_\lambda\to e^-\bar{\nu}_e$ decay,
and specifically through the angular dependence of the $e^-$ spinor.
Hence, the $\theta$ and $\phi$ dependence of $\mathcal{M}_\lambda$ scales as
\begin{eqnarray}
\mathcal{M}_0(\theta,\phi)				&\sim &\sin\theta\,, \\
\mathcal{M}_{L/R}(\theta,\phi) 			&\sim &(1\pm\cos\theta)e^{\mp i\phi}  \,.
\end{eqnarray}
Now, at beam-symmetric experiments such as the LHC, 
since the momenta of quarks are typically larger than of antiquarks,
and since the EW gauge couplings to fermions are chiral, the emission rates of RH, LH, or longitudinal $W,Z$ bosons off initial-state parton lines differ.
The relative emission rates are further skewed by non-Abelian gauge and Higgs couplings.
Ultimately, this leads to asymmetric production of polarized $W^-_\lambda$ for not just single boson production~\cite{Ellis:1991qj} but also in multiboson processes~\cite{COCITE}.
Allowing for an asymmetric production of $W^-_\lambda$, and in the notation of Ref.~\cite{Belyaev:2013nla},
the differential cross section for inclusive $W^-$ production in terms of the angles $\theta$ and $\phi$ can be written as~\cite{Collins:1977iv}:
 \begin{eqnarray}
{1\over\sigma} {d^2\sigma\over d\cos\ths d\phi}
&=&  \frac{3}{16\pi}\left[ (1 + \cos\ths)^2 \,f_L
    +  (1 - \cos\ths)^2 \,f_R
    + 2 \sin^2\ths \, f_0 
     - g_{RL}\sin^2\theta\cos(2\phi)
    \right.
     \nonumber\\
    &-& \sqrt{2}g_{L0}\sin\theta(1+\cos\theta)\cos\phi
  +\left. \sqrt{2}g_{R0}\sin\theta(1-\cos\theta)\cos\phi    \right]\,.
\label{eq:wDiffPolarAzi}
\end{eqnarray}  
Here, $f_\lambda$ can be interpreted as the likelihood of producing $W^-_\lambda$ with helicity $\lambda$ in the inclusive process.
The $g_{\lambda\lambda'}$ result from interference among the different $W^-_\lambda$ helicity polarizations but vanish upon integration over $\phi\in [-\pi,\pi]$.
The precise form of \eq{eq:wDiffPolarAzi} is somewhat arbitrary as trigonometric identities and redefinitions relate coefficients here to other parameterizations; 
see, e.g., Ref.~\cite{Stirling:2012zt}.
After integrating, one obtains the familiar expression:
\begin{equation}
{1\over\sigma} {d\sigma\over d\cos\ths}
\ =\  \frac{3}{8} (1 + \cos\ths)^2 \,f_L
    + \frac{3}{8} (1 - \cos\ths)^2 \,f_R
    + \frac{3}{4} \sin^2\ths \, f_0 \,.
\label{eq:wDiffPolar}
\end{equation}
In the following we use \eq{eq:wDiffPolarAzi} and \eq{eq:wDiffPolar} as guiding relationships to explore the polarization of $W_\lambda^- \to e^-\overline{\nu}_e$ decays in \eq{eq:vbs_sm_ProcDef}.
We also consider the ability to extract the coefficients $f_\lambda$ using the MC methods developed and reported in \sec{sec:polarDef}.
Throughout the following we use the OSP technique as a benchmark to quantify our results.

\begin{table}[!t]
\begin{center}
\begin{tabular}{ c || c || c  c | c  c }
\hline
\hline
			&  	Decay Scheme
			& 	\multicolumn{2}{c|}{Generator-Level Cuts} 
			&	\multicolumn{2}{c}{Analysis-Level Cuts} 
			\tabularnewline
Process		&	& $\sigma$ [fb]		& $f_{\lambda}$	&	$\sigma$ [fb]		& $f_{\lambda}$		\tabularnewline
\hline
$jjW^{+}W^{-}$		& MadSpin		&  3.818		&  $\dots$      	   	& 3.243  			& $\dots$  	    	            \tabularnewline
$jjW^{+}W_{T}^{-}$	& MadSpin		&  3.043		& 79.7\%      		& 2.567			& 79.2\%	   		           \tabularnewline
$jjW^{+}W_{T}^{-}$	& OSP			&  3.041		& 79.6\%    		& 2.568			& 79.2\%			             \tabularnewline
$jjW^{+}W_{0}^{-}$	& MadSpin		&  0.7824		& 20.5\%   		& 0.6527			& 20.1\%			             \tabularnewline
$jjW^{+}W_{0}^{-}$	& OSP			&  0.7797		& 20.4\%    		& 0.6514			& 20.1\%			             \tabularnewline
	    \hline\hline
\end{tabular}
\caption{Cross sections [fb] and polarization fractions $(f_\lambda)$ [$\%$], 
of the pure EW process $pp\to jj W^{+} W^{-}_{\lambda}$, with $W^{+}\rightarrow \mu^{+}\nu_{\mu}$ and $W^{-}_{\lambda}\rightarrow e^{-}\bar{\nu_{e}}$,
in the SM for unpolarized $W^+$ and $W^-$ helicity polarization $\lambda$, defined in the p-CM frame,
assuming generator- and analysis-level cuts of \eq{eq:SMVBScuts} and \eq{eq:decaycuts}, and 
using the MadSpin and OSP decay schemes for $W^\pm$.
}
\label{tab:SMVBSxs}
\end{center}
\end{table}

\subsubsection*{Production and Decay of Polarized $W$ Bosons in \mgamc}
To simulate \eq{eq:vbs_sm_ProcDef} in the SM using \mgms,
we first use the syntax reported in \sec{sec:vbs_bsm} to generate the subprocess $qq'\to qq' W^-_\lambda W^+$.
The \mgamc~commands are 
\begin{verbatim}
import model sm
generate p p > w+ w-{X} j j QED<=4 QCD=0
\end{verbatim} 
One should replace \texttt{X} in the \texttt{generate} command with $\texttt{0}$ or $\texttt{T}$ for $\lambda=0$ or $\lambda=T$.
We consider the polarizations defined in the p-CM frame, and thus in \texttt{run\_card.dat} set \texttt{me\_frame = [1, 2]}.
For comparison, we consider also the unpolarized process, which is simulated by
\begin{verbatim}
generate p p > w+ w- j j
\end{verbatim} 
To regulate infrared poles in the matrix element and enhance the VBS topology over interfering topologies, 
we require events to fulfill the generator-level cuts:
\begin{eqnarray}
p_T(j)>20\GeV,\quad |\eta(j)|<5,\quad \Delta R(jj)>0.4, \nonumber\\ M_{jj}> 120\GeV, \quad M(W^{+}W^{-})>300\GeV. \quad 
\label{eq:SMVBScuts}
\end{eqnarray}
We relax cuts relative to \eq{eq:CHcuts} since we are not strictly interested in isolating the pure VBS topology.
After event generation, unpolarized and polarized $W^+W^-_\lambda$ pairs are decayed using \ms.
As described in \app{sec:codingBits_decay}, the \ms~syntax is the same for unpolarized and polarized $W_\lambda$.
Therefore in the \texttt{madspin\_card.dat} file, one only needs:
 \begin{verbatim}
decay w+ > mu+ vm
decay w- > e- ve~
\end{verbatim}
The process with identical final states and kinematic cuts is also simulated using our implementation of the OSP method.
In \tab{tab:SMVBSxs}, we report generator-level cross sections [fb] and polarization fraction $(f_\lambda)$ [$\%$] for the full $2\to6$ process using both the MadSpin and OSP decay schemes.
We report good agreement in generator-level normalizations\footnote{As implemented, we also report comparable MC  efficiency between the \ms~and~\texttt{OSP} techniques.}.

\begin{figure*}[t!]
\begin{center}
\subfigure[]{\includegraphics[width=.48\textwidth]{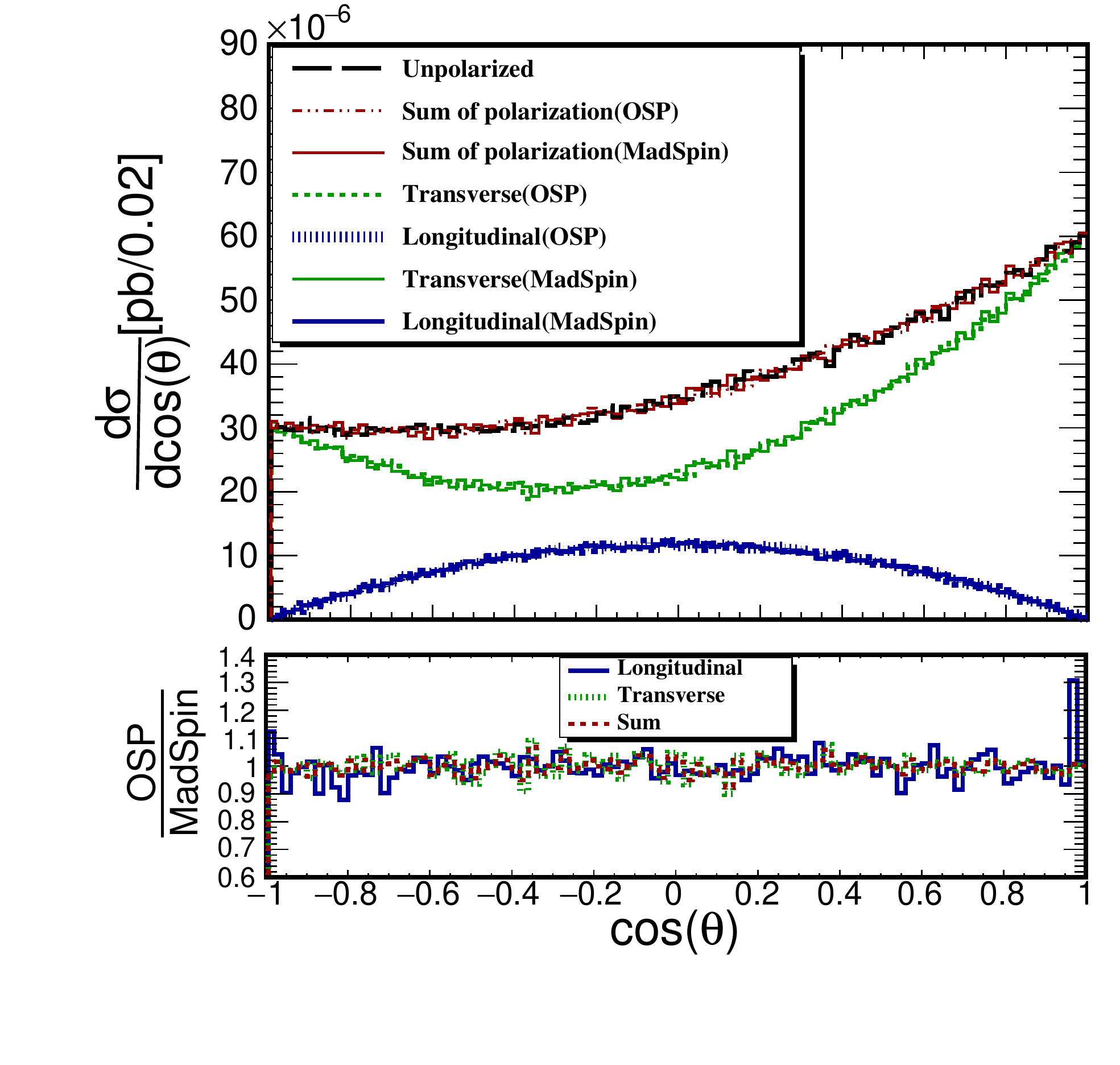}		\label{fig:VBSpolCosTh}		}
\subfigure[]{\includegraphics[width=.48\textwidth]{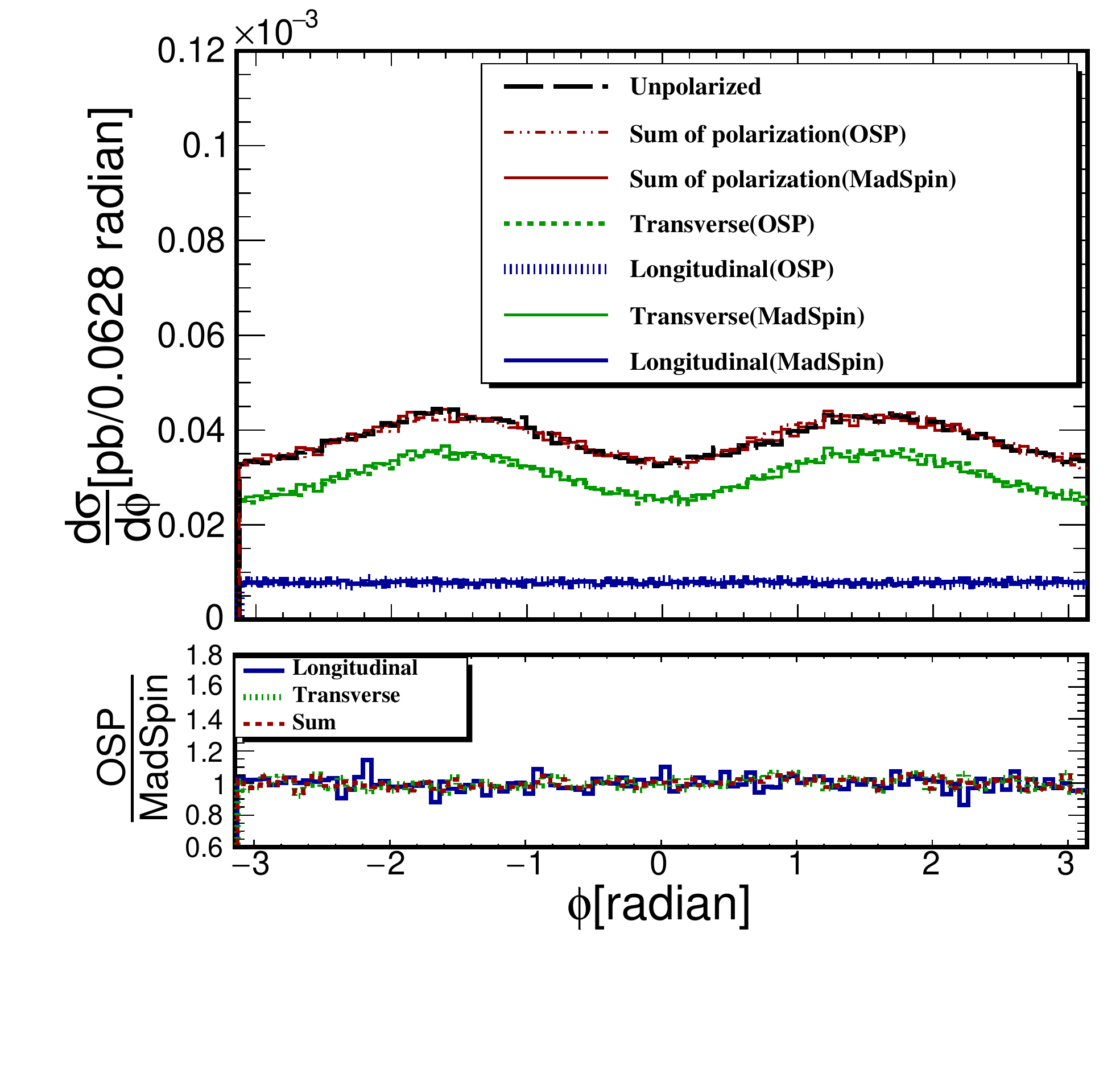}			\label{fig:VBSpolPhi}			}
\\
\subfigure[]{\includegraphics[width=.48\textwidth]{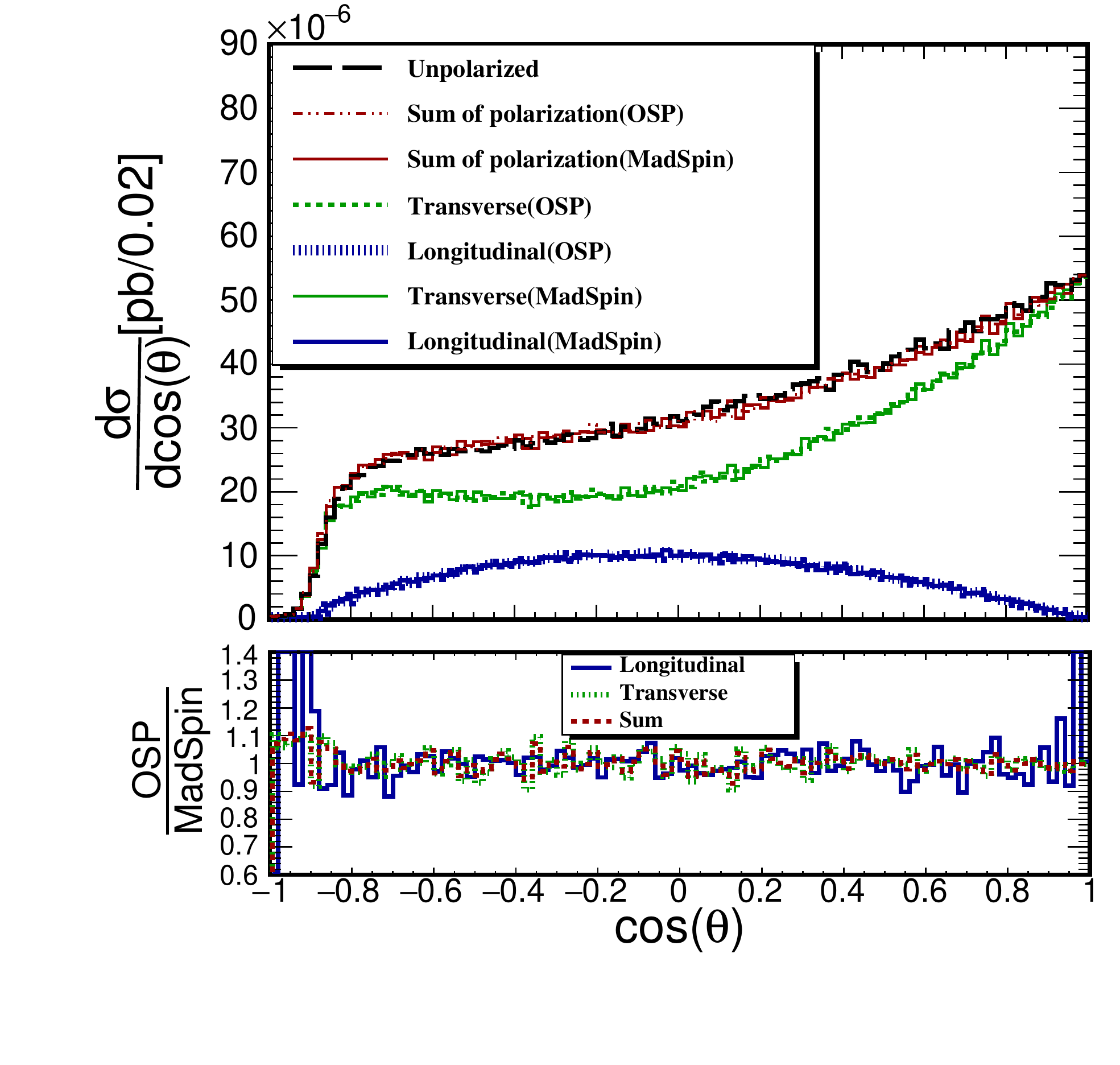}	\label{fig:AngdistDecayCutsTh}		}
\subfigure[]{\includegraphics[width=.48\textwidth]{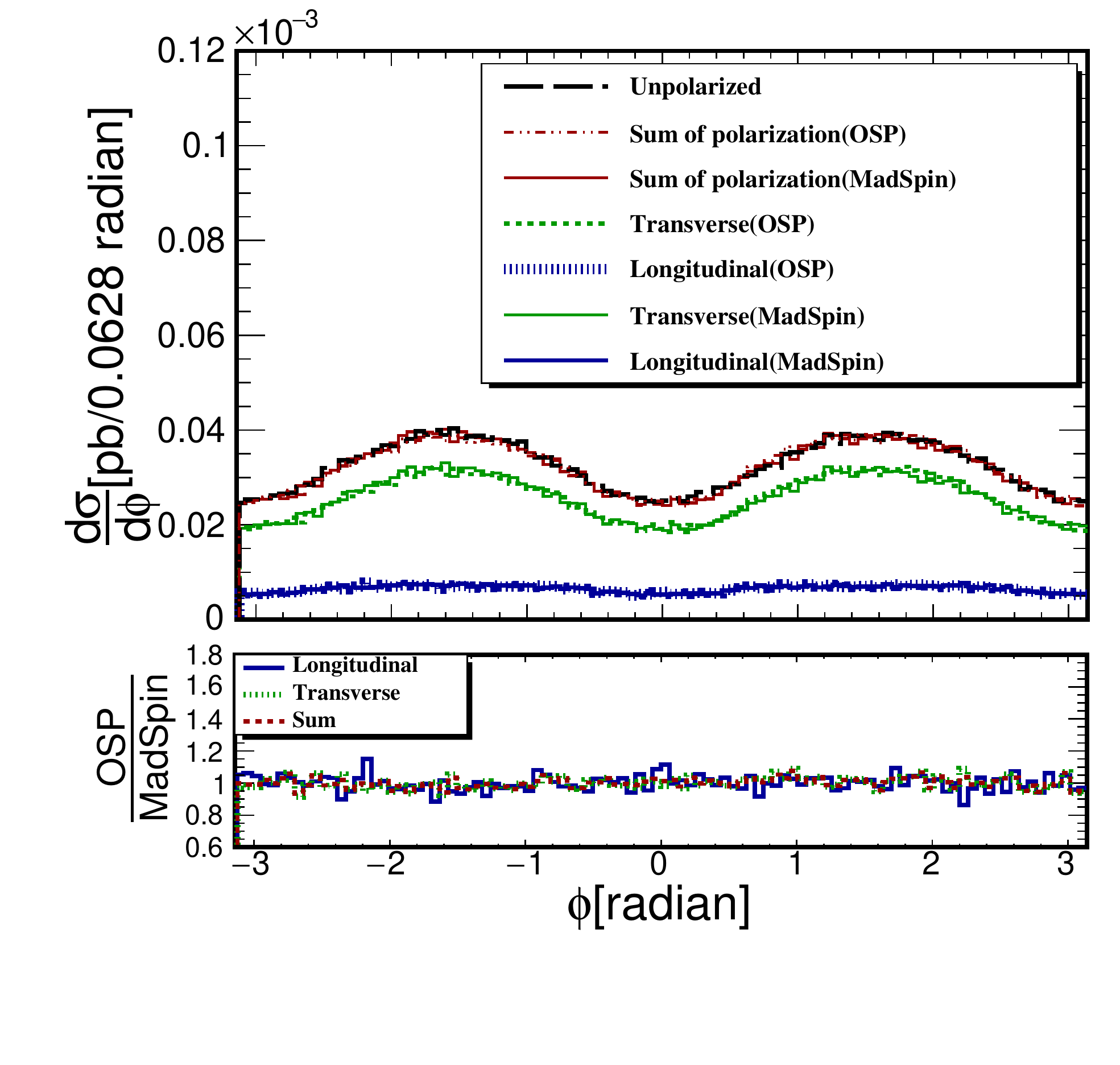}		\label{fig:AngdistDecayCutsPhi}	}
\end{center}
\caption{
Upper Panels:
For (a,c) $\cos\theta$ and (b,d) $\phi$, 
overlapping distributions of unpolarized $jj W^+W^-$ production (black; dash-double dot),
transversely polarized $W_\lambda^-$ production (green),
longitudinally polarized $W_\lambda^-$ production (blue), and
polarization-summed $W_\lambda^-$ production (red),
with $W^+W^-$ decays treated using the MadSpin (solid) and OSP (bar) methods,
assuming (a,b) only generator-level cuts of \eq{eq:SMVBScuts}
and (c,d) both  \eq{eq:SMVBScuts} and analysis-level cuts of \eq{eq:decaycuts}.
Lower Panels: The OSP-to-MadSpin ratio of the polarized and polarization-summed distributions.
}
\label{fig:VBSpol}
\end{figure*}

%%%%%%%%%%%%%%%%%%%%%%%%%%%%%%%%%%%%%%%%%%%%%%%%%%%%
%%%%%%%%%%%%%%%%%%%%%%%%%%%%%%%%%%%%%%%%%%%%%%%%%%%%
\subsubsection*{Leptonic Observables from Polarized $W$ Boson Decays}

We now turn to kinematic observables built from final-state charged leptons in the EW process of \eq{eq:vbs_sm_ProcDef}.
Throughout this section we present in upper panels of plots 
overlapping distributions for unpolarized $jj W^+W^-$ production (black; dash-double dot),
transversely polarized $W_\lambda^-$ production (green),
longitudinally polarized $W_\lambda^-$ production (blue), and
polarization-summed $W_\lambda^-$ production (red).
$W^+W^-$ decays are treated using the MadSpin (solid) and OSP (bar) methods.
For unpolarized production, we only use MadSpin.
To quantify potential disagreement between the two decay techniques
and unless specified,  for each observable we also report in lower panels of plots the OSP-to-MadSpin ratio of the polarized and polarization-summed curves.
In summary, we find good agreement with the shape and normalization between the MadSpin and OSP samples.
Differences are consistent with MC statistics and therefore demonstrate strong checks of both the methods.

\begin{figure*}[t!]
\begin{center}
\subfigure[]{\includegraphics[width=.48\textwidth]{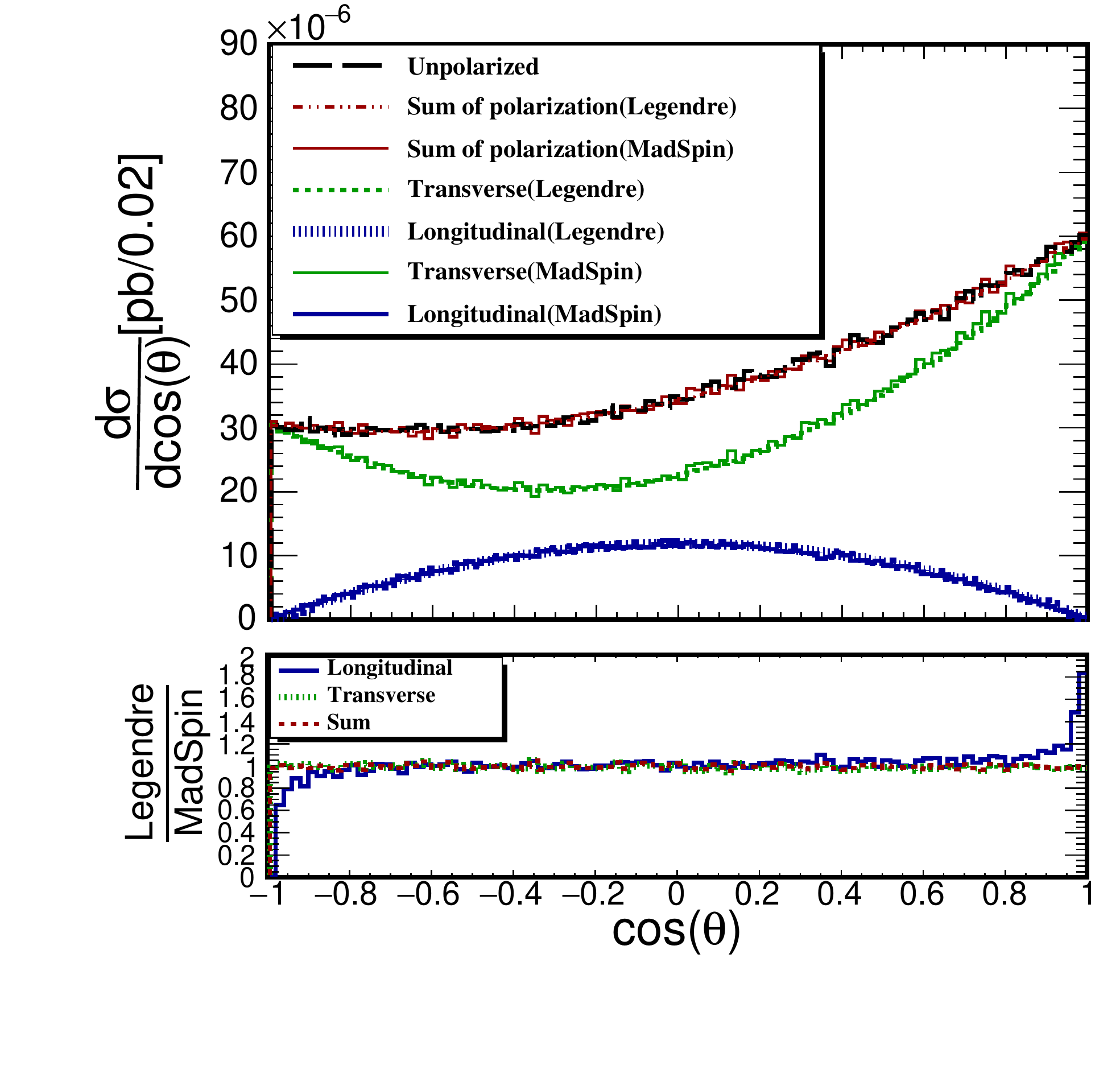}		\label{fig:Legendre}}
\subfigure[]{\includegraphics[width=.48\textwidth]{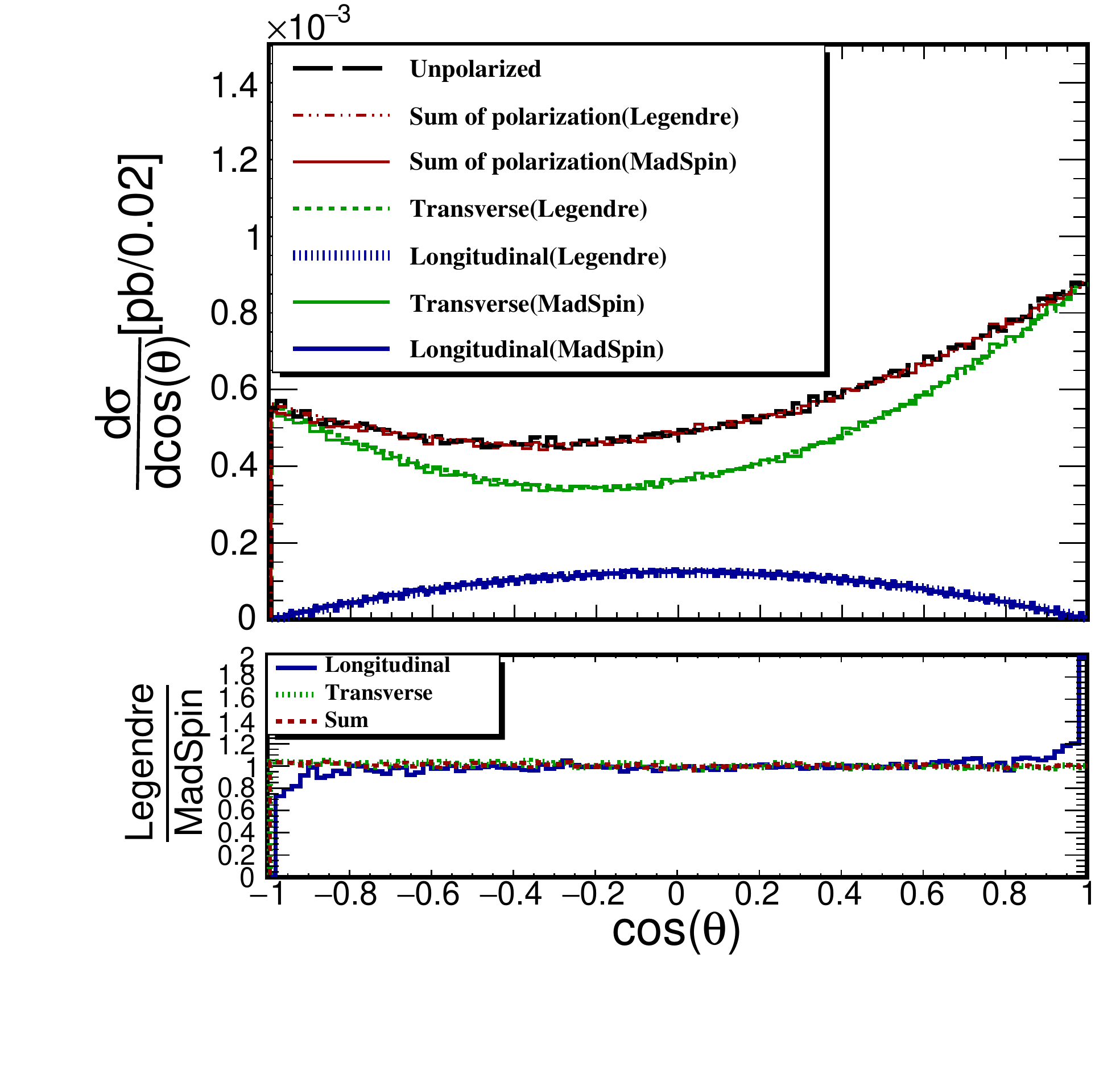}	\label{fig:Legendre_qcd}}
\end{center}
\caption{
Same as \fig{fig:VBSpolCosTh} but for the MadSpin and the Legendre Expansion methods,
assuming (a) pure EW and (b) mixed EW-QCD production of $jj W^+W^-$ at LO.
}
\label{fig:Legendre_jjWW}
\end{figure*}

We start with \fig{fig:VBSpolCosTh}, which shows the polar distribution $\cos\theta$ as defined in \eq{eq:vbs_sm_polarDef}.
As anticipated from \eq{eq:wDiffPolar}, we observe that the longitudinal component of $W^-_\lambda$
exhibits a polar dependence that behaves as  $d\sigma(W_{\lambda=0})/d\cos\theta \sim \sin^2\theta$, 
while the transverse modes are given as a coherent sum of left and right contributions. 
We see a preference for $d\sigma(W_{\lambda=T})/d\cos\theta > 0$, indicating that $f_L > f_R$, 
and consistent with above arguments that the production of $W_{\lambda=-1}^-$ is preferred over $W_{\lambda=+1}^-$ at the LHC.

In \fig{fig:VBSpolPhi} we show the azimuthal distribution $\phi$ as defined in \eq{eq:vbs_sm_aziDef}.
Notably, the longitudinal mode exhibits a flat behavior and the transverse modes oscillate.
This follows from \eq{eq:wDiffPolarAzi}, which shows that the $\lambda=0$ polarization is only 
sensitive to $\phi$ through $\lambda=T$ interference terms; these are neglected in polarized computations.
Consistently, for $\lambda=T$, we observe the $\cos2\phi$ behavior that originates from the $\lambda=\pm1$ interference, which is modeled since $\lambda=\pm1$ modes are summed coherently.
In comparison to the unpolarized sample, which sums all polarizations coherently,
and the semi-incoherent sum of $\lambda=0$ and $T$ polarizations, we find that the difference is small for both the MadSpin and OSP schemes.
This suggests that the $g_{L0}, g_{R0}$ interference terms are small, and that the interference is dominated by $g_{RL}$.
 The difference between the unpolarized curve and the MadSpin sum quantifies the interference between $\lambda=0$ and $T$ modes, and is small.

To extract the polarization fractions $f_{\lambda}$ from the distribution in \fig{fig:VBSpolCosTh},
 we use the Legendre expansion technique as used by Ref.~\cite{Ballestrero:2017bxn} for VBS,
 which is related to the moment method used by Refs.~\cite{Bern:2011ie,Stirling:2012zt} for $pp\to W^\pm_\lambda+nj$.
We start by noting that the polar distribution of \eq{eq:wDiffPolar} can be written in term of first Legendre polynomials, with
\begin{equation}
\frac{1}{\sigma} \frac{d\sigma}{d\cos\theta}
= \sum_{l=0}^2 \alpha_l P_l(\cos\theta) \, ,\quad\text{and}\quad
\alpha_l = \frac{2l+1}{2} \, \int_{-1}^{1} d\cos\theta \,\frac{1}{\sigma}
\frac{d\sigma}{d\cos\theta} P_l (\cos\theta) \,.
\label{eq:LegendreExpansion}
\end{equation}
After explicit integration, the polarization fractions in terms of Legendre coefficients are:
\begin{align}\label{eq:inversion_eq}
f_L = \frac{2}{3}(\alpha_0 + \alpha_1 + \alpha_2), \quad
f_R = \frac{2}{3}(\alpha_0 - \alpha_1 + \alpha_2), \quad
f_0= \frac{2}{3}(\alpha_0 - 2\alpha_2).
\end{align}
We can extract the values of $\alpha_l$ for $l=0,\dots,2$, from our simulated predictions (or from data) by performing a sum over each histogram bin
and by approximating the $\alpha_l$  integral:
\begin{eqnarray}
\alpha_l &= &  \frac{2l+1}{2} \int_{-1}^{1} dx ~ g(x) ~ P_l(x), \qquad g(x)\equiv  \frac{1}{\sigma} \frac{d\sigma}{d\cos\theta}\Bigg\vert_{\cos\theta=x}
\\
&=&  \frac{2l+1}{2} \sum_{\rm bins~k} \int_{x_k}^{x_{k+1}} dx ~g(x)~ P_l (x), 
\\
&\approx&  \frac{2l+1}{2} \sum_{\rm bins~k} ~ g(x^*) ~ \int_{x_k}^{x_{k+1}} dx  ~ P_l (x), 
\\
&=&  \frac{1}{2} \sum_{\rm bins~k} ~ g(x*)~ \left[ P_{l+1} (x)  -  P_{l-1} (x)  \right]^{x_{k+1}}_{x_k}.
\label{eq:LegendreIntegral}
\end{eqnarray}
In the first line, we make the change of variable $x=\cos\theta$ for clarity.
In the second, we partition the integral into a large number of disjoint integrals over continuous ranges (bins), 
such that the bin widths satisfy $\vert x_{k+1}-x_{k}\vert \ll 1$. 
We then factor the normalized histogram weight $g(x)$ using the Mean Value Theorem, and thereby approximate $g(x)$ as a constant $g(x^*)$ for $x^*\in [x_{k},x_{k+1}]$.
This allows us to evaluate the integrals exactly.

Choosing $x^*=x_{k}$, i.e., the bin starting boundary, we obtain the following fractions
\begin{equation}
f_L=0.5264\pm~ 0.3\%,	\quad 
f_R=0.2658\pm~ 0.6\%,	\quad 
f_0=0.2077\pm~ 1\%.
\label{eq:legendreFit}
\end{equation}
The uncertainty we report is statistical, but other theory uncertainties, e.g., scale uncertainties, can be propagated in a straightforward manner.
The reconstructed distribution in $\cos\theta$ is shown in \fig{fig:Legendre}, together with the simulated expression.
For nearly the entire domain of $\cos\theta$, we report a good reproduction of the MC simulation using Legendre polynomials.
A large disagreement is observed at the boundaries, near $\cos\theta=\pm1$, for the $\lambda=0$ distribution.
This can be attributed simply to the fact that the distribution $d\sigma(W^-_{\lambda=0})$ itself vanishes smoothly at the endpoints,
resulting in an ill-defined ratio.

\subsubsection*{Impact of Selection Cuts on Polarized Distributions}

\begin{figure*}[h!]
\begin{center}
\subfigure[]{\includegraphics[width=.48\textwidth]{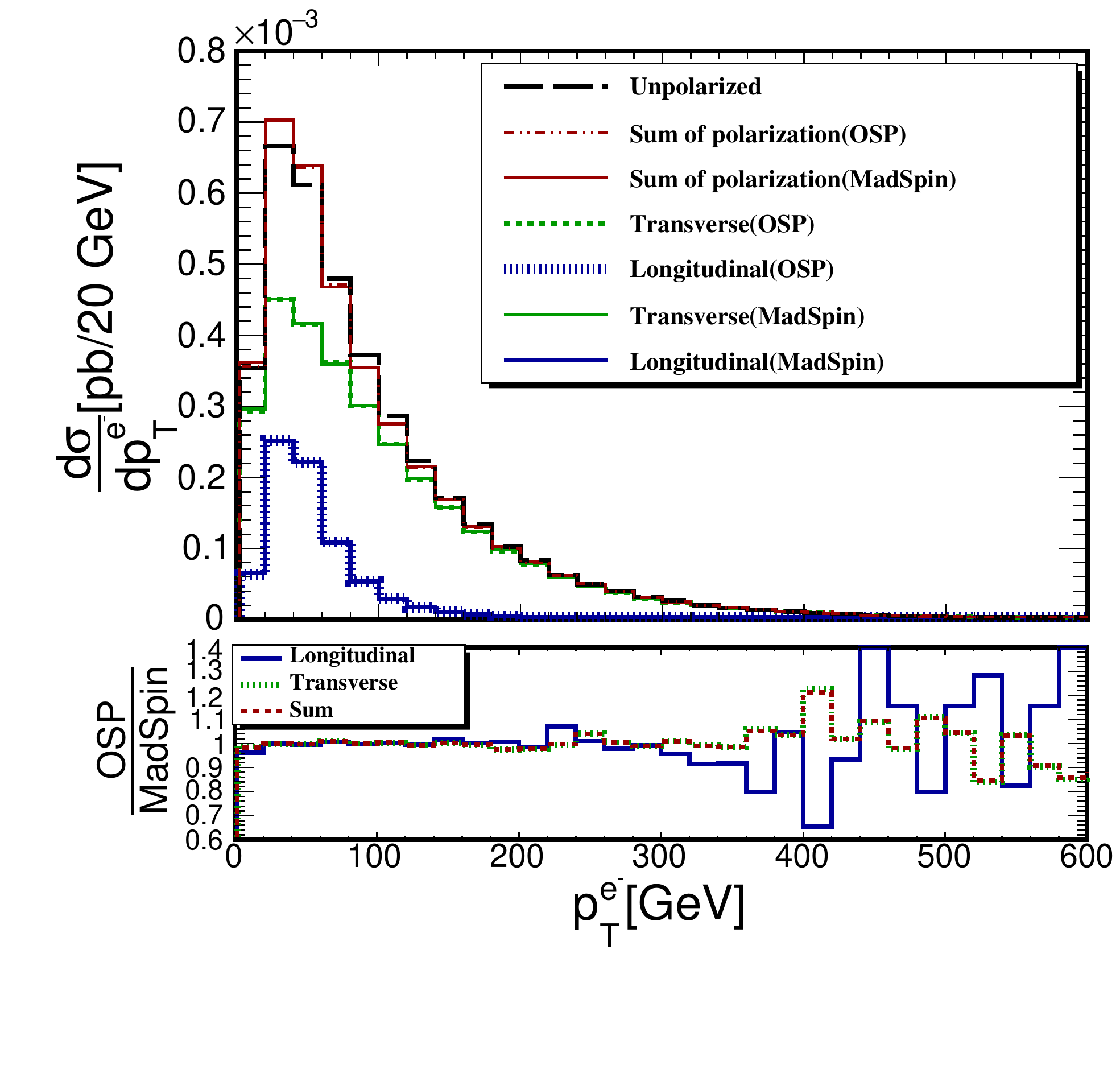}		\label{fig:distPTe}	}
\subfigure[]{\includegraphics[width=.48\textwidth]{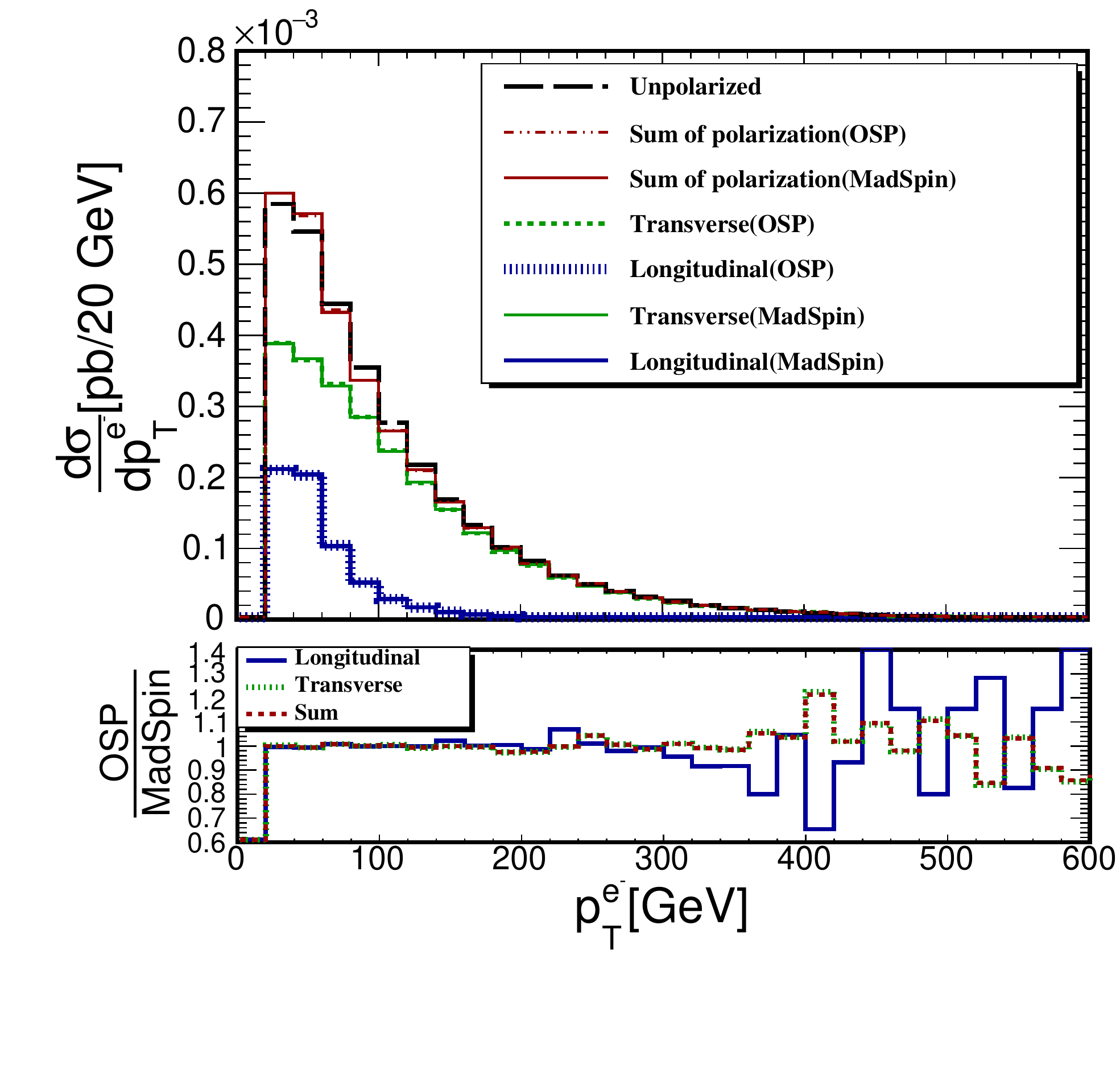}	\label{fig:distPTe}	}
\\
\subfigure[]{\includegraphics[width=.48\textwidth]{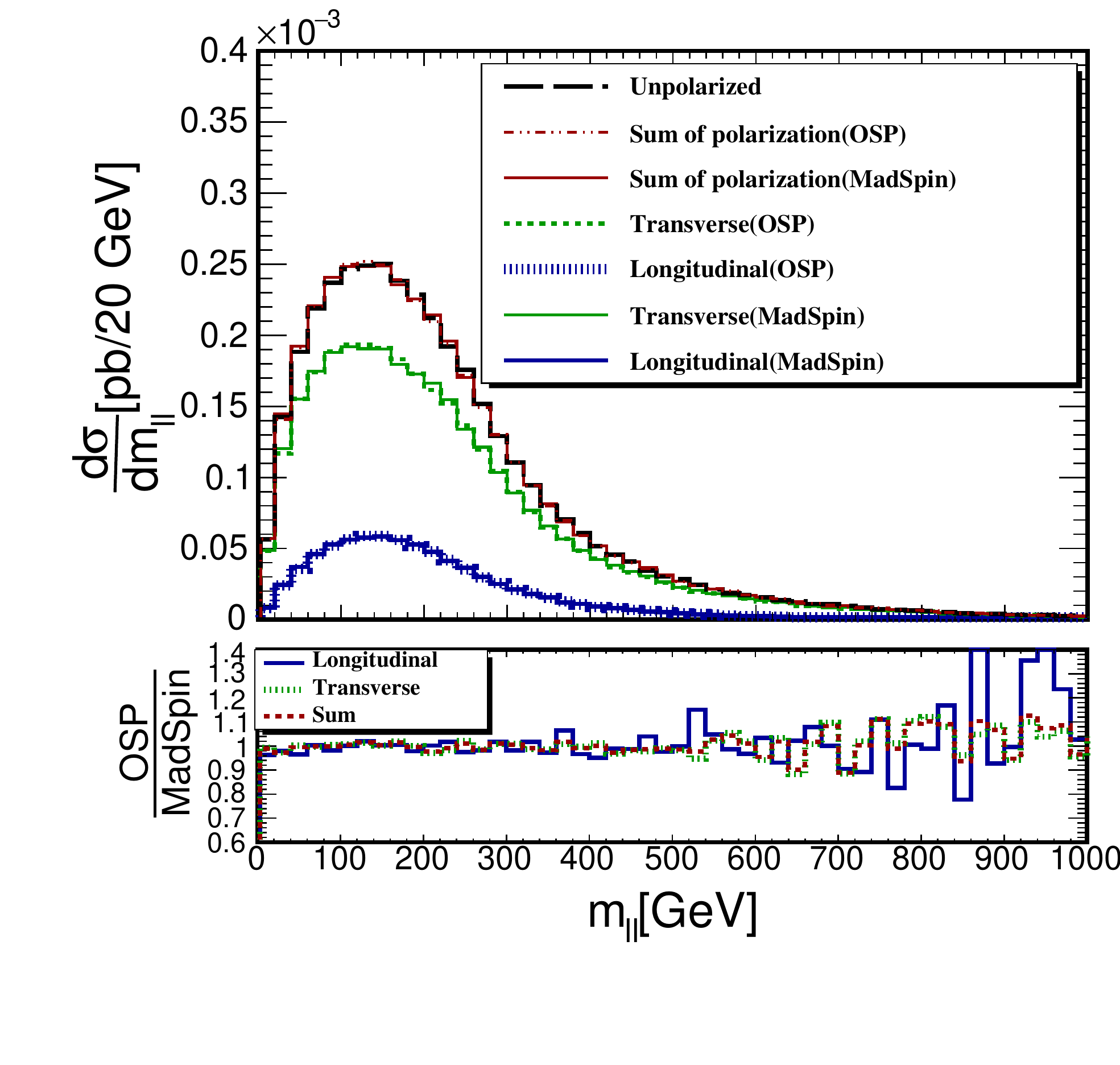}		\label{fig:distmll}	}
\subfigure[]{\includegraphics[width=.48\textwidth]{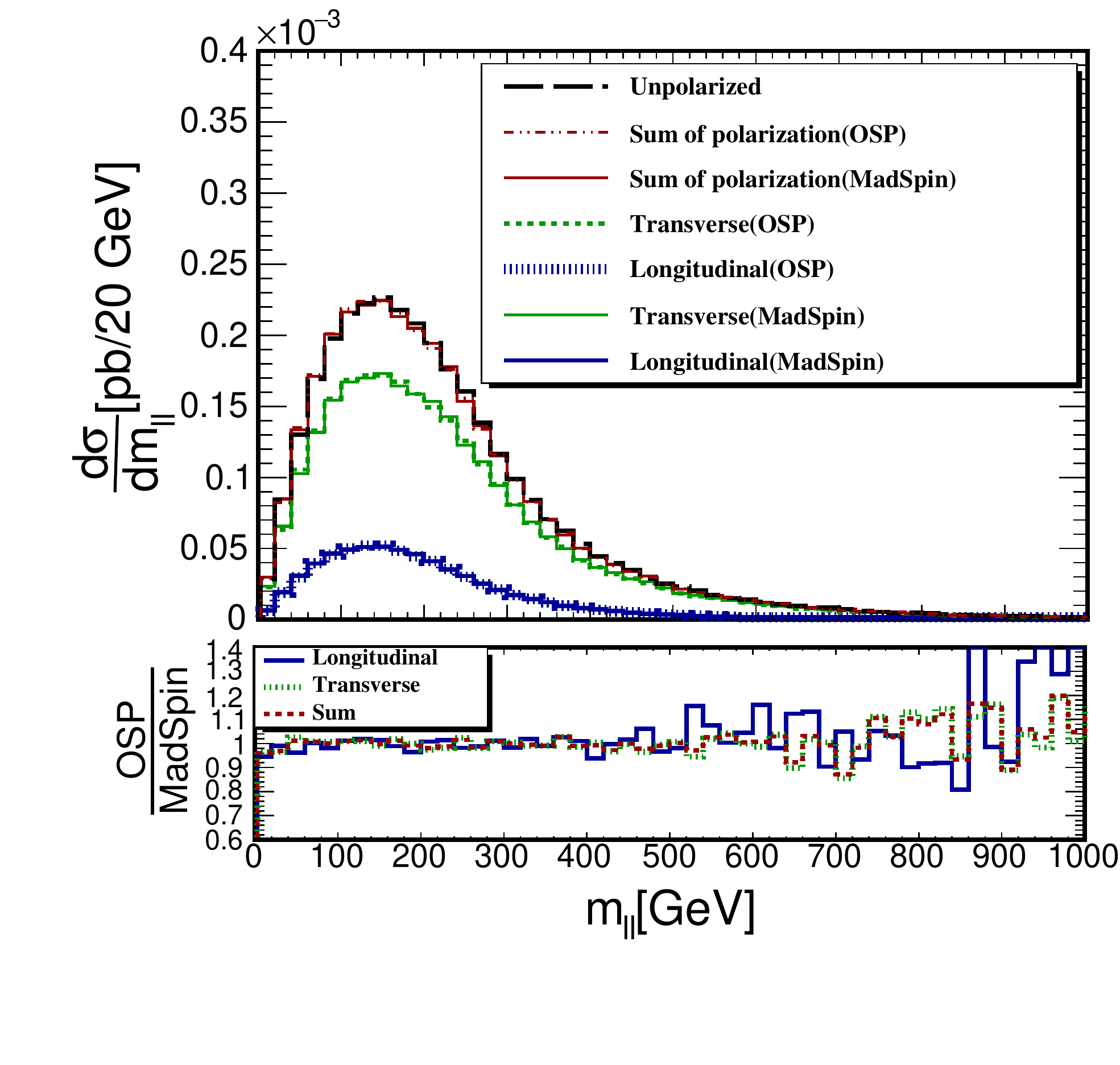}	\label{fig:distmll}	}
\end{center}
\caption{
Same as \fig{fig:VBSpol} but for (a,b) $p_T(e^-)$ and (c,d) $M(\mu^+e^-)$,
assuming (a,c) only generator-level cuts of \eq{eq:SMVBScuts}
and (b,d) both  \eq{eq:SMVBScuts} and analysis-level cuts of \eq{eq:decaycuts}
}
\label{fig:decaydist1}
\end{figure*}

\begin{figure*}[h!]
\begin{center}
\subfigure[]{\includegraphics[width=.48\textwidth]{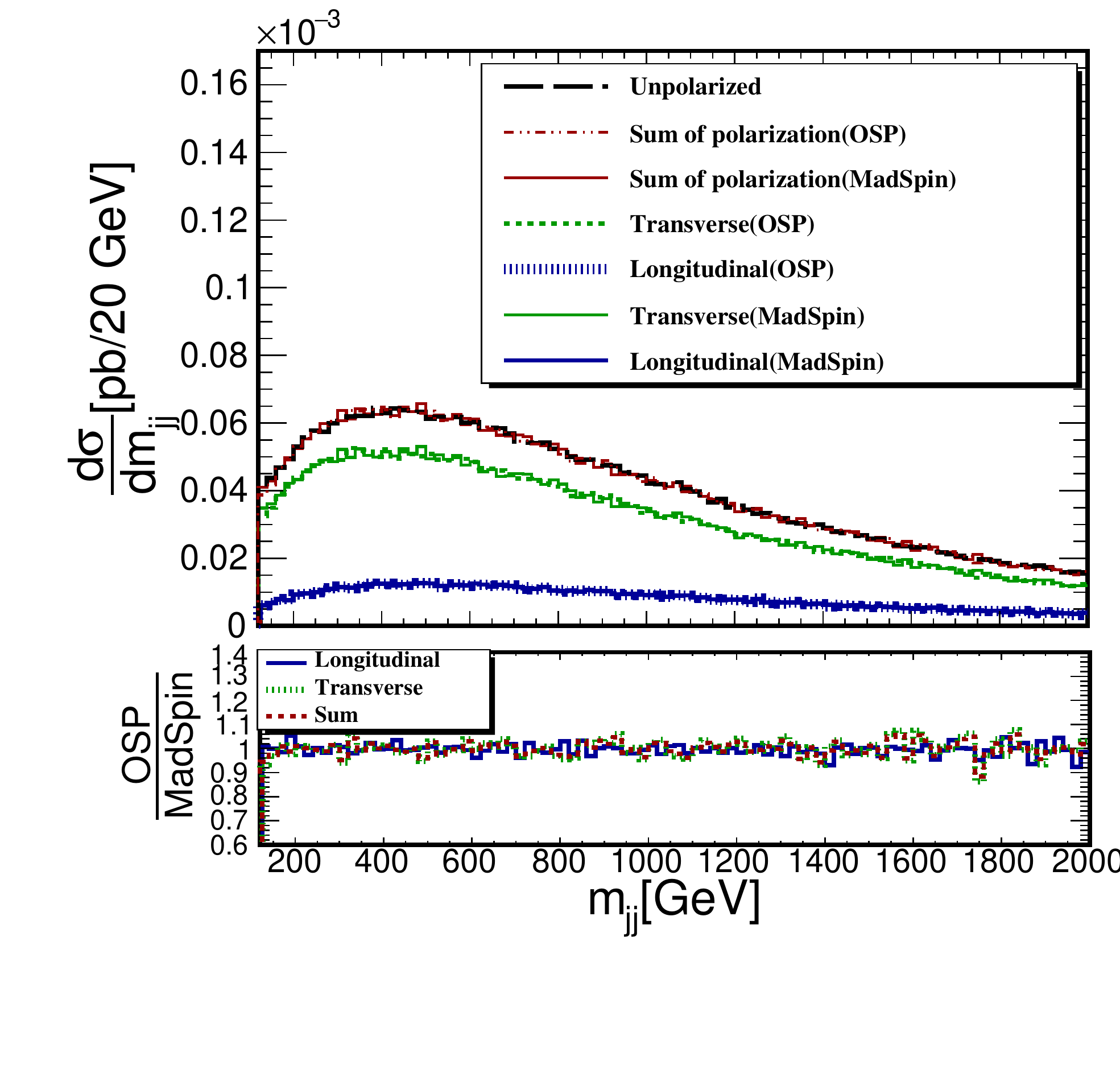}			\label{fig:distmjj_genCuts}		}
\subfigure[]{\includegraphics[width=.48\textwidth]{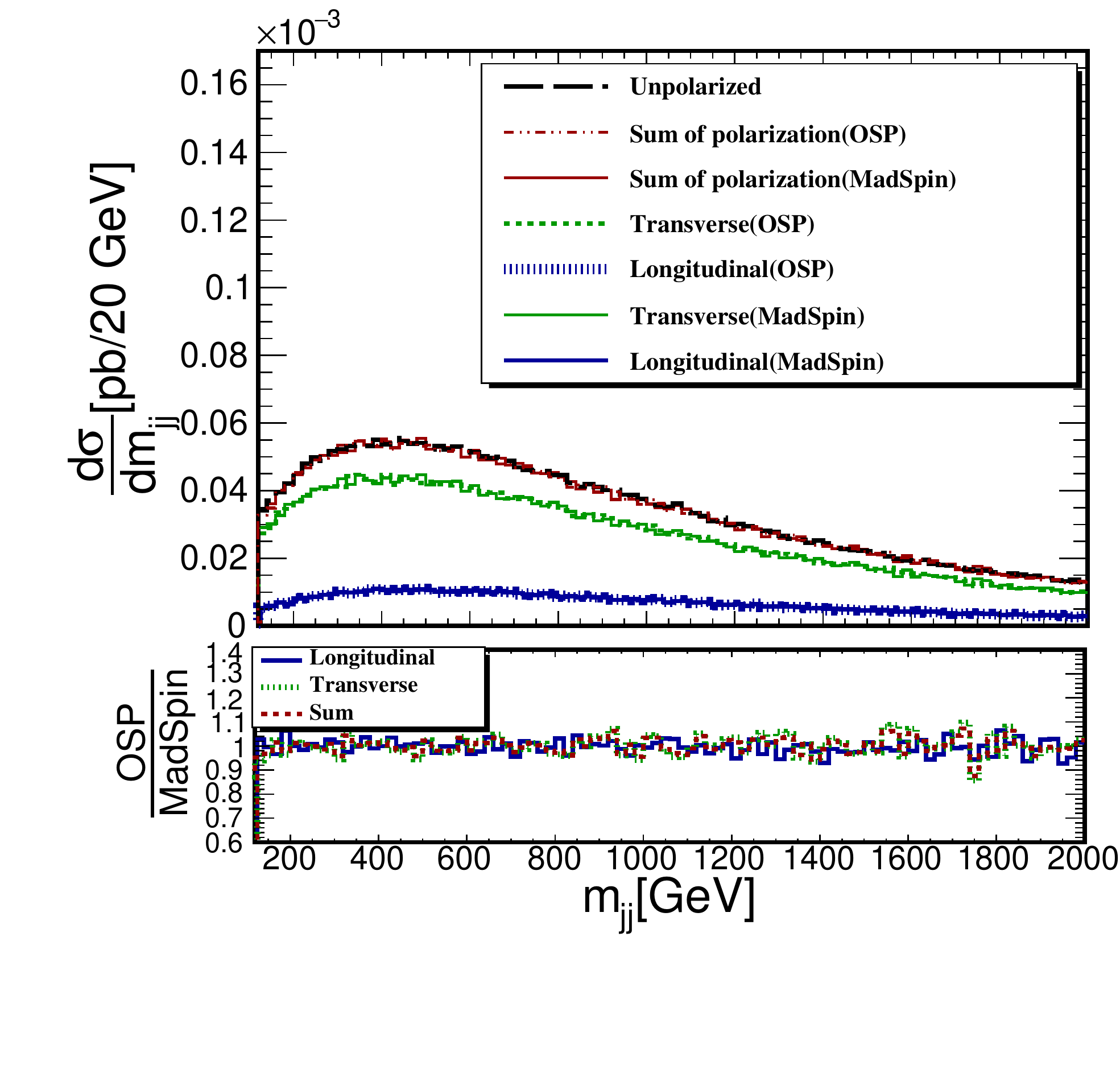}		\label{fig:distmjj_anaCuts}		}
\\
\subfigure[]{\includegraphics[width=.48\textwidth]{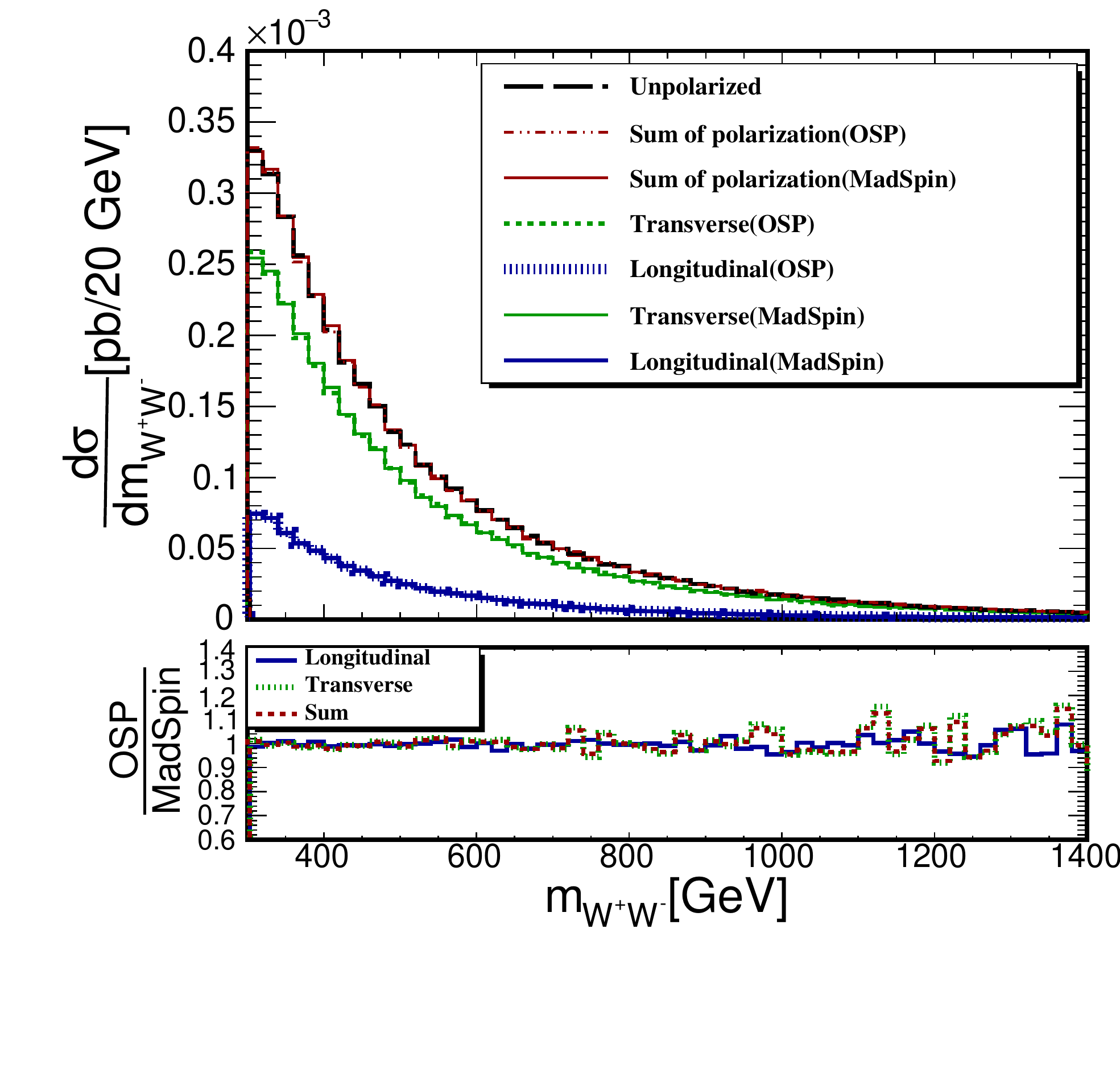}			\label{fig:distMWW_genCuts}	}
\subfigure[]{\includegraphics[width=.48\textwidth]{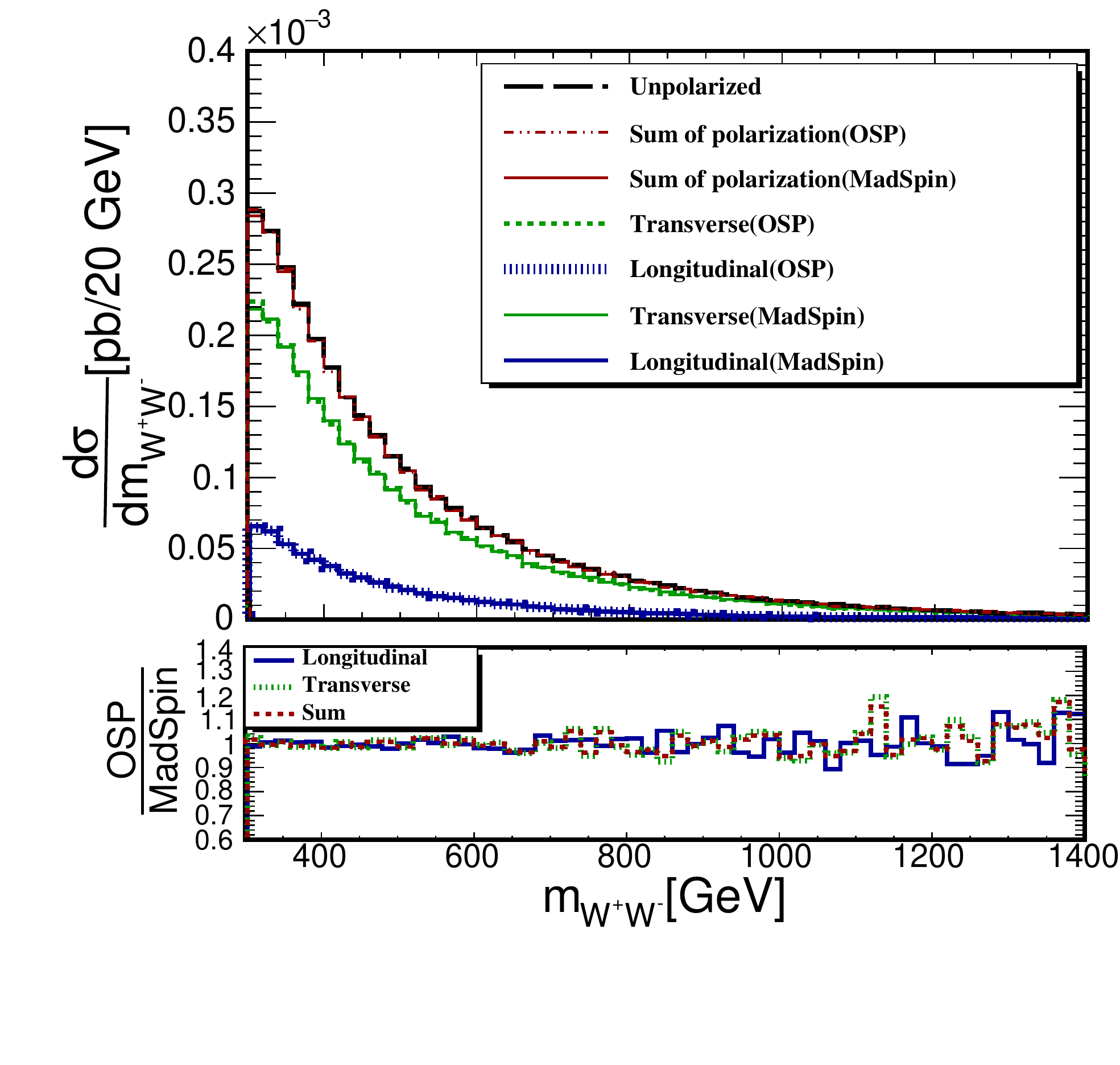}		\label{fig:distMWW_anaCuts}	}
\end{center}
\caption{
Same as \fig{fig:decaydist1} but for (a,b) $M(jj)$ and (c,d)  $M(W^+W^-)$.
}
\label{fig:decaydist2}
\end{figure*}

\begin{figure*}[t!]
\begin{center}
\subfigure[]{\includegraphics[width=.48\textwidth]{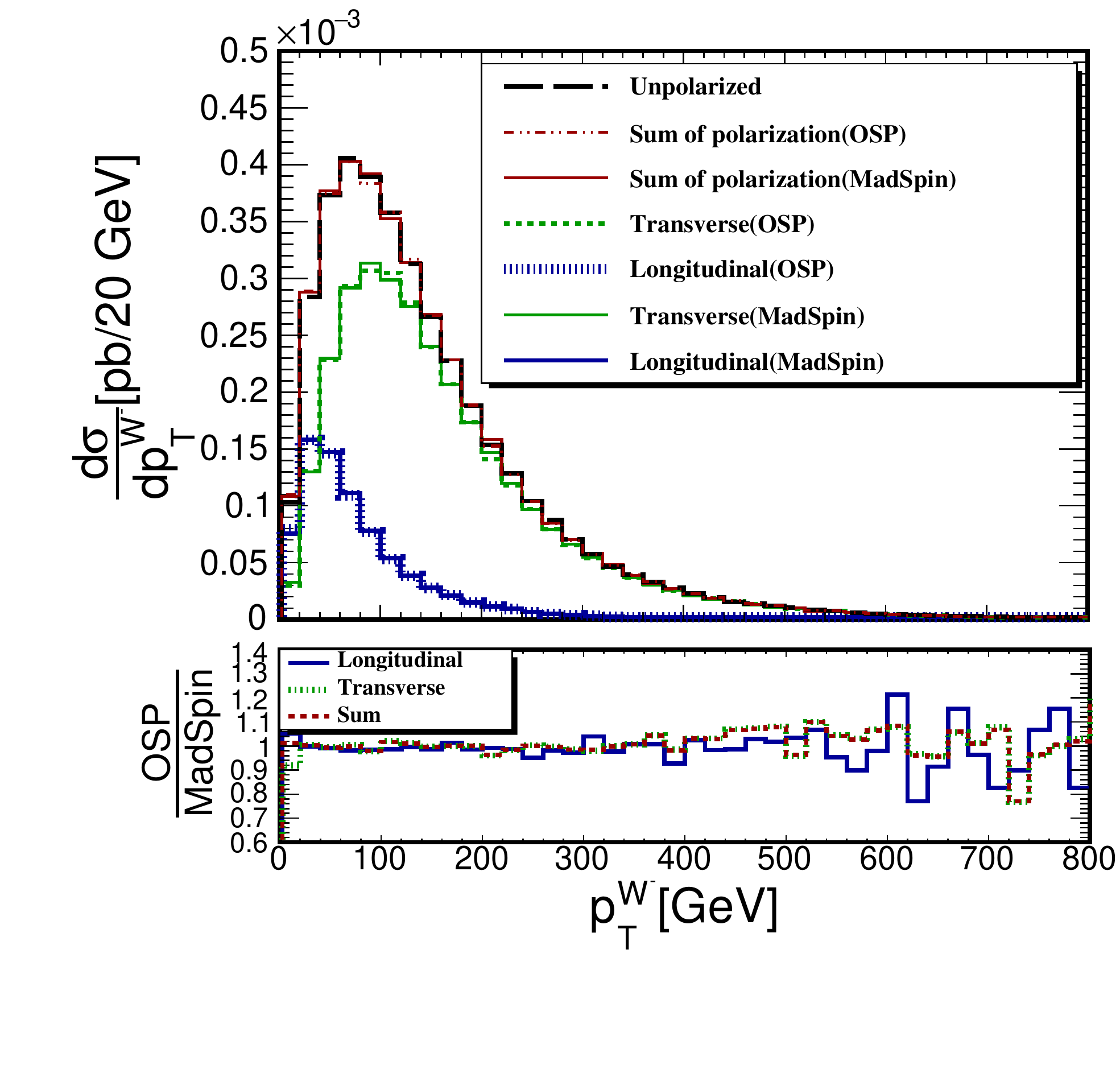}			\label{fig:distPTW_genCuts}	}
\subfigure[]{\includegraphics[width=.48\textwidth]{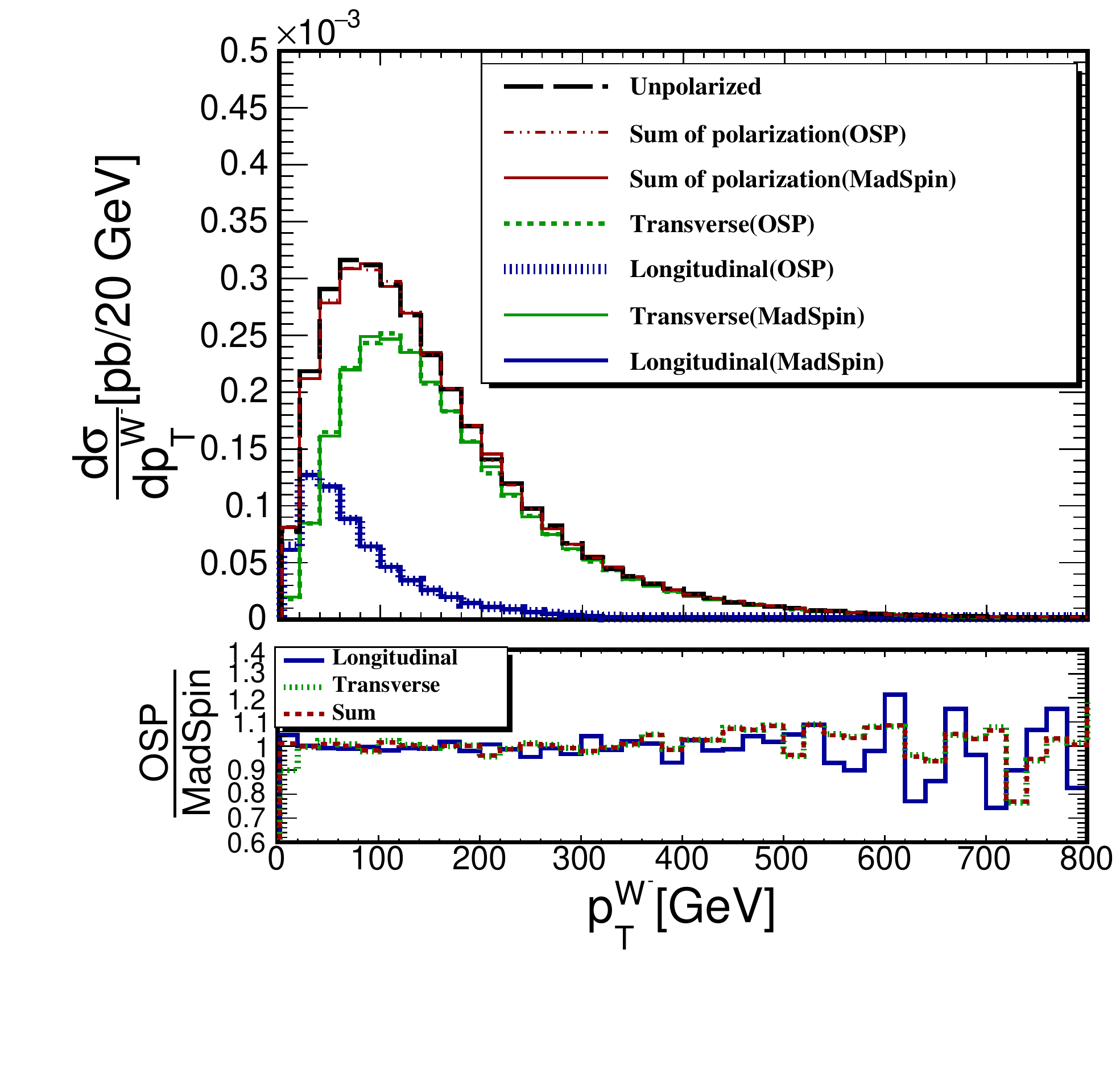}		\label{fig:distPTW_anaCuts}	}
\\
\subfigure[]{\includegraphics[width=.48\textwidth]{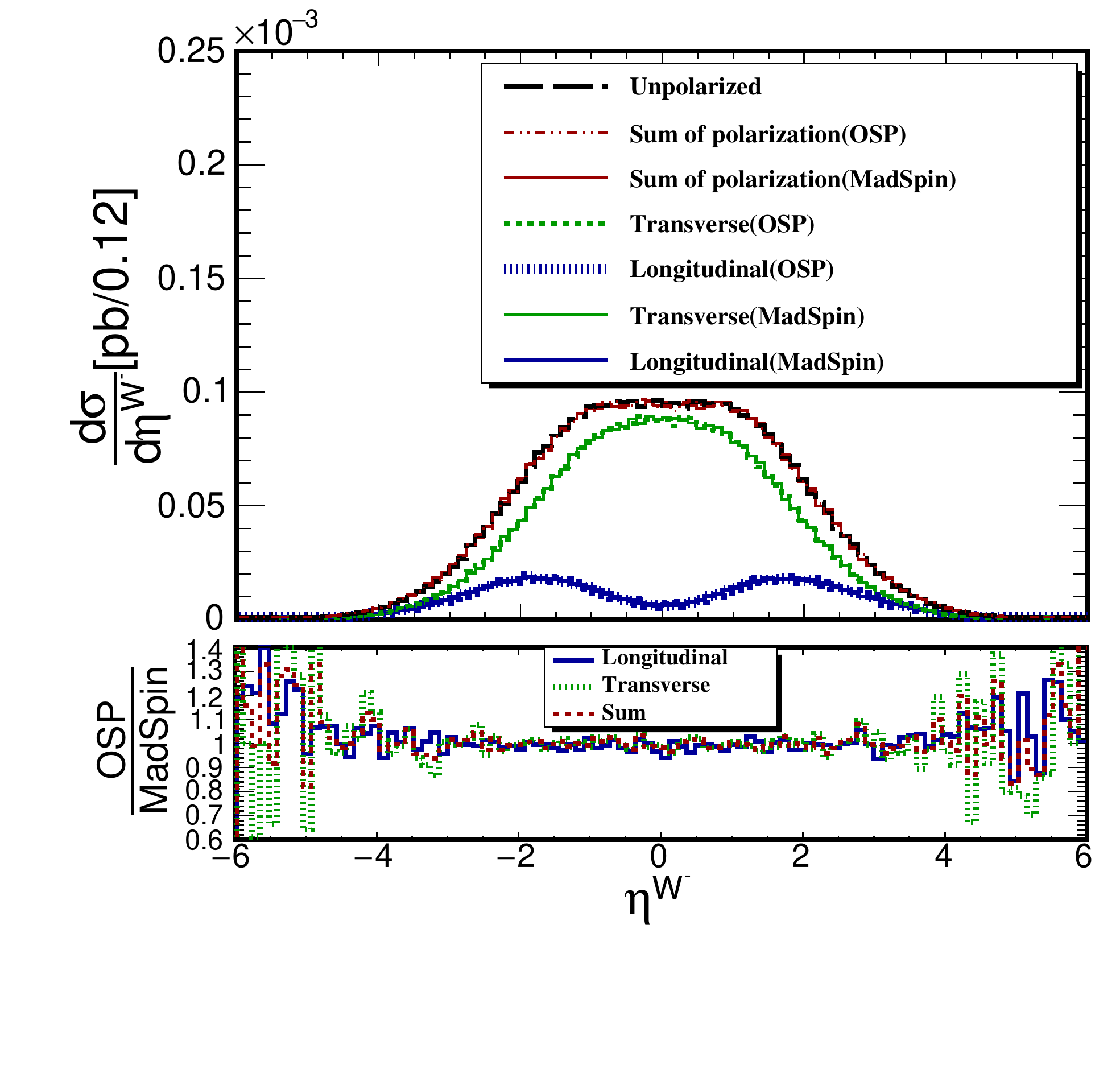}			\label{fig:distETAW_genCuts}	}
\subfigure[]{\includegraphics[width=.48\textwidth]{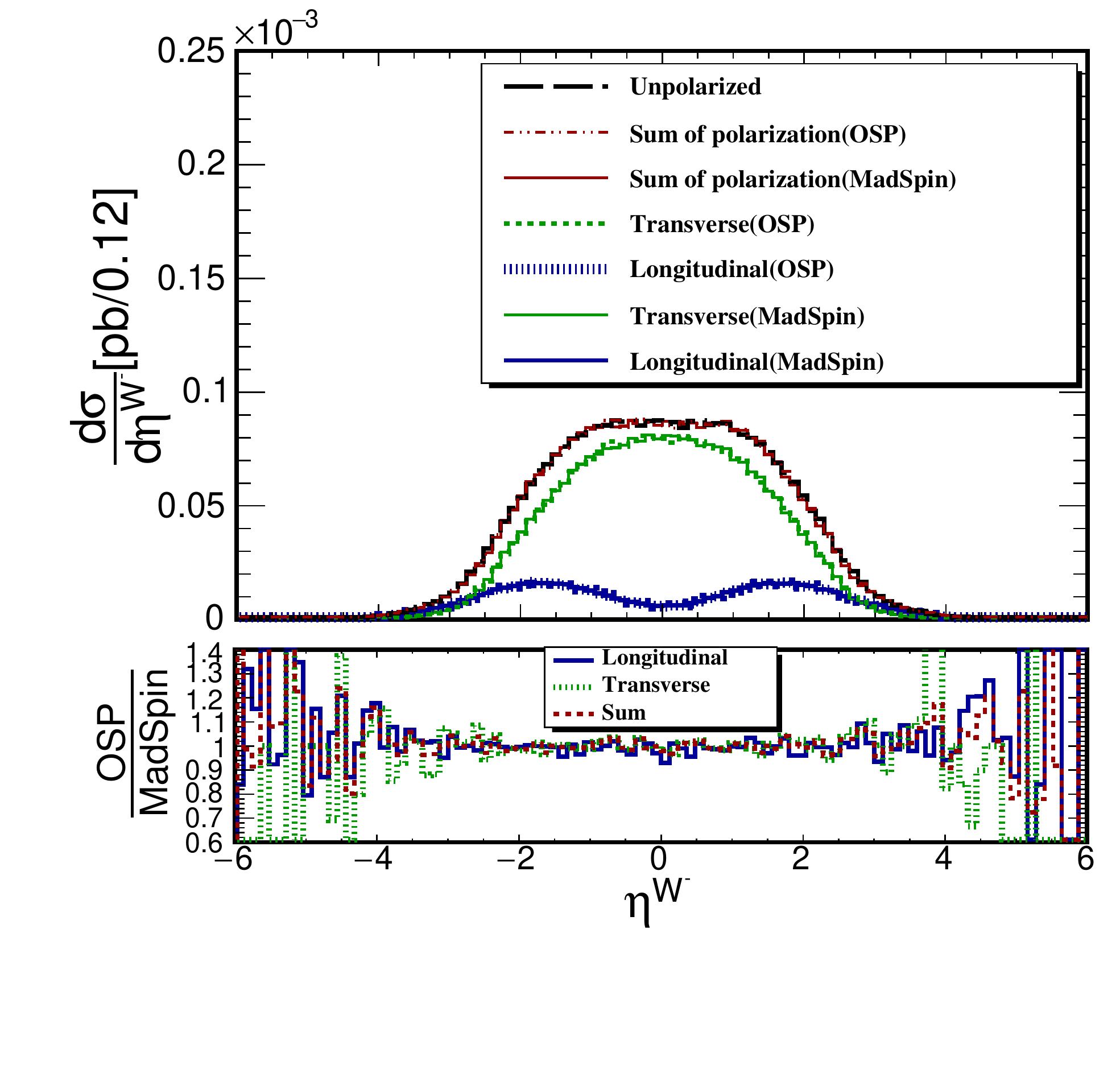}		\label{fig:distETAW_anaCuts}	}
\end{center}
\caption{
Same as \fig{fig:decaydist1} but for (a,b)  $p_T(W^-)$ and (c,d) $\eta(W^-)$.
}
\label{fig:decaydist3}
\end{figure*}

As noted in the double differential distribution in \eq{eq:wDiffPolarAzi},
interference terms between different $W^-_\lambda$ polarizations possess a dependence on the azimuthal angle $\phi$.
Hence, observables integrated over $\phi$ are insensitive to the interference between the different polarization terms. 
For such observables, the incoherent sum of transverse and longitudinal contributions agrees with the unpolarized prediction. 
However, it is known that realistic experimental conditions are not totally inclusive with respect to $\phi$ due to kinematical selection cuts,
which are motivated by detector acceptance or analysis criteria. 

To study the impact of realistic selection cuts on our modeling of polarized matrix elements
 as well as the residual size of possible polarization interference,
 we consider the following selection cuts applied to the decay products of $W^{-}_{\lambda'}$:
\begin{equation}
p_T(e^{-})>20\GeV,\quad |\eta(e^{-})|< 2.5,\quad \Delta R(j, e^{-})>0.4.
\label{eq:decaycuts}
\end{equation}
The selection cuts here are applied at the analysis-level and are in addition to the generator-level cuts of \eq{eq:SMVBScuts}.
We report the resulting cross section for both the MadSpin and OSP samples in \tab{tab:SMVBSxs}.
Observed differences between the two are consistent with MC statistical uncertainty and rounding errors.
We find that while there is a $20-25\%$ reduction in cross section, the $W^-_\lambda$ polarization fractions remain essentially the same.
We do, however, see the emergence of a sub-percent discrepancy between the incoherent sum of  helicity-polarized cross sections
and the full, unpolarized cross section 
when the selection cuts of \eq{eq:decaycuts} are applied to decay products.

In \fig{fig:AngdistDecayCutsTh} and \fig{fig:AngdistDecayCutsPhi}, we show the same polar and azimuthal observables described above and shown in \fig{fig:VBSpolCosTh} and \fig{fig:VBSpolPhi} 
but after applying the selection cuts of \eq{eq:decaycuts}. 
In comparing to the distributions without cuts, a large impact can be noted, namely a total depletion of events in the neighborhood of $\cos\theta=-1$.
This results in an increased forward-backward asymmetry and stems from the softer nature of ``backwards'' flying $e^-$ originating from $W_L^-$ decays,
 which are selected out by the $p_T^l$ cut in \eq{eq:decaycuts}~\cite{Bern:2011ie,Stirling:2012zt}. 
As the concavity of the $\lambda=T$ curve, and hence the helicity polarization-summed curve as well, 
now changes multiple times as a function of $\cos\theta$, the functional form of \eq{eq:wDiffPolar} does not serve as a good description of \fig{fig:AngdistDecayCutsTh}.
Therefore, to recover polarization fits as reported in \eq{eq:legendreFit}, modern unfolding techniques are necessary; 
see, for example, Refs.~\cite{DAgostini:1994fjx,Prosper:2011zz,Balasubramanian:2019itp,Bellagente:2019uyp,Andreassen:2019cjw} and references therein.
Such techniques ``correct'' reconstructed distributions/observables for real, detector-level and analysis-level acceptance efficiencies, enabling more direct comparisons to truth-level, 
MC predictions~\cite{DAgostini:1994fjx,Prosper:2011zz}.
This, however, comes at the cost of introducing systematic uncertainties stemming from imperfect modeling of underlying physics and detector response.
The availability of a polarized MC event generator, which is a main result of this work, can significantly help to reduce such uncertainties.
For example: the ability to generate specific helicity samples provides a means to directly model detector responses
 to kinematic regions that are strongly suppressed in the SM or exhibit ultra low detector acceptance efficiencies, say from a forward-backward asymmetry,
 and subsequently help ameliorate singularities that may otherwise appear in unfolding response matrices.
 
For the azimuthal distribution, we observe  similar shapes to the generator-level cut curves, albeit with larger maxima and minima differences.
Lastly, to reiterate, we observe good agreement between the OSP and MadSpin samples.

In \fig{fig:decaydist1}(a) and \fig{fig:decaydist1}(b)  we show the $p_{T}^{e^{-}}$ distributions before and after the cuts on the decay products (\eq{eq:decaycuts}) respectively. 
We observe a small difference  between the incoherent sum of polarizations with respect to the unpolarized simulation,
which we attribute to the interference between longitudinal and transverse polarizations in some restricted region of phase space. 
In \fig{fig:decaydist1}(c) and \fig{fig:decaydist1}(d)  we show the invariant mass of the di-lepton system $m(e^{-}, \mu^{+})$ before and after the cuts in \eq{eq:decaycuts} respectively.
Unlike the $p_{T}^{e^{-}}$, no difference between unpolarized and polarized samples can be observed. 

Turning to more reconstructed objects, we show in \fig{fig:decaydist2}(a,b) the $m_{jj}$  distribution and in (c,d) the $M(W^{+}W^{-})$ distribution,
assuming only (a,c) generator-level cuts and (b,d) with the cuts of \eq{eq:decaycuts}.
We find that both before and after \eq{eq:decaycuts} the observables are insensitive to interference and that the incoherent polarization sum describes the unpolarized distributions well. 
We attribute this insensitivity to the fact that interference effects appear first at the $W_\lambda^-$ decay level, though the angle $\phi$ as defined in \eq{eq:vbs_sm_aziDef}.
By working at the $W^-_\lambda$ level, we are inclusive with respect to $\phi$, leading to a washout of interference effects.
By identical arguments, an insensitivity to  interference can be found in \fig{fig:decaydist3},
where we show (a,b) the $p_{T}^{W^{-}}$ distribution and in (c,d) $\eta^{W^{-}}$ distribution,
assuming only (a,c) generator-level cuts and (b,d) with the cuts of \eq{eq:decaycuts}.

In all  distributions and cross sections we find good agreement between the \ms~ and OSP method. 
We also find the interference effect between transverse and longitudinal polarization channels to be small, and the incoherent sum of polarization describes the distributions we consider to a good degree. 
The largest difference, although still small, is observed in the $p_T(e^-)$ distributions.
The difference remains negligible even after applying selection cuts defined in \eq{eq:decaycuts}.
Of course, this observation somewhat follows the fact that this process is dominated by transverse modes, 
it is hard to access the effect of the longitudinal bosons, 
and interference with the transverse modes are even less accessible.

%%%%%%%%%%%%%%%%%%%%%%%%%%%%%%%%%%%%%%%%%%%%%%%%%%%%%%%%%%%%%%%%%
%%%%%%%%%%%%%%%%%%%%%%%%%%%%%%%%%%%%%%%%%%%%%%%%%%%%%%%%%%%%%%%%%
\subsection{Polarized $W$ Bosons in Mixed EW-QCD Production of $jj W^+ W^-$}\label{sec:vbs_qcd}

\begin{table}[!t]
\begin{center}
\begin{tabular}{ c || c || c  c | c  c }
\hline
\hline
			&  	Decay Scheme
			& 	\multicolumn{2}{c|}{Generator-Level Cuts}
			&	\multicolumn{2}{c}{Analysis-Level Cuts} 
			\tabularnewline
Process		&	& $\sigma$ [fb]		& $f_{\lambda}$	&	$\sigma$ [fb]		& $f_{\lambda}$		\tabularnewline
\hline
$jjW^{+}W^{-}$		& MadSpin		&  56.61   			&  $\dots$         		& 47.86  			& $\dots$		\tabularnewline
$jjW^{+}W_{T}^{-}$	& MadSpin		&  48.01			& 84.8\%      		& 40.13			& 83.8\%		\tabularnewline
$jjW^{+}W_{T}^{-}$	& OSP			&  47.92			& 84.6\%    		& 40.01			& 83.6\%		\tabularnewline
$jjW^{+}W_{0}^{-}$	& MadSpin		&  8.26	& 14.6\%    		&  7.26	& 15.2\%		\tabularnewline
$jjW^{+}W_{0}^{-}$	& OSP			&  8.28	& 14.6\%    		&  7.29	& 15.2\%		\tabularnewline
	    \hline
\end{tabular}
\caption{Same as \tab{tab:SMVBSxs} but for the mixed EW-QCD process $pp\to jj W^{+} W^{-}_{\lambda}$ at $\mathcal{O}(\alpha_s^2\alpha_{EW}^2)$.
}
\label{tab:SMVBSxs_qcd}
\end{center}
\end{table}

As a final case study, we consider the LO production of the mixed EW-QCD process
\begin{equation}
pp ~\to~ jj W^+ W^-_{\lambda'}, \quad\text{with}\quad W^+_\lambda \to \mu^+\nu_\mu \quad\text{and}\quad W^-_{\lambda'}\to e^-\bar{\nu}_e,
\label{eq:qcd_sm_ProcDef}
\end{equation}
at $\mathcal{O}(\alpha_s^2\alpha_{EW}^2)$.
Aside from its own interesting features, the process is a primary background for the pure EW process $jj W^+ W^-_{\lambda'}$ at $\mathcal{O}(\alpha_{EW}^4)$.
Subsequently, in this section, we discuss the similarities and differences in distributions between EW and mixed EW-QCD production of $jj W^+ W^-_{\lambda'}$
for the same observables discussed in \sec{sec:vbs_ew}.
We use both the MadSpin and OSP methods to treat the decays of $W$ bosons.
The \mgamc~syntax for \eq{eq:qcd_sm_ProcDef} is
\begin{verbatim}
generate p p > w+ w-{X} j j QCD<=2 QED<=2
\end{verbatim}
where the polarization (\texttt{X}) of $W^-_\lambda$ is set to \texttt{T} or \texttt{0}.
The \ms~syntax and all phase space cuts are the same as those reported in \sec{sec:vbs_ew}.

In analogy to \tab{tab:SMVBSxs}, we report in \tab{tab:SMVBSxs_qcd} the cross sections [fb] and helicity fractions $(f_\lambda)$ [$\%$] for the full process $2\to 6$,
assuming generator-level cuts of \eq{eq:SMVBScuts} and analysis-level cuts of \eq{eq:decaycuts}, using both the MadSpin and OSP decay schemes.
Compared to the pure EW process, which shows $\sigma(W^-_{\lambda=T}):\sigma(W^-_{\lambda=0})$ ratio of about $4:1$,
the mixed EW-QCD process here exhibits a bigger difference of about $6:1$.
This difference can be attributed to the fact that most $W_\lambda^\pm$ in the mixed EW-QCD case are emitted off massless fermion legs,
which only contribute to the $W_{\lambda=T}^\pm$ process.
The pure EW process permits $W_{\lambda=0}^\pm$  production through diboson and VBS type of scattering topologies.

\begin{figure*}[t!]
\begin{center}
\subfigure[]{\includegraphics[width=.48\textwidth]{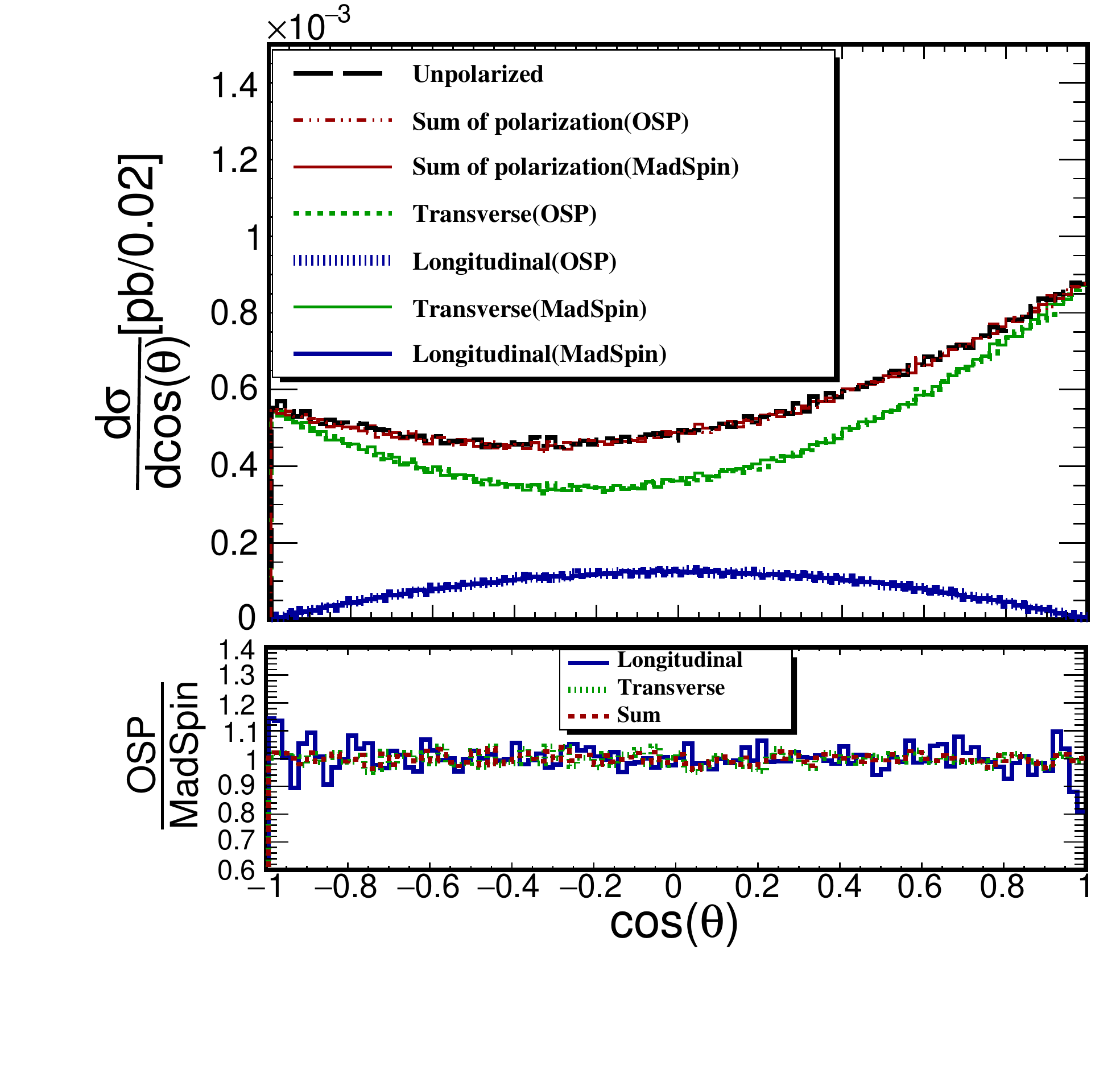}		\label{fig:VBSpolCosTh_qcd}		}
\subfigure[]{\includegraphics[width=.48\textwidth]{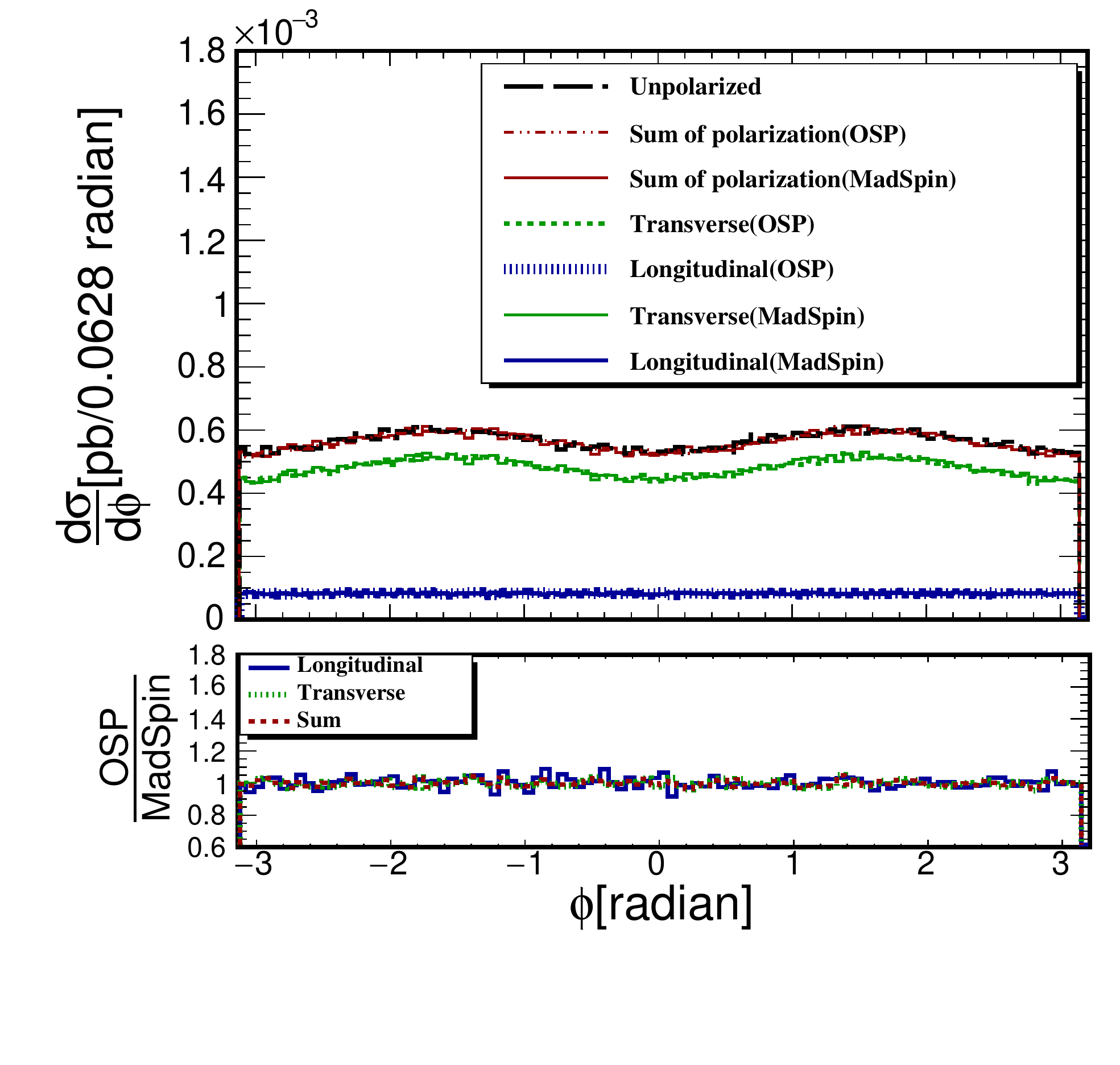}			\label{fig:VBSpolPhi_qcd}			}
\\
\subfigure[]{\includegraphics[width=.48\textwidth]{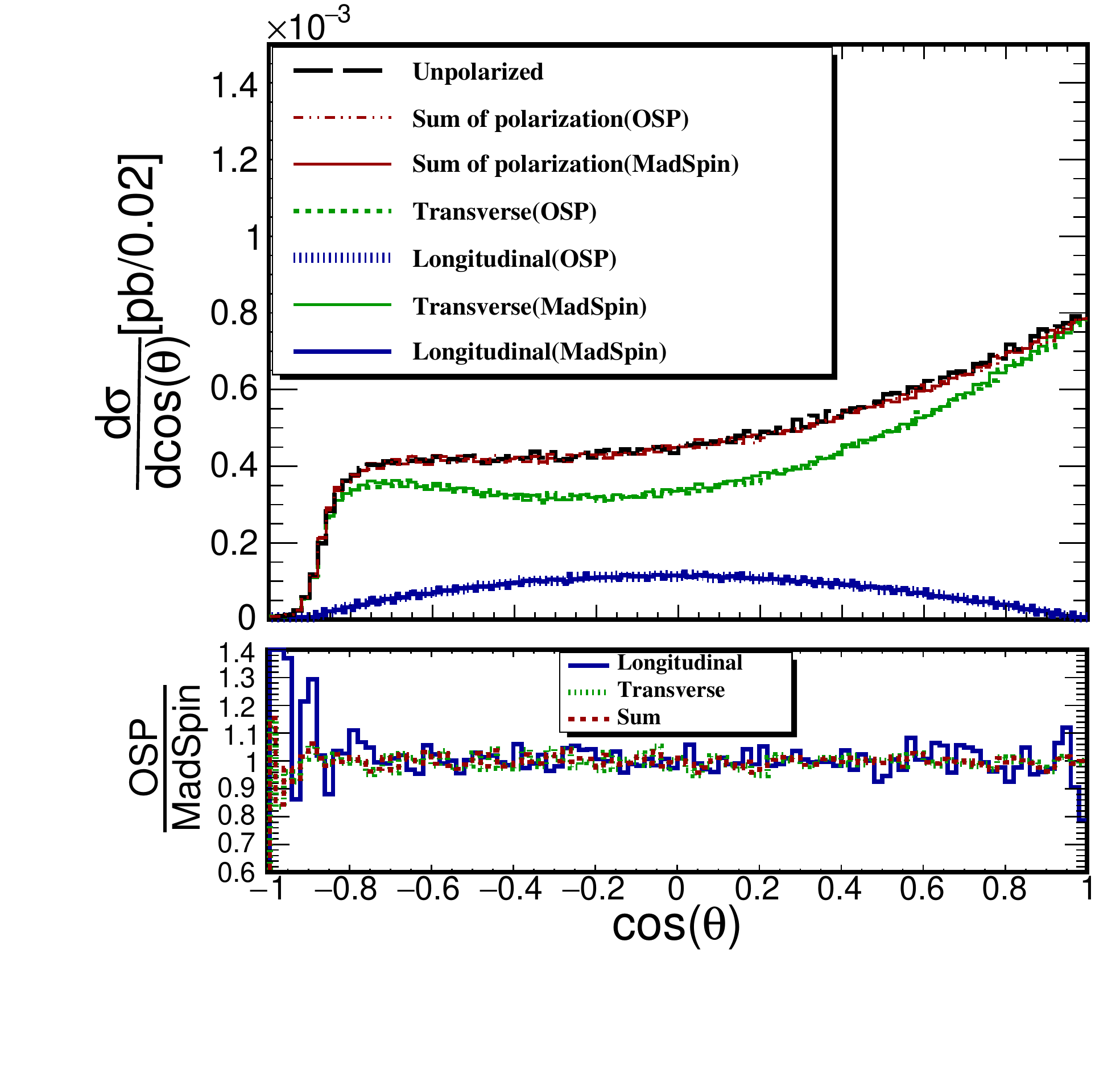}	\label{fig:AngdistDecayCutsTh_qcd}		}
\subfigure[]{\includegraphics[width=.48\textwidth]{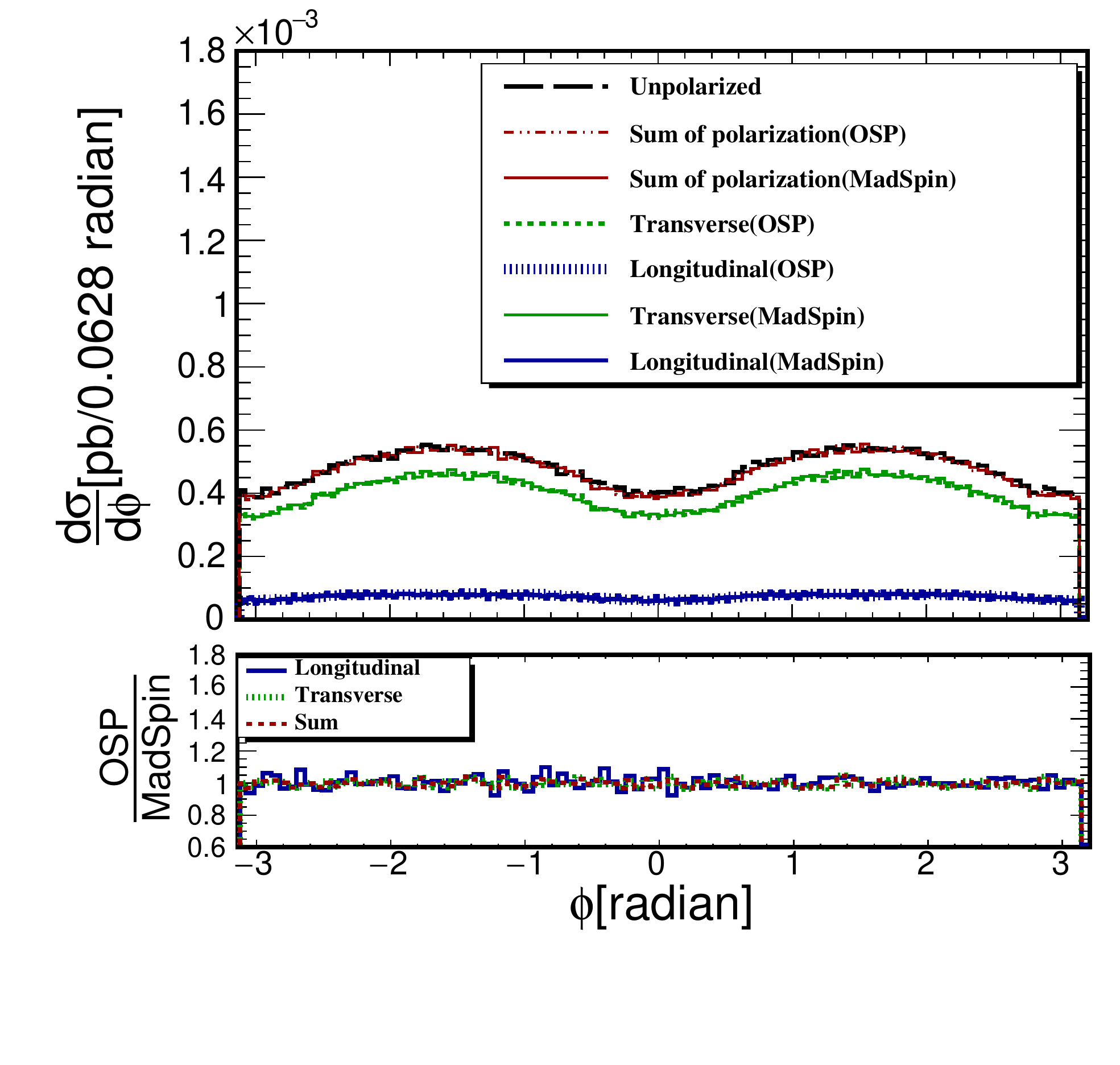}		\label{fig:AngdistDecayCutsPhi_qcd}	}
\end{center}
\caption{
Same as \fig{fig:VBSpol} but for the mixed EW-QCD process $pp\to jj W^{+} W^{-}_{\lambda}$ at $\mathcal{O}(\alpha_s^2\alpha_{EW}^2)$.
}
\label{fig:VBSpol_qcd}
\end{figure*}

We start our investigation of differential observables sensitive to $W^-_\lambda$ polarization with \fig{fig:VBSpol_qcd},
where we show the polar $(\theta)$ and azimuthal $(\phi)$ angular distributions as defined in \eqs{eq:vbs_sm_polarDef}{eq:vbs_sm_aziDef}.
As in \fig{fig:VBSpol}, panels  (a,b) include only generator-level cuts 
while (c,d) include analysis-level cuts listed in \eq{eq:decaycuts}.
In comparison to the pure EW process, we observe a smaller fraction of $W_{\lambda=0}^\pm$ events, consistent with results reported in \tab{tab:SMVBSxs_qcd}.
In contrast to the EW process shown in \fig{fig:VBSpol}, 
we observe a milder $g_{T0}$ interference pattern in the $\phi$ distribution with both sets of phase space cuts.
The smaller interference pattern can be attributed to the smaller $\lambda=0$ contribution.	
Crucially, we note that the shape of distributions are not substantially affected by the production mechanism;
only the normalizations are strongly affected, reflecting the processes' different coupling orders.

\begin{figure*}[t!]
\begin{center}
\subfigure[]{\includegraphics[width=.48\textwidth]{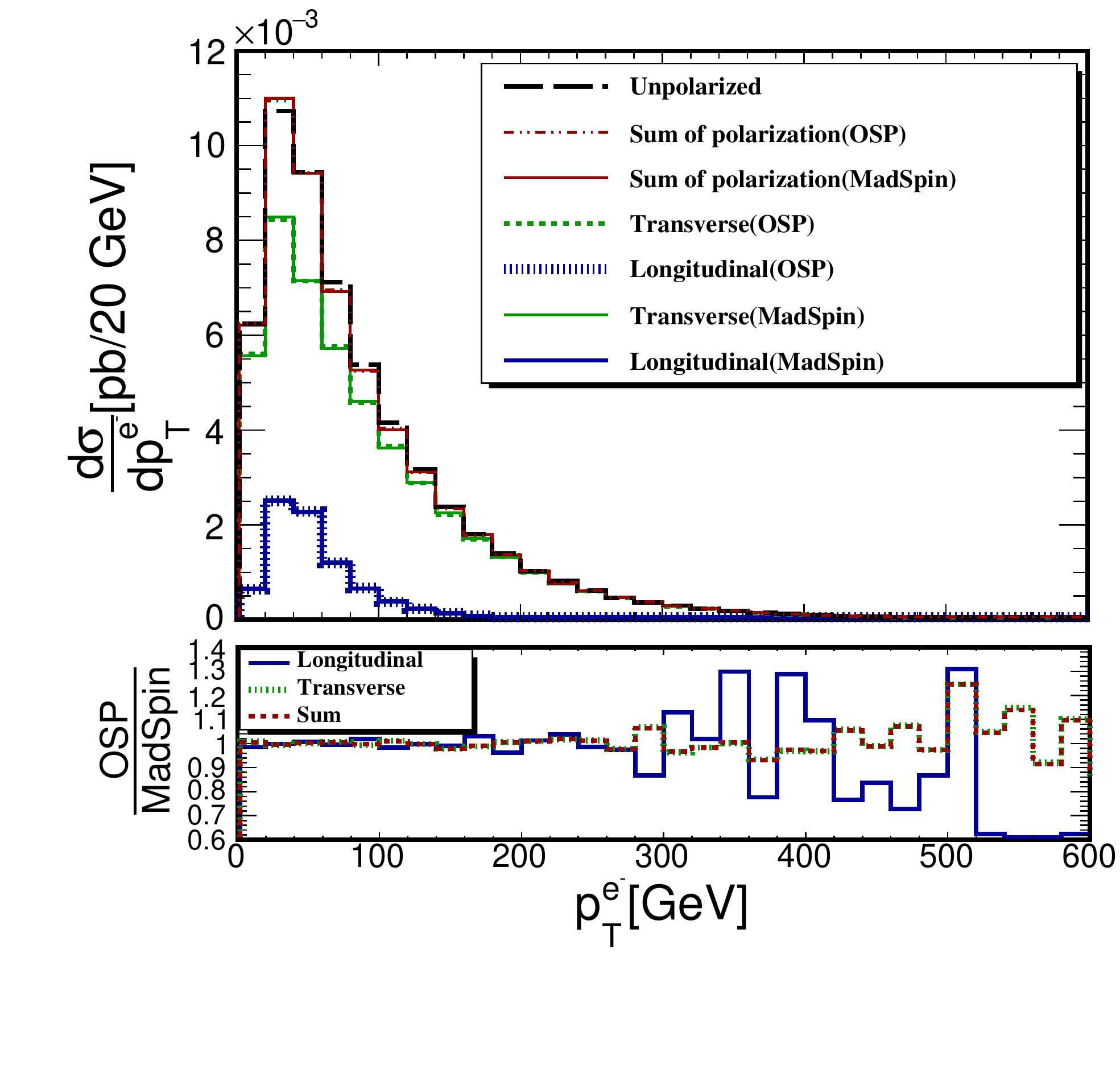}		\label{fig:distPTe_qcd}	}
\subfigure[]{\includegraphics[width=.48\textwidth]{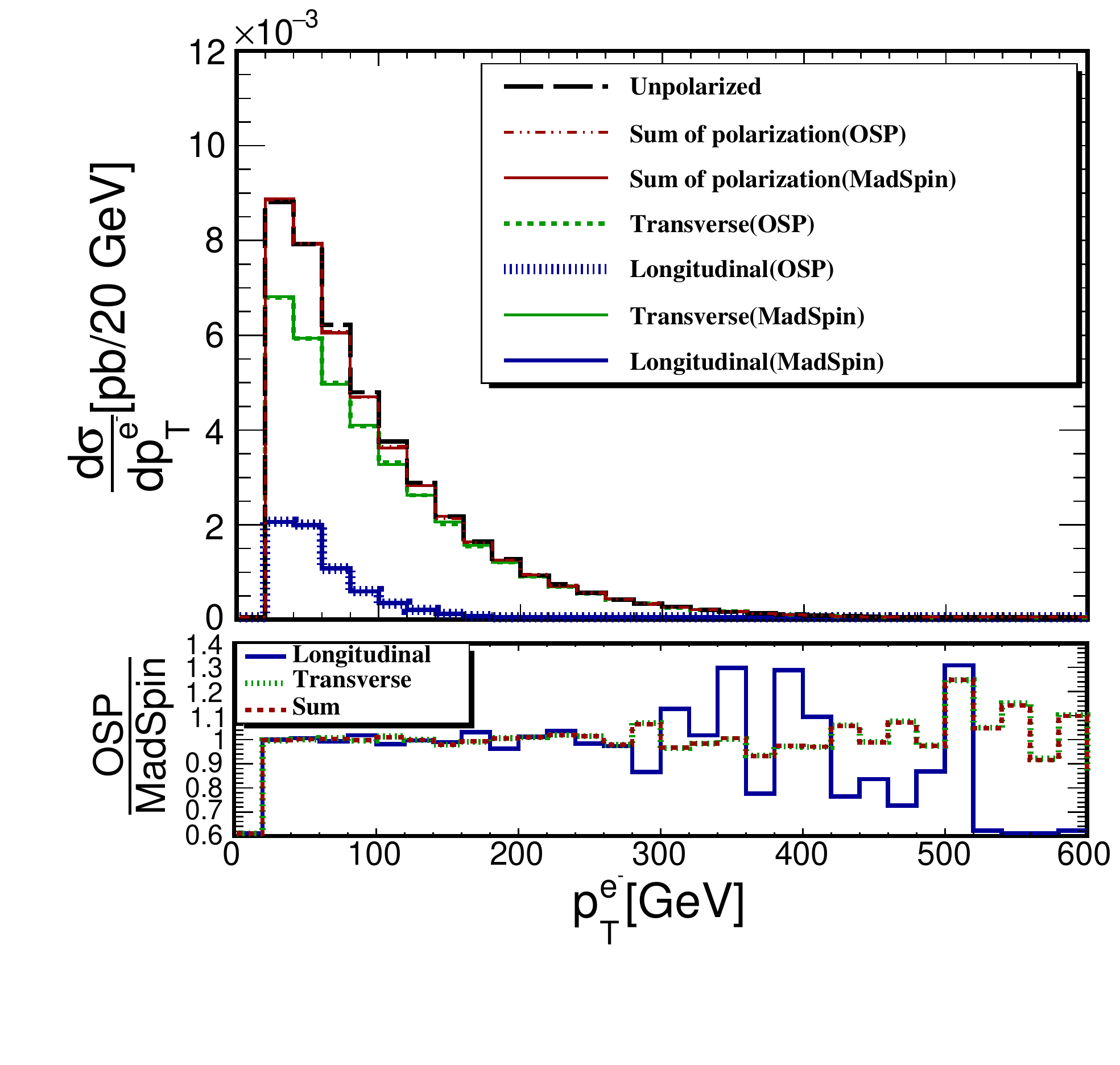}		\label{fig:distPTe_qcd}	}
\\
\subfigure[]{\includegraphics[width=.48\textwidth]{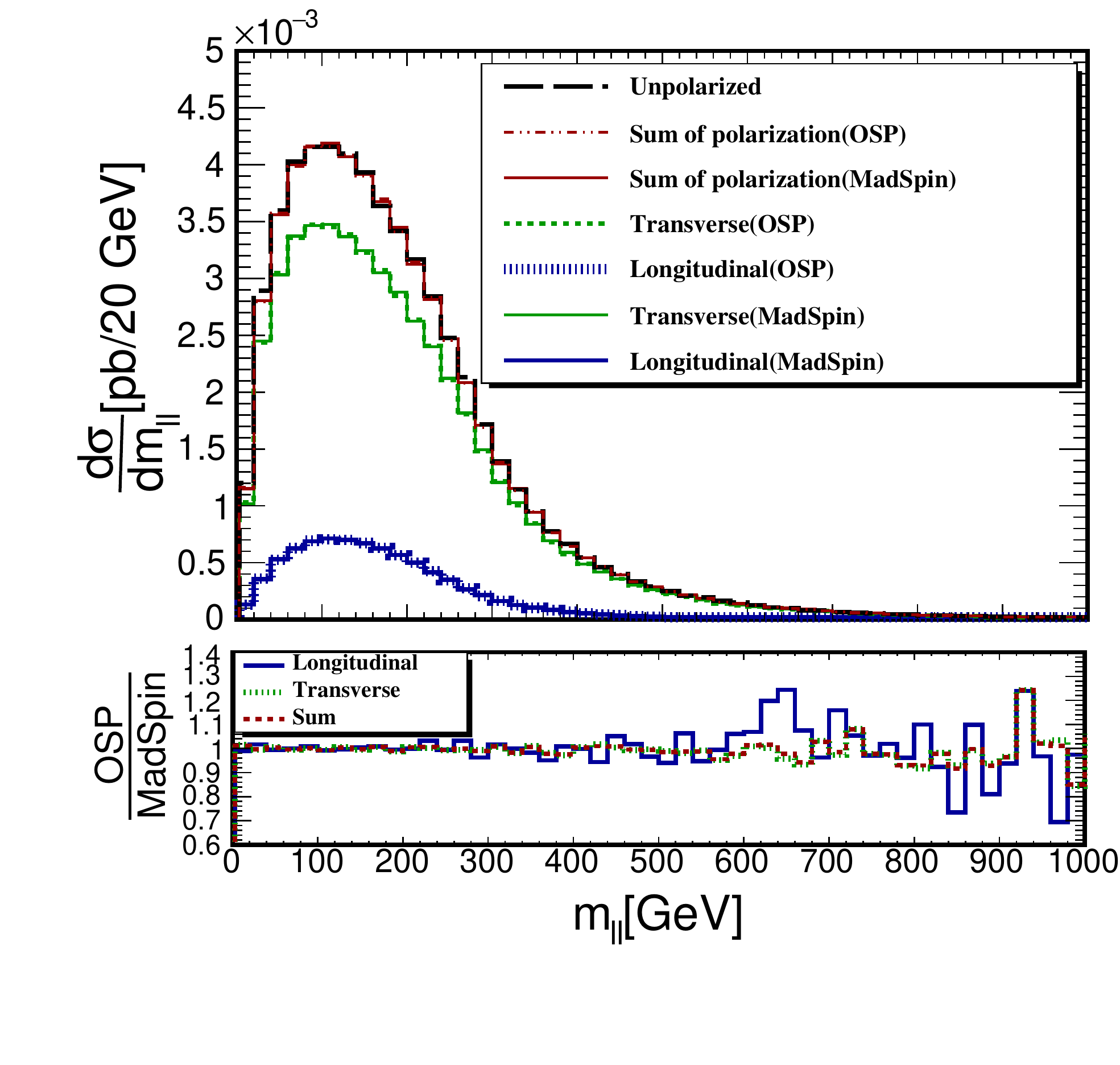}		\label{fig:distmll_qcd}	}
\subfigure[]{\includegraphics[width=.48\textwidth]{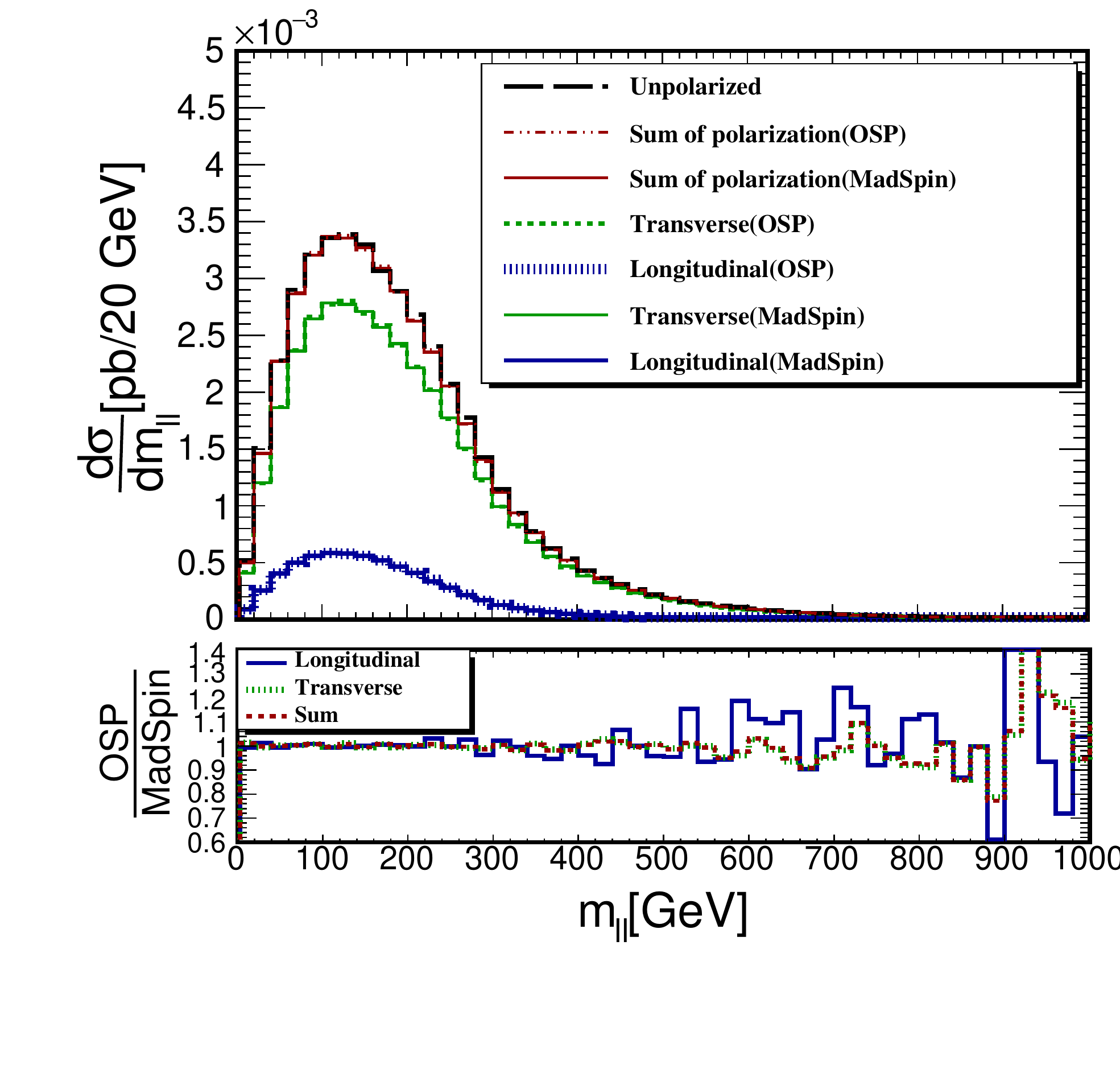}		\label{fig:distmll_qcd}	}
\end{center}
\caption{
Same as \fig{fig:decaydist1} but for the mixed EW-QCD process $pp\to jj W^{+} W^{-}_{\lambda}$ at $\mathcal{O}(\alpha_s^2\alpha_{EW}^2)$.
}
\label{fig:decaydist1_qcd}
\end{figure*}

\begin{figure*}[t!]
\begin{center}
\subfigure[]{\includegraphics[width=.48\textwidth]{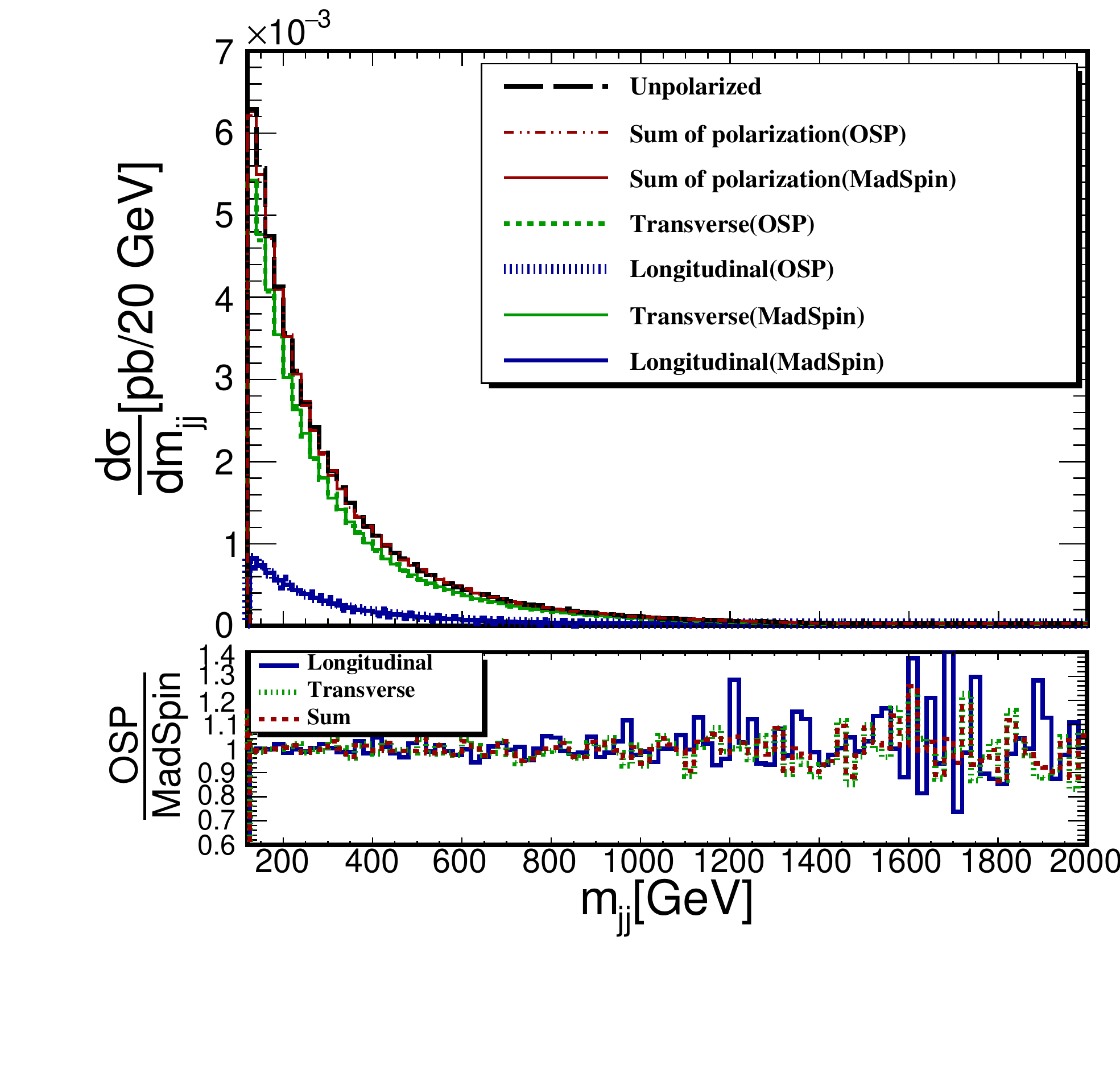}			\label{fig:distmjj_genCuts_qcd}		}
\subfigure[]{\includegraphics[width=.48\textwidth]{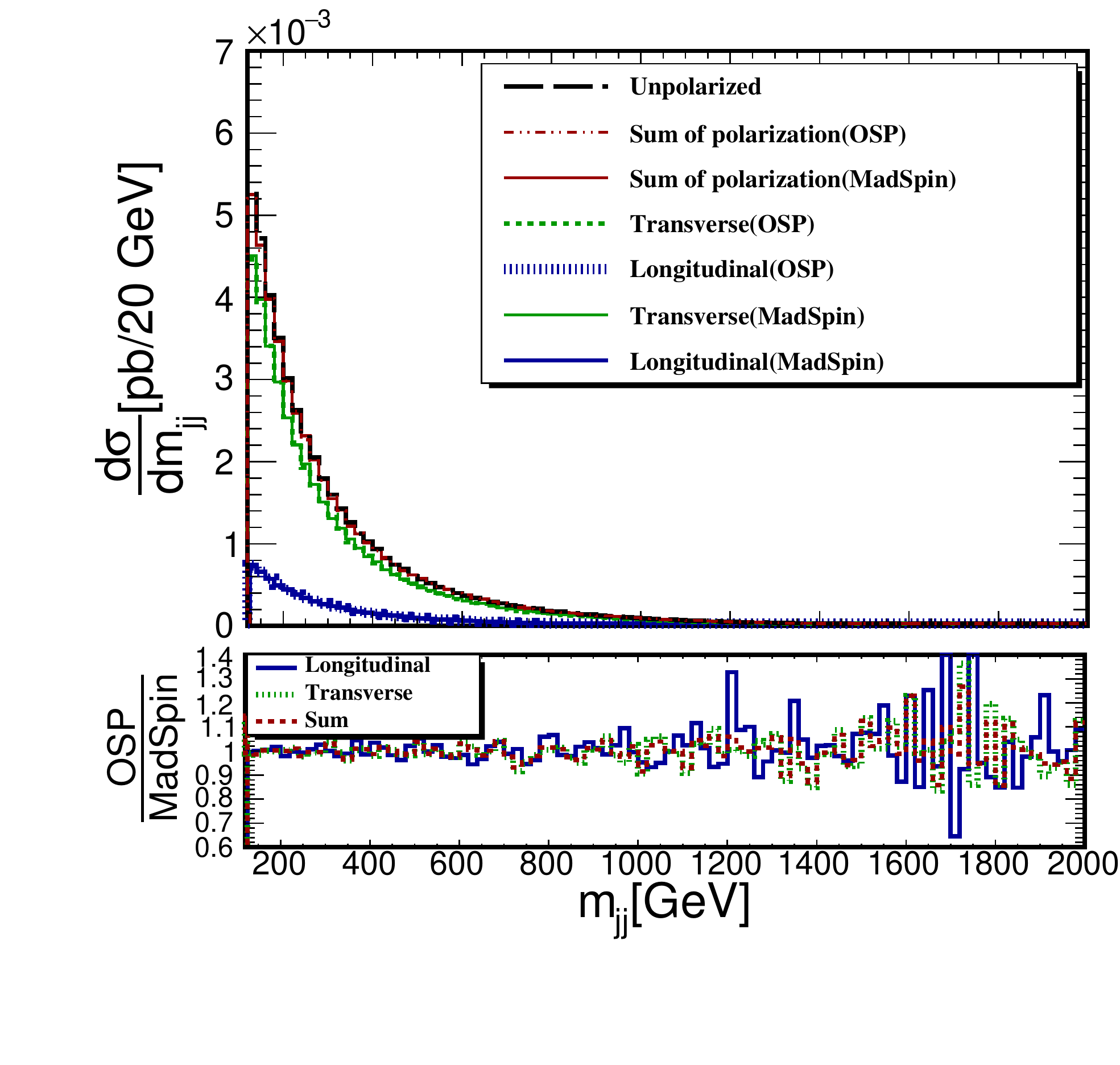}		\label{fig:distmjj_anaCuts_qcd}		}
\\
\subfigure[]{\includegraphics[width=.48\textwidth]{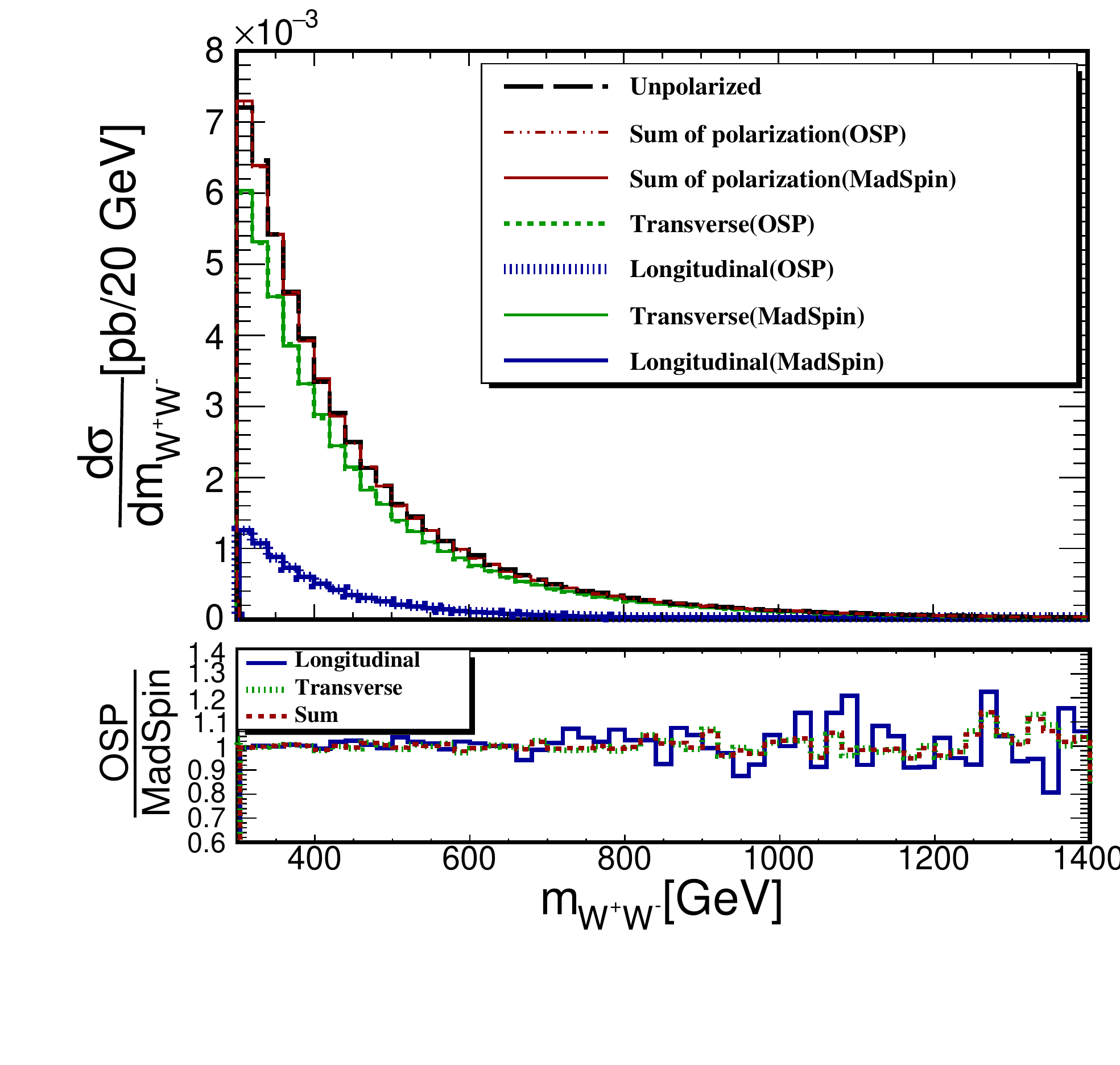}		\label{fig:distMWW_genCuts_qcd}	}
\subfigure[]{\includegraphics[width=.48\textwidth]{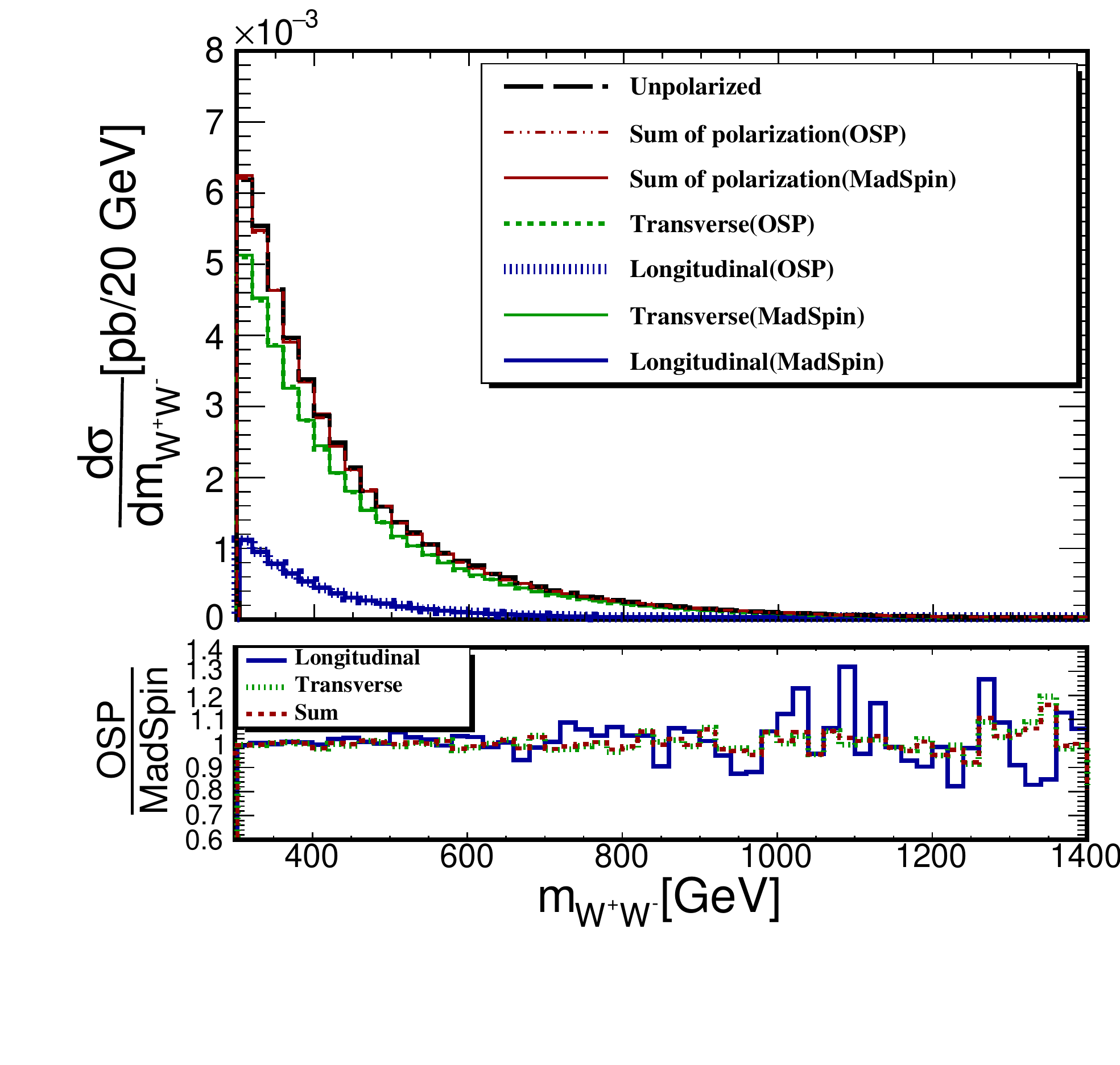}		\label{fig:distMWW_anaCuts_qcd}	}
\end{center}
\caption{
Same as \fig{fig:decaydist2} but for the mixed EW-QCD process $pp\to jj W^{+} W^{-}_{\lambda}$ at $\mathcal{O}(\alpha_s^2\alpha_{EW}^2)$.
}
\label{fig:decaydist2_qcd}
\end{figure*}

We extract the helicity fractions $f_\lambda$ from \fig{fig:VBSpolCosTh_qcd} using the Legendre expansion technique described in \eqs{eq:LegendreExpansion}{eq:LegendreIntegral}.
For the EW-QCD process, we report
\begin{equation}
f_L=0.5248\pm~ 0.3\%,	\quad 
f_R=0.3307\pm~ 0.4\%,	\quad 
f_0=0.1445\pm~ 2\%.
\label{eq:legendreFit_qcd}
\end{equation}
In comparison to the pure EW process, we observe strong similarities for the production of $W_\lambda^-$ possessing LH polarizations, with $f_L\approx50\%$.
For the RH and longitudinal polarizations, we see an increase (decrease) of the $\lambda=+1~(\lambda=0)$ modes of about $\delta f_\lambda\sim+5\%~(-5\%)$.
In \fig{fig:Legendre_qcd} we  show the polar distribution as reconstructed from \eq{eq:legendreFit_qcd}  as well  as using \ms. 
As in the pure EW process, we find good agreement between the Legendre expansion and \ms, except at the boundaries.
There the distribution $d\sigma(W^-_{\lambda=0})$ vanishes and our ratios quantifying disagreements become ill-defined.

We report similar shape behaviors between the EW and EW-QCD  processes in \fig{fig:decaydist1_qcd},
where we show for the EW-QCD process the (a,b) $p_T(e^-)$ and (c,d) $m(\ell\ell)$ distributions, assuming (a,c) only generator-level cuts and (b,d) analysis-level cuts. 
The analogous distributions for the EW process are shown in \fig{fig:decaydist1}.
We find a much smaller $g_{T0}$ interference pattern for the $p_T(e^-)$ curves here, due in part to the smaller $W^-_{\lambda=0}$ component.

\begin{figure*}[t!]
\begin{center}
\subfigure[]{\includegraphics[width=.48\textwidth]{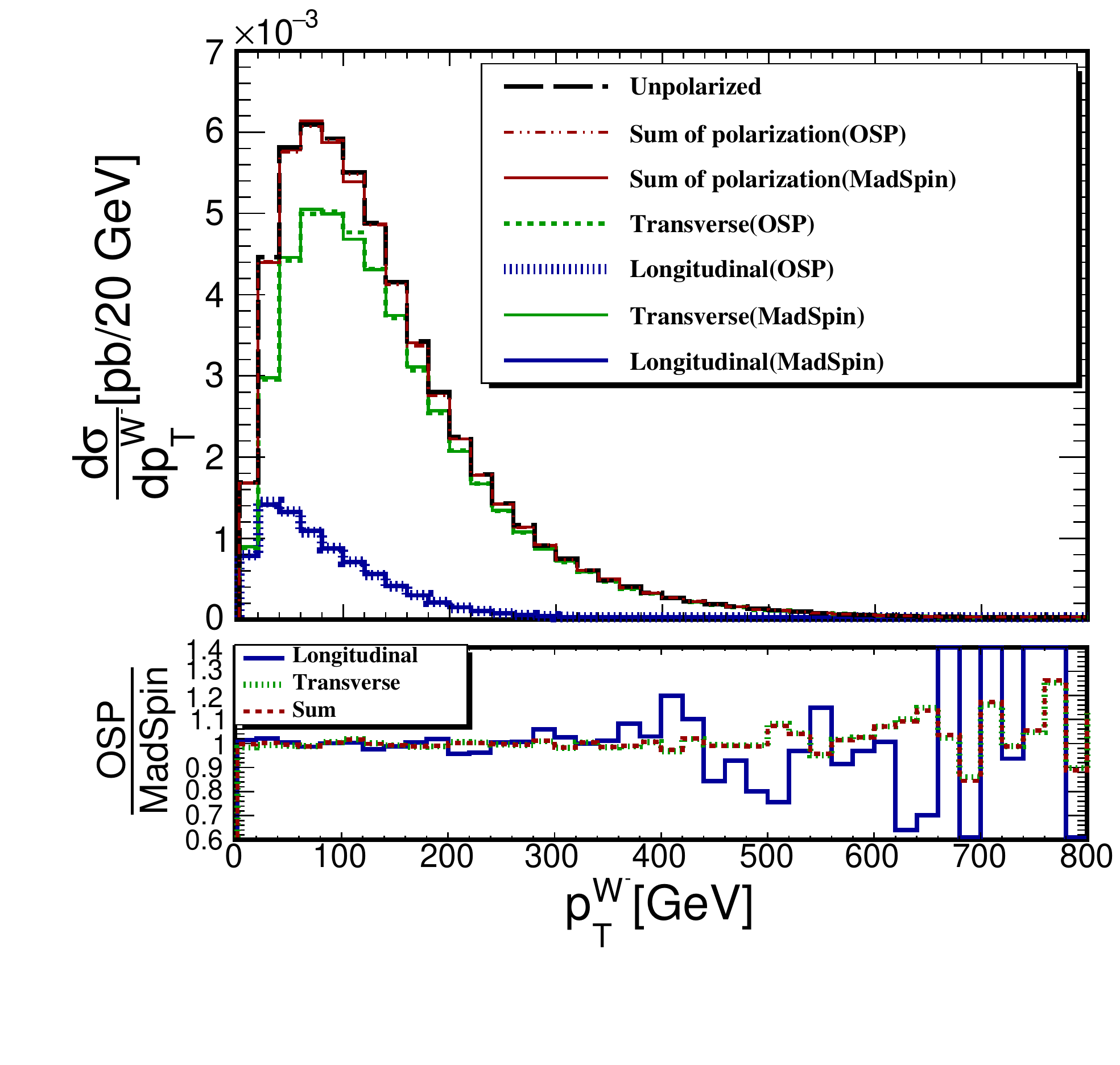}			\label{fig:distPTW_genCuts_qcd}	}
\subfigure[]{\includegraphics[width=.48\textwidth]{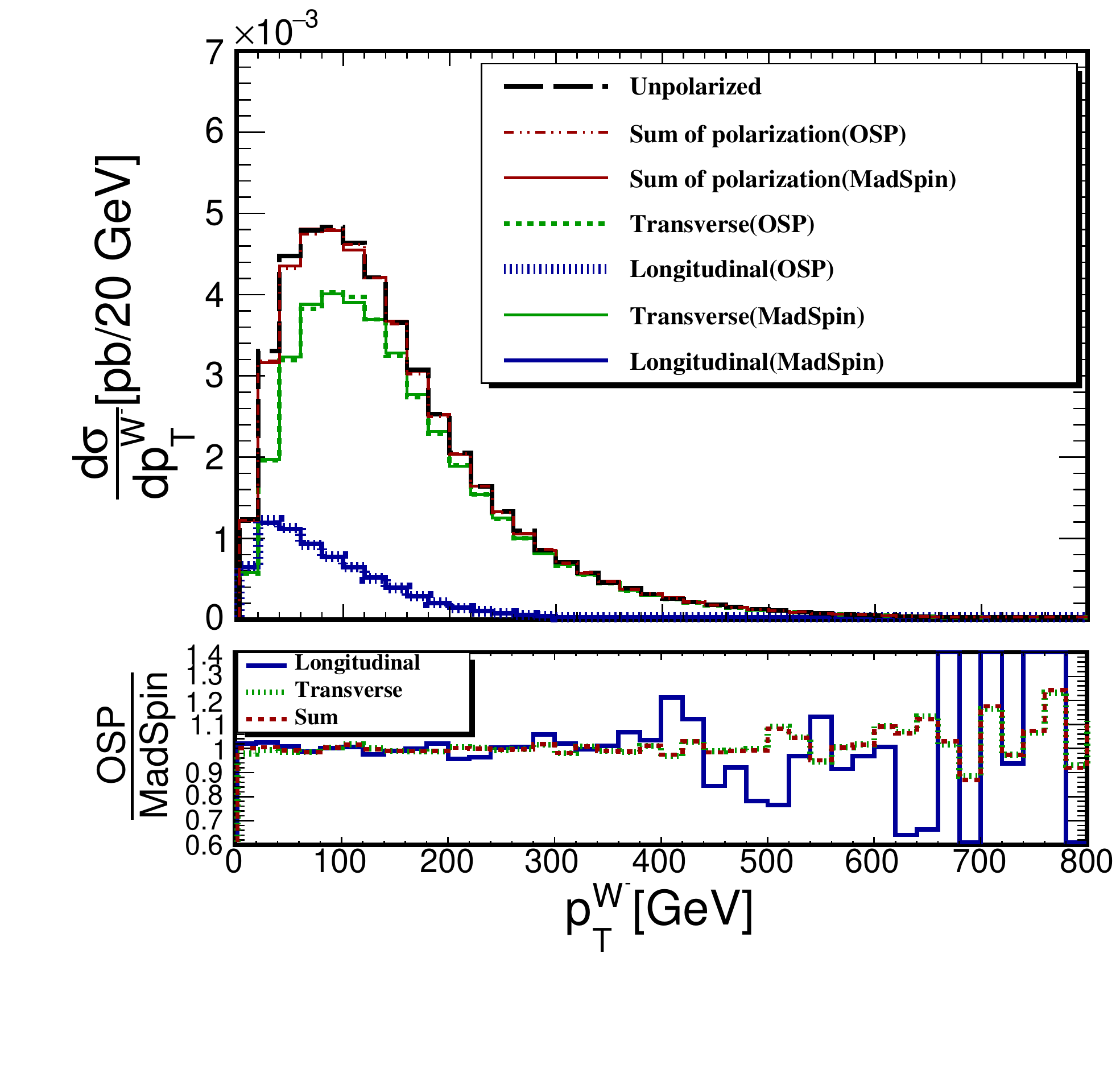}		\label{fig:distPTW_anaCuts_qcd}	}
\\
\subfigure[]{\includegraphics[width=.48\textwidth]{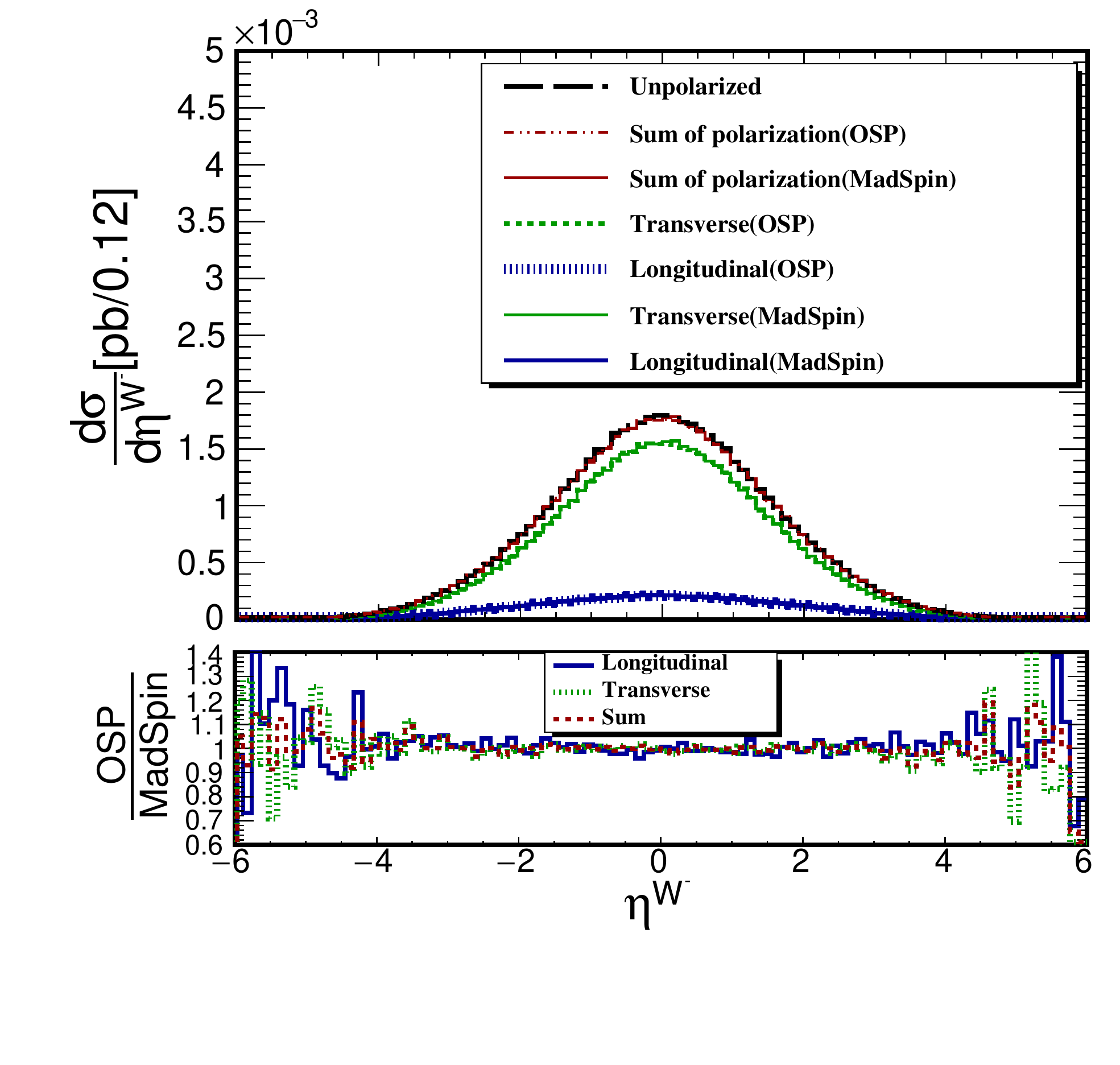}			\label{fig:distETAW_genCuts_qcd}	}
\subfigure[]{\includegraphics[width=.48\textwidth]{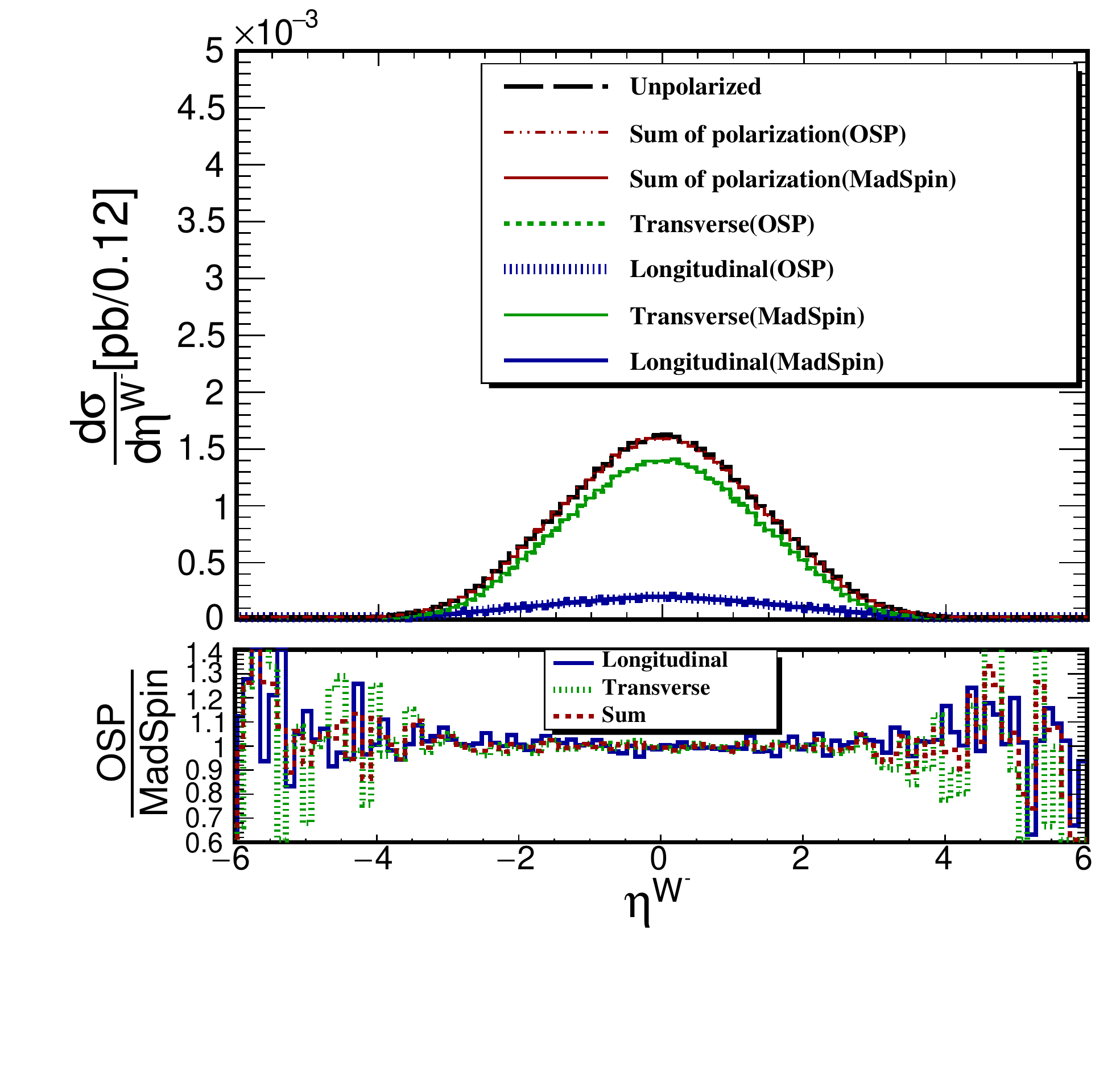}		\label{fig:distETAW_anaCuts_qcd}	}
\end{center}
\caption{
Same as \fig{fig:decaydist3} but for the mixed EW-QCD process $pp\to jj W^{+} W^{-}_{\lambda}$ at $\mathcal{O}(\alpha_s^2\alpha_{EW}^2)$.
}
\label{fig:decaydist3_qcd}
\end{figure*}

Turning to the (a,b) dijet invariant mass $m(jj)$ and (c,d) diboson invariant mass $m(W^+W^-)$ in  \fig{fig:decaydist2_qcd},
we observe large differences with respect to the pure EW process in  \fig{fig:decaydist2}.
With and without analysis-level cuts, we see that both the dijet and diboson spectra here strongly peak toward smaller invariant masses;
the shape is driven mostly by the $\lambda=T$ modes.
The $d\sigma\sim1/m^k(jj)$ behavior is typical of $s$-channel $g^*\to q\overline{q}$ splittings and suggests that the mixed EW-QCD process is not driven by valence-valence scattering.
This is opposed to the EW process which shows a plateau in the dijet spectrum and a softer peaking of the diboson mass,
which are consistent with VBS-like topologies.
In both sets of distributions we find that the impact of the $g_{0T}$ interference is negligible.

Finally, in \fig{fig:decaydist3_qcd} we show in (a,b) the $p_T(W^-)$ distributions and in (c,d) the $\eta(W^-)$, assuming only (a,c) generator-level cuts and (b,d) with analysis-level cuts.
For the unpolarized and the incoherent summation curves, we observe little differences between the mixed EW-QCD process here and the pure EW process in \fig{fig:decaydist3}.
By individual polarizations, however, we observe that the $\lambda=0$ and $\lambda=T$ polarizations in the EW-QCD process possess slightly broader peaks than their pure EW counter parts.
This feature is hidden because the EW process possesses a relatively larger $\lambda=0$ fraction than the mixed process (see \eq{eq:legendreFit} and \eq{eq:legendreFit_qcd}),
and that the narrower peaks of the $\lambda=0$ and $\lambda=T$ polarizations in the EW process are more widely separated than in the EW-QCD process.
This in turns broadens the polarization-summed curve in the EW process.
As a result of this preference for a higher $p_T$, the $\eta(W^-)$ distributions for both $\lambda=0$ and $\lambda=T$ in the mixed process are more central than their pure EW counterpart. 
This is particularly striking when comparing the two $\lambda=0$ curves.
In the EW-QCD case, the broad but central single-bump shape is indicative of a moderate recoil against the dijet system, and consistent with process not being driven by valence-valence scattering.
In the EW case, the forward, double-bump shape is indicative of forward $W^-$ production via VBS.

%%=====================================================================
% !TEX root = mgPolarization_main.tex
\section{Conclusions}\label{sec:conclusions}

The SM of particle physics remains the best description to date of how nature functions at small distances and high momentum-transfer scales.
This is especially the case in light of a SM-like Higgs boson and the multitude of data collected during Runs I and II of the 
LHC~\cite{Khachatryan:2015sga,Sirunyan:2019bez,Aaboud:2019gxl,Aaboud:2019nkz,
Aad:2019udh,CMS:2019mpq,Sirunyan:2017ret,Sirunyan:2017fvv,Aaboud:2018ddq,Sirunyan:2019ksz,Aaboud:2019nmv}.
However, the unambiguous evidence for dark matter and nonzero neutrino masses, 
as well as theoretical demands to understand the origin of flavor and the stability of the Higgs's mass, require extending  the SM.
Among the viable solutions are scenarios that predict 
the production of fermions and EW gauge bosons in high-energy scattering processes that are polarized in a distinctly different manner than that predicted by the SM.
Consequently, searches for the anomalous polarization of SM particles represent an important and well-motivated component of the LHC's program.

To facilitate such studies, we report the development of a method for using polarized matrix elements, 
where some or all external states are in a definite helicity eigenstates and where spin-averaging or spin-summing is truncated or not present,
in the publicly available event generator~\texttt{MadGraph5\_aMC@NLO.}
For an arbitrary reference frame, partonic scattering and decay rates of asymptotic states with fixed helicity polarizations can be computed at LO,
with little impact on runtime, and supports particle spins up to $3/2$ and $2$.
The helicity polarizations of resonances are transmitted to their decay products via modifications to their propagators.
Furthermore, our framework can be used beyond the scope of the LHC from low energy physics to astrophysics.

The scattering formalism underlying our work and main implementation details are given in \sec{sec:polarDef};
technical and usage details are reported in \app{sec:codingBits}.
As case studies, we investigated the production and decay of polarized $W^+_\lambda W^-_{\lambda'}$ pairs in the process $pp\to j j W^+_\lambda W^-_{\lambda'}$,
 with helicity eigenstates $(\lambda,\lambda')$ defined in various reference frames.
We considereded a benchmark Composite Higgs scenario (\sec{sec:vbs_bsm}) as well as SM production at $\mathcal{O}(\alpha^4)$ (\sec{sec:vbs_ew}) and $\mathcal{O}(\alpha^2\alpha_s^2)$ (\sec{sec:vbs_qcd}).
We focused on the helicity polarization decomposition of processes according to their reference frame
as well as investigated the impact of typical generator-level and analysis-level selection cuts.
In all case studies, we found that accounting for interference between LH and RH $W_\lambda$ bosons is much more important than interference between transverse and longitudinal polarizations.
Investigations into the production and decay of polarized EW bosons beyond tree-level are reported in a companion paper~\cite{COCITE}.

%%=====================================================================
\section*{Acknowledgements}
DBF is supported by the Knut and Alice Wallenberg foundation under the grant KAW 2017.0100 (SHIFT project). 
DBF thanks Ezio Maina for useful discussions during VBSCan meeting in Istanbul.
RR thanks Benjamin Fuks and Carlito Tamarit for helpful discussions. OM thanks the full \mgamc~team for helpful discussions.
RR is supported under the UCLouvain ``MOVE-IN Louvain'' scheme, and
the F.R.S.-FNRS ``Excellence of Science'' EOS be.h Project No 30820817.
SS is supported by SAMKHYA: HPC computing facility provided by Institute of Physics, Bhubaneswar. 
SS thanks Manimala Mitra for funding. 
SS also acknowledges the generous hospitality of UCLouvain.
The authors acknowledge the contribution of the VBSCan COST Action CA16108.
Computational resources have been provided by the Consortium des \'Equipements de Calcul Intensif (C\'ECI), funded by the Fonds de la Recherche Scientifique de Belgique (F.R.S.-FNRS) under Grant No. 2.5020.11 and by the Walloon Region. This work has received funding from the European Union's Horizon 2020 research and innovation programme as part of the Marie Sklodowska-Curie Innovative Training Network MCnetITN3 (grant agreement no. 722104). The authors would like to thank the organisers of the Heifei's MadGraph school (sponsored by the University of Science and Technology of China, Institute of High Energy Physics) where this project started.

%%=====================================================================
\appendix

%%=====================================================================
% !TEX root = mgPolarization_main.tex

\section{Polarized Matrix Elements in \mgFull}\label{sec:codingBits}
In this appendix we expand on the formalism reported in Sec.~\ref{sec:polarDef}
and describe the main features introduced into the event generator \mgamc~that allows the modeling of  parton scattering with polarized matrix elements.
That is, matrix elements where some or all external states are in a definite helicity eigenstates and where spin-averaging or spin-summing is truncated or not present.
In~ \app{sec:codingBits_syntax} we describe the new syntax that triggers the creation of scattering amplitudes with a truncated polarization summation.
Decays of polarized resonances are described in \app{sec:codingBits_decay}.
Leading order event generation within a reconstructable reference frame is described in \app{sec:codingBits_evtgen},
while in \app{sec:codingBits_reweight} the possibility of event re-weighting of polarized samples is discussed.

%%%%%%%%%%%%%%%%%%%%%%%%%%%%%%%
\subsection{Syntax for Polarized Matrix Elements}\label{sec:codingBits_syntax}
In order to fix the helicity polarization of particles in \mgamc,
we introduce new syntax commands at the process-definition and event-generation levels.
When specifying a scattering or decay  process using the usual~\cite{Alwall:2014hca} \mgamc~commands,
any particle followed immediately (without spacing) by \texttt{\{X\}} will be polarized in the helicity eigenstate ``X''.
We stress that the notion of helicity polarization is not Lorentz invariant for massive particles.
Consequently, using the polarization syntax requires that a reference frame 
be specified at the time of matrix element evaluation.
For massive spin 1/2, 1, 3/2, and 2 particles, we list in \tab{tb:polarSyntax} the allowed  syntax, 
the corresponding helicity states in the \texttt{HELAS} basis,
and whether the polarization can be transmitted through the propagators of massive particles (see also \sec{sec:polarDef_decay} and \app{sec:codingBits_decay} for details).

At LO, the bracket polarization syntax can be used for any initial-state (IS) or final-state (FS) particle  {in any scattering process.}
Examples of such usage are:
\begin{verbatim}
generate p p > t t~{R}
generate e+{L} e- > w+{0} w-{T}
generate z z{R} > w+ w-{0}
\end{verbatim}
which respectively describe the  Born-level processes:
\begin{equation}
q\overline{q},gg \to t \overline{t}_R, \quad 
e^+_L e^- \to W^+_0 W^-_T, \quad\text{and}\quad 
Z Z_R \to W^+ W^-_0.
\end{equation}
The helicity label \texttt{0} denotes a longitudinally polarized massive vector boson;
\texttt{L} and \texttt{R} represent LH and RH helicity polarizations {for spin 1/2 and 1 particles};
and \texttt{T} models transverse polarizations of spin 1 particles as a coherent sum of
\texttt{L} and \texttt{R} helicities.
Throughout this following, omitting a helicity label expresses an unpolarized particle.
The \texttt{\{X\}} polarization syntax can also be used with multi-particle definitions. 
For example: to model the diboson process $pp\to W^\pm_{T} W^\mp_{0}$,
the following commands are possible:
\begin{verbatim}
define ww = w+ w-
generate p p > ww{T} ww{0}
\end{verbatim}
To avoid polarization definition conflicts, multi-particle definitions consisting of polarized states, e.g., \texttt{define wwX = w+\{T\} w-\{0\}}, is not allowed.

\begin{table*}[!t]
\begin{center}
\resizebox{\textwidth}{!}{
\begin{tabular}{ccc|ccc}
\hline\hline
 Syntax & $\lambda$ in \texttt{HELAS} Basis & Propagator &	 Syntax & $\lambda$ in \texttt{HELAS} Basis  & Propagator \\	\hline 
 \multicolumn{3}{c}{ spin $\frac{1}{2}$} &   \multicolumn{3}{c}{ spin $\frac{3}{2}$}  	\\ \hline
 \texttt{\{L\}} \texttt{\{-\}} 		& -1 (Left) 	& Yes (massive only) 	& \texttt{\{-1\}}	& -1 & No \\
\texttt{\{R\}} \texttt{\{+\}} 		& +1 (Right) 	& Yes (massive only) 	& \texttt{\{1\}}	&  1 & No \\
\multicolumn{3}{c|}{} 								& \texttt{\{3\}}	& 3 & No \\
\multicolumn{3}{c|}{} 								& \texttt{\{-3\}}	& -3 & No \\
\hline\hline
 \multicolumn{3}{c}{spin 1} &   \multicolumn{3}{c}{ spin 2} 	\\ \hline
\texttt{\{0\}	}			& 0 (Longitudinal; massive only) 		& Yes (massive only) 		& \texttt{\{-2\}}	& -2 & No \\
\texttt{\{T\}	}			& 1 and -1 (Transverse; coherent sum) 	& Yes (massive only) 		& \texttt{\{-1\}}	& -1 & No \\
\texttt{\{L\}}  \texttt{\{-\}} 	& -1 					& No 		& \texttt{\{0\}}	& 0 & No \\
\texttt{\{R\}} \texttt{\{+\}} 	& +1 				& No 		& \texttt{\{1\}}	& 1 & No \\
\texttt{ \{A\}} 			&  		& Propagators only 			& \texttt{\{2\}}	& 2 & No \\
\hline
\end{tabular}
} %% resize
\caption{
For a given particle spin, the allowed \mgamc~polarization syntax,
its helicity state in the \texttt{HELAS} basis,
and whether the polarization is transmitted through propagators of massive particles.
}
\label{tb:polarSyntax}
\end{center}
\end{table*}

In standard computations using \mgamc, once a process has been defined, e.g., \texttt{generate p p > t t$\sim$}, 
the \mg~sub-program~\cite{Stelzer:1994ta,Alwall:2011uj} will build all helicity amplitudes 
from \texttt{ALOHA} \cite{deAquino:2011ub} and \texttt{HELAS} \cite{Murayama:1992gi} routines,  
for all contributing  sub-channels, e.g., $gg,q\overline{q}\to t\overline{t}$,
and for all external helicity permutations, e.g., $t_L\overline{t}_L, t_L\overline{t}_R, t_R\overline{t}_L,$ and $t_R\overline{t}_R$.
Next, amplitudes are evaluated numerically, squared, and summed.
For initial states and identical final states, dof. averaging and symmetry multiplicity factors are then incorporated.
When using the polarization features on IS/FS particles, this procedure is changed in two ways:
\begin{itemize}
\item Instead of summing over all helicity polarizations of all external particles, \mgamc~only sums over the polarizations allowed in the process definition.
\item Averaging symmetry factors over initial state polarizations are modified according to the new number of initial states.
\end{itemize}

\begin{table*}[!t]
\begin{center}
\resizebox{\textwidth}{!}{
\begin{tabular}{c c | c c}
\hline\hline
Polarization $(\lambda)$ & Squared Amplitude $(\vert\mathcal{M}_\lambda\vert^2)$ & Polarization $(\lambda)$ & Squared Amplitude $(\vert\mathcal{M}_\lambda\vert^2)$ \\
\hline
$+1$ 	&  1.8377936439613620	& 	$+3$ 	& 2.65E-32 \\
$-1 $		&  1.7456113543927256 	&	$-3$ 		& 2.57E-32 \\
\hline
Unpolarized Avg. $(\overline{\vert\mathcal{M}_\lambda\vert^2})$	& 0.89585124958852191 
& $\sum_{\lambda}\vert\mathcal{M}_\lambda\vert^2$ & $4\times 0.89585124958852191$\\
\hline\hline
\end{tabular}
} %% resize
\caption{
For the gravitino scattering process $g_{rv}(p_1,\lambda) ~+~ g_{rv}(p_2) \to \tau^+(p_3) ~+~ \tau^-(p_4)$,
the squared scattering amplitude $\vert\mathcal{M}\vert^2$ as a function of gravitino helicity $\lambda$,
defined in the partonic c.m.~frame at the phase space point provided in the text,
as well as the unpolarized, spin-averaged squared matrix element $\overline{\vert\mathcal{M}\vert^2}$, 
and the sum of the four squared amplitudes.
}
\label{tb:spin2Amp}
\end{center}
\end{table*}

Special attention is needed for processes with identical particles.
 For initial state particles, the \mgamc~convention is that the order of the particles during process declaration matters. 
 The ordering condition is particularly suited for asymmetric beam experiments, where the first state is associated to one beam and the second state to another beam.
 As an example, consider the production of an unpolarized $\tau^+\tau^-$ pair 
from massless, spin $3/2$ gravitino scattering~\cite{Christensen:2013aua} with fixed external momenta:
\begin{equation}
g_{rv}(p_1,\lambda) ~+~ g_{rv}(p_2) \to \tau^+(p_3) ~+~ \tau^-(p_4).
\label{eq:gravitinoScatt}
\end{equation}
This can be simulated using the syntax
\begin{verbatim}
import model GldGrv_UFO
generate grv{X} grv > ta+ ta-
output standalone Polar_grv_grv_tau_tau; launch -f
\end{verbatim}
In the above, $g_{rv}(p_2)$ is unpolarized and \eq{eq:gravitinoScatt} is not equivalent to $g_{rv}(p_1) g_{rv}(p_2,\lambda) \to \tau^+ \tau^-$.
In order to recover the unpolarized process $g_{rv} g_{rv}  \to \tau^+ \tau^-$, one has to incoherently sum over the four helicity configurations, $\lambda=\pm1, \pm3$.
And as described above, spin-averaging over possible initial states is modified to only account for relevant dof. 
As a check of modeling asymmetrically polarized, initial-state particles that are identical, 
we report in \tab{tb:spin2Amp} for the specific phase space point defined in the partonic c.m.~frame,
\begin{Verbatim}[fontsize=\small]
 pi      E [GeV]       px [GeV]      py [GeV]         pz [GeV]      m [GeV]
 p1   0.5000000E+03  0.0000000E+00  0.0000000E+00  0.5000000E+03  0.0000000E+00
 p2   0.5000000E+03  0.0000000E+00  0.0000000E+00 -0.5000000E+03  0.0000000E+00
 p3   0.5000000E+03  0.1109236E+03  0.4448280E+03 -0.1995517E+03  0.1777000E+01
 p4   0.5000000E+03 -0.1109236E+03 -0.4448280E+03  0.1995517E+03  0.1777000E+01
\end{Verbatim}
the squared scattering amplitude $\vert\mathcal{M}\vert^2$
of \eq{eq:gravitinoScatt} for polarizations $\lambda=\pm1, \pm3$, 
as well as the unpolarized, spin-averaged squared matrix element $\overline{\vert\mathcal{M}\vert^2}$, 
and the sum of the four squared amplitudes.
One sees precisely that the difference between the spin-averaged result and the summed result is the
symmetry factor $(2s_{g_{gv}}+1)=4,$ for $s_{g_{gv}}=3/2$.

For identical, final-state particles, the \mgamc~convention  demands that each identical particle has a specified polarization.
For example, the process 
\begin{equation}
pp\to Z_0 Z_T
\end{equation}
implies a sum over all interfering diagrams  with one transversely polarized $Z$ and one longitudinally polarized $Z$.
In this sense, $p p \to Z_0 Z_T$ is equivalent to $p p \to Z_T Z_0$.
To recover the full, unpolarized process, $p p \to ZZ$, ones must sum the three helicity configurations: $Z_0 Z_T$, $Z_0 Z_0$, and $Z_T Z_T$. 
For identical, final-state particles, a mixed syntax where some identical particles are polarized and others are not, {\it e.g.,} $pp\to Z_0 Z$,  is not supported.

%%%%%%%%%%%%%%%%%%%%%%%%%%%%%%%%%%%%%%%%%%%%%%%%
\subsection{Decays of Polarized Resonances with \mgFive~and \ms}\label{sec:codingBits_decay}

As described in \sec{sec:polarDef_decay}, the helicity polarization features introduced into \mgamc~extend to unstable resonances.
After specifying a hard scattering or decay process for a massive, polarized final state at LO,
one can steer the decay of a resonance to the desired final state in the usual manner~\cite{Alwall:2008pm}.
For example: the syntax 
to model the production and decay of $t_L$ or $W^+_0 W^-_T$ pairs at LO is
\begin{verbatim}
generate p p > t t~{L}, t~ > b~ w-
generate e+{L} e- > w+{0} w-{T}, w+ > e+ ve, w- > e- ve~
\end{verbatim}
The \texttt{\{X\}} syntax changes the standard \mgamc~decay protocol
by inserting the spin-truncated propagators defined in \eqs{eq:polarDef_decay_FXProp}{eq:polarDef_decay_FBProp}, \eq{eq:polarDef_decay_VXProp} and \eq{eq:auxprop},
instead of a normal BW propagator.
Special care has been taken for the case where the transverse momentum of a spin 1 boson is vanishing in order to consistently adhere to the limit employed by \texttt{HELAS}.

The inclusion of {polarized propagators} is possible through the extension of the \texttt{ALOHA} package \cite{deAquino:2011ub} 
to support non-Lorentz invariant quantities and the auxiliary polarization $\lambda=A$, defined in \eq{eq:auxprop}. 
The polarization can be called explicitly using the syntax
\begin{verbatim}
generate p p > z{T} z{A}, z > e+ e-,
\end{verbatim}
which describes resonant diboson production $q\overline{q}\to Z_T Z_A$, $Z_\lambda \to e^+ e^-$.
In principle, the $\lambda=A$ polarization vector is needed to recover unpolarized events from polarized event samples, particularly in the off-shell region.
However, its kinematical structure leads to a highly suppressed or vanishing contribution in practical applications.

In the presence of identically polarized, final state particles, the handling of symmetry factors and optimization of phase space integration requires care.
As such, two polarization \textit{modes} has been implemented:
(i) If the user specifies exactly one decay for each polarized particle, like in the following:
\begin{verbatim}
generate p p > z{X} z{Y}, z > e+ e-, z > mu+ mu-
generate p p > z{X} z{Y}, z > l+ l-, z > j j
\end{verbatim}
then \mg~enters an \textit{ordered mode} where the decays of \texttt{z\{X\}} and \texttt{z\{Y\}} are steered according to the order of the decay chains.
In the first instance, \texttt{z\{X\}} will be decayed to \texttt{e+e-} and  \texttt{z\{Y\}} to \texttt{mu+mu-};
in the second instance, \texttt{z\{X\}} will be decayed to \texttt{l+l-} and  \texttt{z\{Y\}}  to \texttt{jj}.
This case is similar to the ordered syntax for initial state particles.
(ii) If the number of polarized particles is different from the specified decays, like in the following:
\begin{verbatim}
generate p p > z{X} z{Y}, Z > l+ l-
generate p p > z{X} z{Y}, Z > e+ e-, Z > mu+ mu-, Z > ta+ ta-
\end{verbatim}
then \mg~enters an \textit{unordered mode} and all possible decay permutations are modeled.

In Table \ref{table-idenpart}, we present the total cross section for the $p p \to ZZ$ process into different decay channels. 
We show the unpolarized cross section and the decomposition into different helicity configurations, together with their incoherent sum. 
The ``correct'' decomposition depends on the mode. 
In the \textit{ordered mode} one needs to sum over all orders of helicity configurations. (In the example, this sums to four configurations since $Z_T Z_0$ and $Z_0Z_T$ are treated differently.) 
In the \textit{unordered mode} permutations are equivalent and should not be double counted. (In the example, only three configurations sum to the unpolarized result.)

\begin{table*}[!t]
\begin{center}
\resizebox{\textwidth}{!}{
\begin{tabular}{c c | c c}
\hline\hline
syntax  & cross (pb) & syntax & cross (pb) \\
\hline\hline
p p > Z Z, Z > e+ e- & 0.011 & p p > Z Z, Z > l+ l- & 0.042 \\
\hline
p p > Z\{0\} Z\{0\}, Z > e+ e- & 6.4e-4 & p p > Z\{0\} Z\{0\}, Z > l+ l- &  0.0026\\
p p > Z\{0\} Z\{T\}, Z > e+ e- &0.0025 & p p > Z\{T\} Z\{0\}, Z > l+ l- &0.010\\
p p > Z\{T\} Z\{T\}, Z > e+ e- & 0.0075 & p p > Z\{T\} Z\{T\}, Z > l+ l- &0.030\\
sum                                      & 0.011  &sum &0.042 \\
\hline
p p > Z Z , Z > e+ e- , z > mu+ mu- & 0.021 & p p > Z Z, Z > l+ l-, Z > j j & 0.66\\
\hline
p p > Z\{0\} Z\{0\}, Z > e+ e-, Z > mu+ mu- & 0.0013 & p p > Z\{0\} Z\{0\}, Z > l+ l-, Z > j j &0.040\\
p p > Z\{0\} Z\{T\}, Z > e+ e-, Z > mu+ mu- &0.0025 & p p > Z\{0\} Z\{T\}, Z > l+ l-, Z > j j & 0.079 \\
p p > Z\{T\} Z\{0\}, Z > e+ e-, Z > mu+ mu- & 0.0025& p p > Z\{T\} Z\{0\}, Z > l+ l-, Z > j j & 0.079 \\
p p > Z\{T\} Z\{T\}, Z > e+ e-, Z > mu+ mu- & 0.015 & p p > Z\{T\} Z\{T\}, Z > l+ l-, Z > j j & 0.47 \\
sum &0.021 &sum &0.67\\
\hline\hline
\end{tabular}
} %% resize
\caption{
Decomposition of the un-polarized sample into a sum of polarized samples. Depending of the syntax used one needs 
to sum either three or four different configurations. 
The sample with the auxiliary/scalar component are here not included since they are negligible.
}
\label{table-idenpart}
\end{center}
\end{table*}

Aside from the LO \mgFive~syntax just described, it is also possible to decay unstable, polarized, spin 1/2 and 1  resonances using \ms~\cite{Artoisenet:2012st}.
When called, \ms~automatically sets up the computation in the frame selected for event generation and employs the modified BW propagators 
described in \sec{sec:polarDef_decay} and above for decaying polarized resonance, with the same support limitations listed in \tab{tb:polarSyntax}.
The syntax for \ms~remains unchanged and ignores polarization information included in production-level Les Houches event files (LHEF).
To clarify, \ms~uses production-level information in the LHEF banner to modify unstable propagators accordingly.
To model the decay of both a polarized or unpolarized $W^+$ boson, one simply uses:
\begin{verbatim}
decay w+ > e+ ve
\end{verbatim}
The  \texttt{\{X\}} command is also supported by \ms~itself.
This allows one to force some particles in a decay chain into a fixed helicity polarization that is defined in the same frame as the original, production-level events.
Such steering can be called using the commands:
\begin{verbatim}
decay t > w+{T} b, w+ > e+ ve
\end{verbatim}
This describes the decay of a top quark $t$ into an unpolarized $b$ quark and a transversely polarized $W^+_T$ boson, 
which in turns decays to electron-flavored leptons.\footnote{While possible, we discourage using polarization features with special modes of \ms. 
For the \texttt{spinmode=none} case (no spin correlation and no off-shell effects), the polarization of particles will be defined in the rest-frame of the primary decay particle.
For \texttt{spinmode=onshell} (no off-shell effect but full spin correlation), 
the frame will be the one associated to the produced event but the phase-space sampling will be optimized according to rest-frame of the primary decay particle.
This can lead to inconsistent results.}

%%%%%%%%%%%%%%%%%%%%%%%%%%%%%%%
\subsection{Event Generation with Polarized Partons}\label{sec:codingBits_evtgen}

As stressed throughout this text, scattering particles with fixed helicity polarizations requires one to fix a reference frame 
in order to meaningfully define individual polarization vectors and spinors.
For LO processes, this is possible at the event-generation level using 
the new ``matrix element frame'' parameter \texttt{me\_frame} in the \texttt{run\_card.dat} steering file.
The parameter is displayed by default only if at least one  massive particle is polarized but can technically be used for any processes.

\begin{table*}[!t]
\begin{center}
\resizebox{\textwidth}{!}{
\begin{tabular}{c|c|c|c||c|c|c|c}
\hline\hline
Process & \texttt{polbeam1} & $\sigma^{\rm Gen.}$ [pb]  & $\sigma^{\rm Expected}$ [pb]  & Process & \texttt{polbeam1} & $\sigma^{\rm Gen.}$ [pb]  & $\sigma^{\rm Expected}$ [pb] \\
\hline\hline
$e^+ e^- \to t t~$ & 0 & 0.1664 		& $\dots$	& 	\multicolumn{4}{c}{$\dots$} \\
\hline
$e^+ e^- \to t t~$ & 100 & 0.2296 	& $\dots$	&	$e^+ e^- \to t t~$ & -100 & 0.1033 	& $\dots$	 \\
\hline
$e^+ e^- \to t t~$ & 25 & 0.1821 & 0.1822 	&	$e^+_R e^- \to t t~$ & 25 & 0.1433 & 0.1435 \\
$e^+ e^- \to t t~$ & 50 & 0.1983 & 0.1980 	&	$e^+_R e^- \to t t~$ & 50 & 0.1719  & 0.1722 \\
$e^+ e^- \to t t~$ & 75 & 0.2137 & 0.2138	&	$e^+_R e^- \to t t~$ & 75 & 0.2008  & 0.2009\\
\hline
$e^+_L e^- \to t t~$ & 0 & 0.1033  	& 0.1033 	& 	$e^+_R e^- \to t t~$ & 0 & 0.2293 & 0.2296\\
$e^+_L e^- \to t t~$ & 100 		& 0 	& 0 	& 	$e^+_R e^- \to t t~$ & 100 & 0.2296 & 0.2296 \\
$e^+_L e^- \to t t~$ & -100 & 0.1036 & 0.1033 	&	$e^+_R e^- \to t t~$ & -100 & 0 		& 0\\
\hline\hline
\end{tabular}
} %% resize
\caption{Cross sections [pb] for the process $e^+ e^- \to t \overline{t}$ at $\sqrt{s}=1000$ GeV,
assuming unpolarized particles, totally and partially polarized beams in the partonic c.m.~frame using the \texttt{polbeam1} steering commands (\texttt{polbeam2=0}),
totally polarized IS particles in the partonic c.m.~frame using the polarization \texttt{\{X\}} syntax, and the anticipated cross section as derived from the \texttt{polbeam1}  results. 
{Here we report a statistical error are at $2.\,10^{-4}$pb.}}
\label{tb:polarEETT}
\end{center}
\end{table*}

For an arbitrary scattering process defined by the \mgamc~syntax
\begin{verbatim}
generate 1 2 > 3 4 ... N
\end{verbatim}
the option to set the frame \texttt{me\_frame} appears in the \texttt{run\_card.dat} file as, 
\begin{Verbatim}[fontsize=\small]
#*********************************************************************
# Frame where to evaluate the matrix element (not the cut!) 
# for particle polarization {X}
#*********************************************************************
  [1,2]  = me_frame ! list of particles to sum-up to define the rest frame
                    ! in which to evaluate the matrix element
\end{Verbatim}
For \texttt{me\_frame = [1,2]}, matrix elements and helicity polarizations are defined in the $(p_1+p_2)$ c.m.~frame,
and is equivalent to setting  \texttt{me\_frame = [3,4,...,N]}, which is also supported.
If, for example, particle \texttt{4} is a massive state, then setting \texttt{me\_frame = [4]} leads to 
evaluating matrix elements and polarizations in the rest frame of particle \texttt{4}.

While the new polarization syntax allows one to simulate a fully polarized beam, it does not support partial beam polarization.
For LO computations, however, support this option is  already available 
via the \texttt{polbeam} entries in the \texttt{run\_card.dat} file:
\begin{Verbatim}[fontsize=\small]
#*********************************************************************
# Beam polarization from -100 (left-handed) to 100 (right-handed)    *
#*********************************************************************
     0     = polbeam1 ! beam polarization for beam 1
     0     = polbeam2 ! beam polarization for beam 2
\end{Verbatim}
Beam polarization tuning in the partonic c.m.~frame remains available and can be used with the new polarization features.
For a comparison we show in \tab{tb:polarEETT} cross sections [pb] for the process $e^+ e^- \to t \overline{t}$ at $\sqrt{s}=1000$ GeV with different polarization configuration of $e^+$.
Polarizations are set either via \texttt{polbeam1} or via the polarization \texttt{\{X\}} syntax, with \texttt{e+\{R\}} and  \texttt{e+\{L\}}.  
$e^-$ is kept unpolarized with \texttt{polbeam2=0}.
In the first line of the table we show the cross section $\sigma^{\rm Gen.}$ obtained assuming unpolarized beams. 
In the second line we show the corresponding rate for a fully polarized $e^+_R$ beam $\sigma_{\rm RH}$ (\texttt{polbeam1=100}) and a fully polarized $e^+_L$ beam $\sigma_{\rm LH}$ (\texttt{polbeam1=-100}).
Other configurations can be extracted by a linear combination of these numbers. 
For example: The unpolarized cross section in the first line is the averaged sum $\sigma_{\rm unpol.}=0.5[\sigma_{\rm RH}+\sigma_{\rm LH}]$.
Likewise, the 25\% RH polarized $e^+$ beam in third row is given by $0.25\sigma_{\rm RH}+0.75\sigma_{\rm unpol.}$.
Cross sections extracted from LH and RH polarizations are displayed as $\sigma^{\rm Expected}$ while the numbers obtained from simulation are displayed as $\sigma^{\rm Gen.}$.
{As expected, rates vanish in the instances where the $e^+$ helicity is fixed via the polarization \texttt{\{X\}} but the beam is polarized with the opposite helicity.}

%%%%%%%%%%%%%%%%%%%%%%%%%%%%%%%
\subsection{Event Re-weighting for Arbitrary Reference Frames}\label{sec:codingBits_reweight}

\begin{table*}[!t]
\begin{center}
\resizebox{\textwidth}{!}{
\begin{tabular}{c || l}
\hline\hline
Attribute & Description \\ \hline
{\bf status} 	&	Returns $-1~(1)~[2]$ for an initial (final) [intermediate] state particle \\
{\bf mother1}	&	Returns the first progenitor particle object.	\\
{\bf mother2}	&	Returns the second progenitor particle object if both progenitors have {\bf status}$=-1$,\\
			&	otherwise returns {\bf mother1}. \\
{\bf color1}	&	 First color index for the leading color associated to the particle \\
{\bf color2}	&	 Second color index for the leading color associated to the particle 	\\
{\bf px}		&	 $p^x$ component of the momenta*	\\
{\bf py}		&	 $p^y$ component of the momenta*	\\
{\bf pz}		&	 $p^z$ component of the momenta*	\\
{\bf E}		&	 $p^0$ component of the momenta*	\\
{\bf mass}		&	 Invariant mass of the particle	\\
{\bf vtim}		&	 Displaced vertex information	\\
{\bf helicity}	&	 Helicity polarization*		\\
\hline\hline
\end{tabular}
} %% resize
\caption{
List of common Les Houches event file attributes available to the \texttt{Re-Weighting} module; see Ref.~\cite{Alwall:2006yp} for further details.
* denotes that the quantity is defined in the lab frame by default.
}
\label{tb:reweightSyntax}
\end{center}
\end{table*}

A key feature of the \texttt{Re-Weighting} module~\cite{Mattelaer:2016gcx} in \mgamc~is the ability to take an event sample defined by one process definition
and, within reason, generate a new event sample defined by a second process definition through matrix element re-weighting.
It is therefore also possible to use new the polarization syntax in conjunction with the \texttt{Re-Weighting} module, allowing one to study the impact of polarization via re-weighting methods.

In order to have meaningful helicity polarizations one needs a definite reference frame as in previous considerations.
By default, the \texttt{Re-Weighting} module will use
 the frame defined in the \texttt{run\_card.dat} file but also allows a user to define an alternative frame.
However, since the module can interface with a generic Les Houches event file \cite{Alwall:2006yp,Alwall:2007mw,Andersen:2014efa}, 
we have designed a specific syntax for building new frames.
The user must simply provide a \texttt{python}-based lambda function that selects the particles to include in the Lorentz-boost definition.
 The fundamental ideas and effects are the same as simulating polarized particle scattering following \app{sec:codingBits_syntax};
 only the procedure for defining a reference frame differs.
Particles whose momentum are to be included in the frame definition can be identified through any of the preexisting Les Houches event file attributes~\cite{Alwall:2006yp}.
A list of common attributes that can be used is given in \tab{tb:reweightSyntax}.
Some examples (and their impact) include:
\begin{Verbatim}[fontsize=\small]
change boost True # use to lab-frame
change boost lambda p: p.status==-1 # go to partonic-center-of mass frame
change boost lambda p: p.pid in [24,-24] # go to the ww rest-frame
\end{Verbatim}

%%=====================================================================
\bibliography{mgPolarization_refs}

%%=====================================================================
\end{document}